\documentclass[bibyear]{aa}

\usepackage{aas_macros}

\usepackage{rotating}
\usepackage[section]{placeins}
\usepackage{pdfpages}
\usepackage{array}
\usepackage[T1]{fontenc}
\usepackage{ae,aecompl}

\DeclareOldFontCommand{\sf}{\normalfont\sffamily}{\mathsf} 

\usepackage{color}

\DeclareGraphicsExtensions{.eps,.png,.pdf}

\usepackage{mathtools}

\usepackage{graphicx}	
\usepackage{amsmath}	
\usepackage{amssymb}	

\usepackage{adjustbox}
\usepackage{pdflscape}
\usepackage{orcidlink}

\newcommand{\myemail}{doris.stoppacher@us.es}

\newcommand{\PUC}{Instituto de Astrof\'{i}sica, Pontificia Universidad Cat\'{o}lica de Chile, Campus San Joaqu\'{i}n, Avda. Vicu\~{n}a Mackenna 4860, Santiago, Chile}

\newcommand{\AIPUC}{Centro de AstroIngenier\'ia, Pontificia Universidad Cat\'{o}lica de Chile, Campus San Joaqu\'{i}n, Avda. Vicu\~{n}a Mackenna 4860, Santiago, Chile}

\newcommand{\UAM}{Departamento de F\'{i}sica Te\'{o}rica, M\'{o}dulo 15, Facultad de Ciencias, Universidad Aut\'{o}noma de Madrid, Cantoblanco, 28049 Madrid, Spain}

\newcommand{\US}{Facultad de F\'{i}sicas, Universidad de Sevilla, Campus de Reina Mercedes,  Avda.\ Reina Mercedes s/n, 41012 Sevilla, Spain}

\newcommand{\DIPC}{Donostia International Physics Centre (DIPC), Paseo Manuel de Lardizabal 4, 20018 Donostia-San Sebastian, Spain}

\usepackage{epsfig}
\usepackage{epstopdf} 
\usepackage{fancyhdr}
\usepackage{multirow} 
\usepackage{multicol}
\usepackage{rotating}
\usepackage{url,times,amsfonts,color}
\usepackage{txfonts,float}
\usepackage{titlesec}

\newcolumntype{M}[1]{>{\centering\arraybackslash}m{#1}}
\usepackage{float}
\usepackage[nodayofweek,level]{datetime}
\usepackage{xspace}

\usepackage{blindtext} 				
\usepackage{lineno}					

\usepackage{epigraph}				
\usepackage{parskip}  
\usepackage{breakurl}

\usepackage{natbib,twoopt}
\bibpunct{(}{)}{;}{a}{}{,}             
\makeatletter
  \newcommandtwoopt{\citeads}[3][][]{\href{http://adsabs.harvard.edu/abs/#3}%
    {\def\hyper@linkstart##1##2{}%
     \let\hyper@linkend\@empty\citealp[#1][#2]{#3}}}
  \newcommandtwoopt{\citepads}[3][][]{\href{http://adsabs.harvard.edu/abs/#3}%
    {\def\hyper@linkstart##1##2{}%
     \let\hyper@linkend\@empty\citep[#1][#2]{#3}}}
  \newcommandtwoopt{\citetads}[3][][]{\href{http://adsabs.harvard.edu/abs/#3}%
    {\def\hyper@linkstart##1##2{}%
     \let\hyper@linkend\@empty\citet[#1][#2]{#3}}}
  \newcommandtwoopt{\citeyearads}[3][][]%
    {\href{http://adsabs.harvard.edu/abs/#3}
    {\def\hyper@linkstart##1##2{}%
     \let\hyper@linkend\@empty\citeyear[#1][#2]{#3}}}
\makeatother

\newcommand{\eagle}{\texttt{EAGLE}\xspace}
\newcommand{\IllTNG}{\texttt{Illustris}-\textsc{TNG}\xspace}

\newcommand{\lsb}{\texttt{LSB}\xspace}
\newcommand{\lsbs}{\texttt{LSBs}\xspace}
\newcommand{\hsb}{\texttt{HSB}\xspace}
\newcommand{\hsbs}{\texttt{HSBs}\xspace}

\newcommand{\logT}{{\ifmmode{\log_{10}}\else{$\log_{10}$\xspace}\fi}}

\newcommand{\Mbnd}{{\ifmmode{M_{\mathrm{bnd}}}\else{$M_{\mathrm{bnd}}$}\fi}}
\newcommand{\Mfof}{{\ifmmode{M_{\mathrm{fof}}}\else{$M_{\mathrm{fof}}$}\fi}}
\newcommand{\Mmean}{{\ifmmode{M_{\mathrm{200m}}}\else{$M_{\mathrm{200m}}$}\fi}}
\newcommand{\MBN}{{\ifmmode{M_{\mathrm{BN98}}}\else{$M_{\mathrm{BN98}}$}\fi}}
\newcommand{\Mc}{{\ifmmode{M_{\mathrm{200c}}}\else{$M_{\mathrm{200c}}$\xspace}\fi}}
\newcommand{\Mvir}{{\ifmmode{M_{\mathrm{vir}}}\else{$M_{\mathrm{vir}}$\xspace}\fi}}

\newcommand{\rc}{{\ifmmode{R_{\mathrm{200c}}}\else{$R_{\mathrm{200c}}$\xspace}\fi}}
\newcommand{\Vz}{{\ifmmode{V_{\mathrm{eff}}}\else{$V_{\mathrm{eff}}$\xspace}\fi}}

\newcommand{\ltsima}{$\; \buildrel < \over \sim \;$}
\newcommand{\gtsima}{$\; \buildrel > \over \sim \;$}
\newcommand{\lsim}{\lower.5ex\hbox{\ltsima}}
\newcommand{\gsim}{\lower.5ex\hbox{\gtsima}}

\newcommand{\Mstar}{{\ifmmode{M_*}\else{$M_*$\xspace}\fi}}
\newcommand{\Mhalo}{{\ifmmode{M_{\mathrm{Halo}}}\else{$M_{\mathrm{Halo}}$\xspace}\fi}}
\newcommand{\Mhot}{{\ifmmode{M_{\mathrm{Hot}}}\else{$M_{\mathrm{Hot}}$\xspace}\fi}}
\newcommand{\Mcold}{{\ifmmode{M_{\mathrm{Cold}}}\else{$M_{\mathrm{Cold}}$\xspace}\fi}}
\newcommand{\Mzgas}{{\ifmmode{M_{\mathrm{Z_{cold}}}}\else{$M_{\mathrm{Z_{cold}}}$\xspace}\fi}}
\newcommand{\Mbh}{{\ifmmode{M_{\mathrm{BH}}}\else{$M_{\mathrm{BH}}$\xspace}\fi}}

\newcommand{\cgf}{{\ifmmode{M_{\mathrm{Cold}}/M_{\mathrm{*}}}\else{$M_{\mathrm{Cold}}/M_{\mathrm{*}}$\xspace}\fi}}
\newcommand{\bheffc}{{\ifmmode{\Mbh/\Mc}\else{$\Mbh/\Mc$\xspace}\fi}}
\newcommand{\bheff}{{\ifmmode{\Mbh/\Mvir}\else{$\Mbh/\Mvir$\xspace}\fi}}
\newcommand{\fatom}{{\ifmmode{f_{\mathrm{atom}}}\else{$f_{\mathrm{atom}}$\xspace}\fi}}

\newcommand{\MHI}{{\ifmmode{M_{\mathrm{H_{I}}}}\else{$M_{\mathrm{H_{I}}}$\xspace}\fi}}
\newcommand{\MHII}{{\ifmmode{M_{\mathrm{H_{II}}}}\else{$M_{\mathrm{H_{II}}}$\xspace}\fi}}
\newcommand{\MHTwo}{{\ifmmode{M_{\mathrm{H_{2}}}}\else{$M_{\mathrm{H_{2}}}$\xspace}\fi}}
\newcommand{\Mgas}{{\ifmmode{M_{\mathrm{gas}}}\else{$M_{\mathrm{gas}}$\xspace}\fi}}
\newcommand{\Mgaskpc}{{\ifmmode{M_{\mathrm{gas,30kpc}}}\else{$M_{\mathrm{gas,30kpc}}$\xspace}\fi}}
\newcommand{\Mgasopt}{{\ifmmode{M_{\mathrm{gas,1.5R_{opt}}}}\else{$M_{\mathrm{gas,1.5R_{opt}}}$\xspace}\fi}}
\newcommand{\MHIHTwo}{{\ifmmode{M_{\mathrm{H_I+H_{2},30kpc}}}\else{$M_{\mathrm{H_I+H_{2},30kpc}}$\xspace}\fi}}

\newcommand{\vrot}{{\ifmmode{v_{\mathrm{rot}}}\else{$v_{\mathrm{rot}}$\xspace}\fi}}
\newcommand{\vdisp}{{\ifmmode{\sigma}\else{$\sigma$\xspace}\fi}}
\newcommand{\vmax}{{\ifmmode{v_{\mathrm{max}}}\else{$v_{\mathrm{max}}$\xspace}\fi}}
\newcommand{\vc}{{\ifmmode{v_{\mathrm{200c}}}\else{$v_{\mathrm{200c}}$\xspace}\fi}}
\newcommand{\vsat}{{\ifmmode{v_{\mathrm{max_{\mathrm{sat}}}}}\else{$v_{\mathrm{max_{\mathrm{sat}}}}$\xspace}\fi}}

\newcommand{\Zcold}{{\ifmmode{Z_{\mathrm{Cold}}}\else{$Z_{\mathrm{Cold}}$\xspace}\fi}}
\newcommand{\DHI}{{\ifmmode{D_{\mathrm{H_{I}}}}\else{$D_{\mathrm{H_{I}}}$\xspace}\fi}} 

\newcommand{\HI}{{\ifmmode{H_{\mathrm{I}}}\else{$H_{\mathrm{I}}$\xspace}\fi}}
\newcommand{\HTwo}{{\ifmmode{H_{\mathrm{2}}}\else{$H_{\mathrm{2}}$\xspace}\fi}}

\newcommand{\rhalfstars}{{\ifmmode{R_{\mathrm{1/2,*}}}\else{$R_{\mathrm{1/2,*}}$\xspace}\fi}}
\newcommand{\rhalfmass}{{\ifmmode{R_{\mathrm{1/2,DM}}}\else{$R_{\mathrm{1/2,DM}}$\xspace}\fi}}
\newcommand{\rhalfgas}{{\ifmmode{R_{\mathrm{1/2,gas}}}\else{$R_{\mathrm{1/2,gas}}$\xspace}\fi}}
\newcommand{\rhalfgasproj}{{\ifmmode{R_{\mathrm{1/2,gas,2D}}}\else{$R_{\mathrm{1/2,gas,2D}}$\xspace}\fi}}
\newcommand{\ropt}{{\ifmmode{R_{\mathrm{opt}}}\else{$R_{\mathrm{opt}}$\xspace}\fi}}
\newcommand{\roptc}{{\ifmmode{R^{\dagger}_{\mathrm{opt}}}\else{$R^{\dagger}_{\mathrm{opt}}$\xspace}\fi}}
\newcommand{\onefiveropt}{{\ifmmode{\mathrm{1.5}\timesR_{\mathrm{opt}}}\else{$\mathrm{1.5}R_{\mathrm{opt}}$\xspace}\fi}}
\newcommand{\rhalfstopt}{{\ifmmode{R_{\mathrm{1/2,*,1.5}R_{\mathrm{opt}}}}\else{$R_{\mathrm{1/2,*,1.5}R_{\mathrm{opt}}}$\xspace}\fi}}
\newcommand{\rhalfstkpc}{{\ifmmode{R_{\mathrm{1/2,*,30kpc}}}\else{$R_{\mathrm{1/2,*,30kpc}}$\xspace}\fi}}
\newcommand{\rhalfstkpcproj}{{\ifmmode{R_{\mathrm{1/2,*,30kpc,2D}}}\else{$R_{\mathrm{1/2,*,30kpc,2D}}$\xspace}\fi}}
\newcommand{\reffgasopt}{{\ifmmode{R_{\mathrm{eff,gas,1.5}R_{\mathrm{opt}}}}\else{$R_{\mathrm{eff,gas,1.5}R_{\mathrm{opt}}}$\xspace}\fi}}
\newcommand{\reffdiskopt}{{\ifmmode{R_{\mathrm{eff,gas,disk}}}\else{$R_{\mathrm{eff,gas,disk,1.5}R_{\mathrm{opt}}}$\xspace}\fi}}
\newcommand{\rvmax}{{\ifmmode{R_{\vmax}}\else{$R_{\vmax}$\xspace}\fi}}
\newcommand{\rbulgevsdisk}{{\ifmmode{R_{\mathrm{bulge/disk}}}\else{$R_{\mathrm{bulge/disk}}$\xspace}\fi}}

\newcommand{\VvStwoR}{{\ifmmode{(\vrot/\vdisp)_{2\rhalfstars}}\else{$(\vrot/\vdisp)_{2\rhalfstars}$\xspace}\fi}}

\newcommand{\DvTstars}{{\ifmmode{D/T_{\mathrm{*}}}\else{$D/T_{\mathrm{*}}$\xspace}\fi}}
\newcommand{\DvTgas}{{\ifmmode{D/T_{\mathrm{Gas}}}\else{$D/T_{\mathrm{Gas}}$\xspace}\fi}}

\newcommand{\MB}{{\ifmmode{M_B}\else{$M_B$\xspace}\fi}}

\newcommand{\kpc}{{\ifmmode{\mathrm{kpc}}\else{$\mathrm{kpc}$\xspace}\fi}}
\newcommand{\Mpc}{{\ifmmode{\mathrm{Mpc}}\else{$\mathrm{Mpc}$\xspace}\fi}}
\newcommand{\Gpc}{{\ifmmode{\mathrm{Gpc}}\else{$\mathrm{Gpc}$\xspace}\fi}}

\newcommand{\hkpc}{{\ifmmode{h^{-1}\mathrm{kpc}}\else{$h^{-1}\mathrm{kpc}$\xspace}\fi}}
\newcommand{\hMpc}{{\ifmmode{h^{-1}\mathrm{Mpc}}\else{$h^{-1}\mathrm{Mpc}$\xspace}\fi}}
\newcommand{\hGpc}{{\ifmmode{h^{-1}\mathrm{Gpc}}\else{$h^{-1}\mathrm{Gpc}$\xspace}\fi}}

\newcommand{\MpcCu}{{\ifmmode{\mathrm{Mpc}^3}\else{$\mathrm{Mpc}^3$\xspace}\fi}}
\newcommand{\MpcV}{{\ifmmode{\mathrm{Mpc}^{-3}}\else{$\mathrm{Mpc}^{-3}$\xspace}\fi}}

\newcommand{\hMsun}{{\ifmmode{h^{-1}\mathrm{M_{\odot}}}\else{$h^{-1}\mathrm{M_{\odot}}$}\fi}}
\newcommand{\Msun}{{\ifmmode{\mathrm{M_{\odot}}}\else{$\mathrm{M_{\odot}}$\xspace}\fi}}

\newcommand{\Msunyr}{{\ifmmode{\mathrm{M_{\odot}yr^{-1}}}\else{$\mathrm{M_{\odot}yr^{-1}}$\xspace}\fi}}
\newcommand{\Gyr}{{\ifmmode{\mathrm{Gyr}}\else{$\mathrm{Gyr}$\xspace}\fi}}
\newcommand{\yr}{{\ifmmode{\mathrm{yr}}\else{$\mathrm{yr}$\xspace}\fi}}
\newcommand{\yrmo}{{\ifmmode{\mathrm{yr}^{-1}}\else{$\mathrm{yr}^{-1}$\xspace}\fi}}

\newcommand{\kms}{{\ifmmode{\mathrm{kms}^{-1}}\else{$\mathrm{kms}^{-1}$\xspace}\fi}}
\newcommand{\Zsolar}{{\ifmmode{\mathrm{Z}_{\odot}}\else{$\mathrm{Z}_{\odot}$\xspace}\fi}}

\newcommand{\magarcs}{{\ifmmode{\mathrm{mag~arcsec}^{-2}}\else{$\mathrm{mag~arcsec}^{-2}$\xspace}\fi}}

\newcommand{\Mr}{{\ifmmode{M_{\mathrm{r}}}\else{$M_{\mathrm{r}}$\xspace}\fi}}
\newcommand{\rp}{{\ifmmode{{{r_{\mathrm{p}}}}\else{${{r_{\mathrm{p}}}}$\xspace}\fi}}
\newcommand{\pim}{{\ifmmode{{{$\pi$_{\mathrm{max}}}}}\else{${{\pi_{\mathrm{max}}}}}$\xspace}\fi}}

\newcommand{\Mcut}{{\ifmmode{M_{\mathrm{cut}}}\else{$M_{\mathrm{cut}}$\xspace}\fi}}
\newcommand{\Mmin}{{\ifmmode{M_{\mathrm{min}}}\else{$M_{\mathrm{min}}$\xspace}\fi}}
\newcommand{\Muno}{{\ifmmode{M_{\mathrm{1}}}\else{$M_{\mathrm{1}}$\xspace}\fi}}
\newcommand{\sigM}{{\ifmmode{\sigma_{\log_{\mathrm{10}}\Mstar}}\else{$\sigma_{\log_{\mathrm{10}} \Mstar}$\xspace}\fi}}
\newcommand{\con}{{\ifmmode{C_{\mathrm{NFW}}}\else{$C_{\mathrm{NFW}}$\xspace}\fi}}

\newcommand{\jstaropt}{{\ifmmode{j_{\mathrm{*}}}\else{$j_{\mathrm{*}}$\xspace}\fi}}
\newcommand{\jgasropt}{{\ifmmode{j_{\mathrm{gas}}}\else{$j_{\mathrm{gas}}$\xspace}\fi}}

\newcommand{\tempSF}{{\ifmmode{T_{\mathrm{SF}}}\else{$T_{\mathrm{SF}}$\xspace}\fi}}
\newcommand{\tempNSF}{{\ifmmode{T_{\mathrm{NSF}}}\else{$T_{\mathrm{NSF}}$\xspace}\fi}}

\newcommand{\SBgas}{{\ifmmode{\mu_{\mathrm{eff,gas,disk,1.5R_{opt}}}}\else{$\mu_{\mathrm{eff,gas,disk,1.5R_{opt}}}$\xspace}\fi}}
\newcommand{\SBopt}{{\ifmmode{\mu_{\mathrm{opt}}}\else{$\mu_{\mathrm{opt}}$\xspace}\fi}}
\newcommand{\SBstopt}{{\ifmmode{\mu_{\mathrm{eff,*,1.5R_{opt}}}}\else{$\mu_{\mathrm{eff,*,1.5R_{opt}}}$\xspace}\fi}}
\newcommand{\SBstkpc}{{\ifmmode{\mu_{\mathrm{eff,*,30kpc}}}\else{$\mu_{\mathrm{eff,*,30kpc}}$\xspace}\fi}}
\newcommand{\SBB}{{\ifmmode{\mu_{\mathrm{B}}}\else{$\mu_{\mathrm{B}}$\xspace}\fi}}
\newcommand{\SBr}{{\ifmmode{\mu_{\mathrm{r}}}\else{$\mu_{\mathrm{r}}$\xspace}\fi}}
\newcommand{\mSBB}{{\ifmmode{\langle\mu_{B}\rangle}\else{$\langle\mu_{B}\rangle$\xspace}\fi}}
\newcommand{\mSBm}{{\ifmmode{\langle\mu_{m}\rangle}\else{$\langle\mu_{m}\rangle$\xspace}\fi}}

\newcommand{\lamst}{{\ifmmode{\lambda_{\rhalfstars}}\else{$\lambda_{R\rhalfstars}$\xspace}\fi}}
\newcommand{\lamTwost}{{\ifmmode{\lambda_{2\rhalfstars}}\else{$\lambda_{2\rhalfstars}$\xspace}\fi}}
\newcommand{\ellst}{{\ifmmode{\epsilon_{\rhalfstars}}\else{$\epsilon_{\rhalfstars}$\xspace}\fi}}
\newcommand{\ellTwost}{{\ifmmode{\epsilon_{2\rhalfstars}}\else{$\epsilon_{2\rhalfstars}$\xspace}\fi}}

\newcommand{\thalf}{{\ifmmode{t_{\text{50\%}}}\else{$t_{\text{50\%}}$\xspace}\fi}}
\newcommand{\tseventy}{{\ifmmode{t_{\text{70\%}}}\else{$t_{\text{70\%}}$\xspace}\fi}}

\newcommand{\tLB}{{\ifmmode{\bar t_{\text{LB}}}\else{$\bar t_{\text{LB}}$\xspace}\fi}}

\newcommand{\thalfhaloc}{{\ifmmode{t_{\text{50\%,halo}}}\else{$t_{\text{50\%,halo}}$\xspace}\fi}}
\newcommand{\thalfstars}{{\ifmmode{t_{\text{50\%,*}}}\else{$t_{\text{50\%,*}}$\xspace}\fi}}
\newcommand{\thalfbh}{{\ifmmode{t_{\text{50\%,BH}}}\else{$t_{\text{50\%,BH}}$\xspace}\fi}}

\newcommand{\delhaloc}{{\ifmmode{t_{\text{50\%-70\%,halo}}}\else{$t_{\text{50\%-70\%,halo}}$\xspace}\fi}}
\newcommand{\delstars}{{\ifmmode{t_{\text{50\%-70\%,*}}}\else{$t_{\text{50\%-70\%,*}}$\xspace}\fi}}
\newcommand{\delbh}{{\ifmmode{t_{\text{50\%-70\%,BH}}}\else{$t_{\text{50\%-70\%,BH}}$\xspace}\fi}}

\newcommand{\delstha}{{\ifmmode{t_{\text{50\%,halo}}-t_{\text{50\%,*}}}\else{$t_{\text{50\%,halo}}-t_{\text{50\%,*}}$\xspace}\fi}}
\newcommand{\delbhst}{{\ifmmode{t_{\text{50\%,*}}-t_{\text{50\%,BH}}}\else{$t_{\text{50\%,*}}-t_{\text{50\%,BH}}$\xspace}\fi}}
\newcommand{\delbhha}{{\ifmmode{t_{\text{50\%,halo}}-t_{\text{50\%,BH}}}\else{$t_{\text{50\%,halo}}-t_{\text{50\%,BH}}$\xspace}\fi}}

\newcommand{\agest}{{\ifmmode{\mathrm{age}_{\rhalfstars}}\else{$\mathrm{age}_{\rhalfstars}$\xspace}\fi}}

\newcommand{\bhacc}{{\ifmmode{\dot{M}_{\mathrm{BH}}}\else{$\dot{M}_{\mathrm{BH}}$\xspace}\fi}}

\newcommand{\Deljstars}{{\ifmmode{\Delta_{\textrm{max-min,j}_*}}\else{$\Delta_{\textrm{max-min,j}_*}$\xspace}\fi}}

\newcommand{\zjstars}{{\ifmmode{z^{\star}_{j_*}}\else{$z^{\star}_{j_*}$\xspace}\fi}}
\newcommand{\zrhalf}{{\ifmmode{z^{\dagger}_{\rhalfstars}}\else{$z^{\dagger}_{\rhalfstars}$\xspace}\fi}}
\newcommand{\zSBr}{{\ifmmode{z^{\dagger}_{\SBr}}\else{$z^{\dagger}_{\SBr}$\xspace}\fi}}
\newcommand{\ztempSF}{{\ifmmode{z^{\ddagger}_{\tempSF}}\else{$z^{\ddagger}_{\tempSF}$\xspace}\fi}}
\newcommand{\zrvmax}{{\ifmmode{z^{\dagger}_{\rvmax}}\else{$z^{\dagger}_{\rvmax}$\xspace}\fi}}
\newcommand{\zSFR}{{\ifmmode{z^{*}_{\mathrm{SFR}}}\else{$z^{*}_{\mathrm{SFR}}$\xspace}\fi}}

\def\lesssim{\mathrel{\hbox{\rlap{\hbox{\lower4pt\hbox{$\sim$}}}\hbox{$<$}}}}
\def\gtrsim{\mathrel{\hbox{\rlap{\hbox{\lower4pt\hbox{$\sim$}}}\hbox{$>$}}}}

\newcommand{\Tab}[1]{Table~\ref{#1}}
\newcommand{\Sec}[1]{Section~\ref{#1}}

\newcommand{\Eq}[1]{Eq.~(\ref{#1})}

\newcommand{\Fig}[1]{Fig.~\ref{#1}}

\newcommand{\beq}{\begin{equation}}
\newcommand{\eeq}{\end{equation}}

\def\beqa{\begin{eqnarray}}
\def\eeqa{\end{eqnarray}}
\def\Mpc{$\mathrm{Mpc}$}
\def\kpc{$\mathrm{kpc}$}
\def\hMpc{$h^{-1}\,\mathrm{Mpc}$}

\begin{document}

\title{Hidden Figures in the Sky}

\subtitle{Evolution of low surface brightness galaxies from a hydro-dynamical perspective}

\author{D.\ Stoppacher\orcidlink{0000-0002-3281-9956}\inst{1}\fnmsep\inst{2}\fnmsep\inst{3}\fnmsep\inst{4}\fnmsep\thanks{\email{\myemail}}\and
        P.\ Tissera\orcidlink{0000-0001-5242-2844}\inst{2}\fnmsep\inst{3}\and
        Y.\ Rosas-Guevara\orcidlink{0000-0003-2579-2676}\inst{5}\and
        G.\ Galaz\orcidlink{0000-0002-8835-0739}\inst{2}\and
        J.\ O\~norbe\orcidlink{0000-0002-8723-1180}\inst{4}
      }
\institute{\UAM\
\and\PUC\
\and\AIPUC\
\and\US\
\and\DIPC}

\date{Received Month Day, 202x; accepted Month Day, 202x}

\abstract
{Low-surface brightness galaxies (LSBGs) are defined as galaxies with central surface brightness levels fainter than the night sky, making them challenging to observe. A key open question is whether their faint appearance arises from intrinsic properties or from stochastic events in their formation histories.}
{We aim to trace the formation histories of LSBGs to assess whether their evolutionary paths differ from those of high-surface brightness galaxies (HSBGs), and to identify the key physical drivers behind these differences.}
{We present a fast and efficient method to estimate stellar surface brightness densities in hydro-dynamical simulations and a statistically robust exploration of over 150 properties in the reference run \textsc{Ref-L0100N1504} of the \eagle\ simulation. To minimise biases, we carefully match the stellar and halo mass distributions of the selected LSB and HSB samples.}
{At $z=0$, LSBGs are typically extended, rotation-supported systems with lower stellar densities, older stellar populations, reduced star formation activity, and higher specific stellar angular momenta ($j_*$) than their HSBG counterparts. They also exhibit larger radii of maximum circular velocity (\rvmax). We identify key transition redshifts that mark the divergence of LSBG and HSBG properties: $j_*$ diverges at $z\sim5$–7 and \rvmax\ at $z\sim2$–3. Star formation activity and large-scale environment seems to play only a minimal role in the development of LSB features.}
{LSBGs follow mass-dependent evolutionary pathways, where early rapid formation and later slowdowns, combined with their distinct structural properties, influence their response to external factors like mergers and gas accretion. Their LSB nature emerges from intrinsic dynamical and structural factors rather than environmental influences, with angular momentum as a key driver of divergence at high redshifts.}

\maketitle
\nolinenumbers    
\section{Introduction} \label{sec:introduction}

In the late 70s, researchers predicted the existence of a population of galaxies with surface brightness below of the night sky with $\sim21.65$ mag arcsec$^{-2}$\footnote{This threshold has been pushed towards higher numbers thanks to the advancement in technology and observation strategies in recent years.} in the $B$-band \citet{Freeman70_LSBGs} of the Johnson-Morgan photometric system \citep{Johnson+Morgan53_UBV}. Those galaxies are referred to as "low-surface brightness galaxies (LSBGs)" \citep{Bothun97_LSBGs_rev}. Thanks to advancements in astronomical instrumentation, the existence of LSBGs was finally confirmed in large numbers more than two decades later \citep[e.g.][]{Bothun90_LSBGs_Malin2,Trachternach06_LSBGs} fundamentally altering our understanding of galaxies and their formation and evolution. LSBGs represent a unique population of galaxies that are challenging to observe due to their faint luminosities and susceptibility to observational selection effects \citep[see][and citation therein]{Tanoglidis21_LSBGs_DES}. These systems are thought to contain the largest baryon reservoirs in the Universe and may even dominate the volume density of galaxies \citep{ONeil+Bothun00}. Understanding the conditions under which LSBGs form and evolve is crucial for improving our knowledge of late-time galaxy formation. It is suspected that a significant fraction of galaxies belong to this population \citep{Dalcanton97_LSBGs_ndensity,Martin19_LSBGs} and that they account for about $\sim$15\% of the total dynamical mass density today \citep{Driver99,Minchin04_LSBGs_cos}\footnote{Dynamical mass refers to the total mass of a galaxy that is derived from its rotational curve or velocity dispersion. This mass includes baryonic matter and dark matter.}. LSBGs are considered distinct laboratories offering extraordinary potential to simultaneously test the properties of dark and baryonic matter in a cosmological context \citep{DiPaolo+Salucci20_rev}.

We now understand that LSBGs exhibit a wide range of sizes and shapes, from giant spirals and irregular galaxies to faint dwarf galaxies \citet[see e.g.\ the review by][]{Das13_rev}. This diversity makes it even more challenging to construct a consistent model of their formation and evolution. The most significant difference between LSBGs and high-surface brightness galaxies (HSBGs), apart from their stellar densities, is that LSBGs have gas discs with low-surface densities, approximately $\sim 5~M_{\odot}\mathrm{pc}^{-2}$ or even lower, and can exhibit up to 50 times higher neutral gas-to-luminosity ratios in the $B$-band than their HSBG counterpart \citep[e.g][]{vanDerHulst93_LSBGs_SFRs-HI}. Additionally, LSBGs are diffuse, with an extreme luminosity-size relation -- some ultra-diffuse LSBGs can even have sizes exceeding 1.5 kpc while possessing the stellar mass content of a dwarf spiral \citep[see e.g.][and citations therein]{Carleton19_UDGs}. Dwarf and irregular LSBGs make up the majority of this population, whereas disc-LSBGs are less common. Dwarf-LSBGs are thought to dominate the faint end of the luminosity function in the local Universe \citep{Geller12_LSBGs} while more diffuse LSBGs are typically found in under-dense environments \citep{Rosenbaum09_LSBGs_env,Ceccarelli12_LSBGs_voids}. Disc-LSBGs often have a prominent bulge, which can lead to confusion with early-type galaxies, although their correct morphological type is a late-type spiral. The first giant LSBG detected was \texttt{Malin 1} with a stellar mass of $\Mstar\sim10^{11}$ \Msun\ and an extraordinary disc size of approximately 200 kpc in diameter \citep{Bothun87_LSBGs,Pickering97_Malin1,Galaz15_Malin1,Galaz22_Malin1,Johnston24_Malin1_center}.

Despite their diversity, LSBGs share several common characteristics such as lower star formation rates (SFRs) of around $\sim 0.1~M_{\odot}yr^{-1}$ at low redshifts \citep{vanDerHulst93_LSBGs_SFRs-HI,vanZee97_LSBGs_gas,vanDenHoek00_LSBGs_SFH,Schombert11_LSBGs_obs,Lei19_LSBGs_SFR}. However, H$\alpha$ observations of larger LSB disc galaxies, particularly giant LSBGs, reveal localised regions of active star formation \citep[e.g.][]{Auld06_LSBGs_SFR}. They also exhibit lower metallicities of the gas \citep[e.g.\ oxygen and carbon monoxide (CO) abundances in the interstellar medium, see ][]{deBlock+vanDerHulst98_OII,deBlock+vanDerHulst98_CO,KuzioDeNaray04_LSBGs_OII}, diffuse stellar discs, extended HI gas disc \citep{Impey+Bothun97_LSBGs_rev}, high gas fraction and substantial total HI-masses, \citep{Burkholder01_L+HSBGs_LC,ONeil04_massive_LSBGs,Huang14_LSBGs_HI,Du15_LSBGs_obs}, and low dust content \citep{Hinz07_LSBGs_obs,Rahman07_GLSBGs_Malin1}. However,  molecular hydrogen (H$_2$) is found to be less abundant \citep{vanDerHulst93_LSBGs_SFRs-HI,Braine00_LSBGs_CO}. The scarcity of both H$_2$ and dust explains the low SFRs in giant LSBGs and contributes to their slow evolution, as these components are crucial for gas cooling \citep{Das13_rev}.

In addition, LSBGs are suggested to be dark matter-dominated galaxies \citep{Lee04_LSBGs,Chequers16_LSBGs_DM,Perez-Montano+Cervantes-Sodi19_LSBGs_env} typically exhibiting slowly rising rotation curves. Some authors propose that LSBGs form at the centres of high-angular momentum dark matter haloes, as indicated by their spin parameter $\lambda$ \citep{Jimenez98_LSBGs_evol-halo,Hernandez+Gilmor98_LSBGs_rot,Verheijen+deBlok00_LSBGs_kin,deBlok01b_LSBGs}. This hypothesis is supported by measurements showing a median spin parameter of $\lambda=0.060$ for LSBGs -- 40\% higher than that of "normal" galaxies in a similar stellar mass range \citep{Zhu23_TNG100_gLSBGs}. In this scenario, the proto-galaxy acquires its angular momentum through tidal torques exerted by neighbouring over-densities in the early Universe \citep{Hoyle51_TTT,Peebles69}. Within this framework, high-angular momentum influences properties such as the scale length of the discs, their stellar surface density, as well as their colour, thickness of the disc, and bulge-to-disc mass ratio \citep{Fall+Efstathiou80_angMom,Hernandez+Cervantes-Sodi06_lambda,Pedrosa+Tissera15_angMom}.

Some authors claim that massive LSBGs are typically found in isolation or under-dense large-scale environments such as voids \citep{Ceccarelli12_LSBGs_voids,Peper21_LSBGs_voids}. Their nearest neighbours are typically found at distances 1.7 times greater than those of HSBGs \citep{Bothun93_LSBGs_env}. This finding is supported by the lower clustering amplitude of LSBGs compared to a reference sample \citep{Mo94_LSBGs_clustering,Tanoglidis21_LSBGs_DES}. More recent studies have reported a scarcity of companions for LSBGs on scales less than 2 Mpc, in contrast to their HSB counterparts $<2$Mpc \citep{Galaz11_LSBGs_SDSS}. However, some researchers argue that the environment plays only a secondary role in the formation of LSBGs, with their dynamical state (e.g., pressure- or rotations-supported systems) and interaction and merger with companions, being the primary drivers \citep{Shao15_LSBGs_env}. \citet{Perez-Montero24_LSBGs_envr_angM} studied the large-scale environmental effects on LSBGs using the \texttt{IllustrisTNG} simulation \citep{Nelson19_Illustris-TNG}. Their findings revealed no significant differences in halo concentration between LSBGs and HSBGs, nor clear indication that the large-scale environment influences their evolution. Instead, their study suggests that local environmental factors such as mergers, gas and stellar mass accretion, and tidal stripping are crucial in shaping the properties of LSBGs within the simulation. 

Furthermore, \citet[][hereafter K20]{Kulier20_LSBGs_Eagle} examined massive LSBGs using the \eagle\ simulation \citep{Schaye15_EAGLE} and found that LSBGs evolve through secular processes (e.g., star formation and growth of their stellar discs) or mergers, depending on their stellar mass. \citet{DiCintio19_LSBGs_NIHAO}, using the zoom-in simulation suite \texttt{NIHAO} \citep{Wang15_NIHAO}, found that LSBGs form due to co-planar, co-rotating mergers and aligned gas accretion in the early stages of galaxy formation, while perpendicular mergers and misaligned gas accretion lead to higher surface brightness densities. \citet{Martin19_LSBGs} studied LSBGs within the \texttt{HorizonAGN} simulation \citep{Dubois14_HorizonAGN} and discovered that these galaxies originate from the same progenitors as HSBGs at $z>2$. However, tidal perturbations broaden their stellar distribution and heat their cold gas, leading to the diffuse, gas-poor LSBGs observed today.

This work explores the redshift evolution and star formation histories of low-surface brightness (\lsb) and high-surface brightness (\hsb) galaxy populations to identify systematic differences in their formation mechanisms. We focus on intermediate- to high-mass central galaxies ($10^9<\Mstar/\Msun<10^{11}$) from the \eagle\ hydro-dynamical simulation \citep{Schaye15_EAGLE}. To trace their evolutionary histories and assembly pathways, we follow their progenitors across approximately 20 snapshots in cosmic history, reaching back to $z\sim9$. In particular, we seek to identify the key processes that determine whether a galaxy of a given stellar or halo mass evolves into an \lsb\ or \hsb\ system. Additionally, we investigate whether the large-scale environment the galaxy resides in -- specifically, under- or over-dense regions of the cosmic web -- plays a role in the development of LSB features in central galaxies. 

This work is organised as follows. We build upon previous studies by incorporating multiple catalogues from our collaborations \citep{Lagos15_EAGLE_Masses-Hydrogen_ref,Lagos18_dynamics,Rosas-Guevara22_EAGLE_voids,Varela-Lavin22_Eagle}, along with additional data extracted from the \eagle\ database\footnote{\url{https://icc.dur.ac.uk/Eagle/}}. In \hyperref[sec:sample]{\Sec{sec:sample}}, we describe the adopted simulation, the parent catalogue, and the detailed classification and mass-matching procedures for \lsbs\ and \hsbs. In more detail, we aim to minimise potential selection effects -- such as biases related to stellar and halo mass -- and randomly downsample the number densities of \lsbs\ and \hsbs\ to ensure full stellar and halo mass matching at $z=0$. This procedure is described in \hyperref[sec:sample_dens]{\Sec{sec:sample_dens}} and results in a generation of a robust catalogue that enables the exploration and fair comparison of approximately 150 galaxy and halo properties of \lsbs\ and \hsbs in various large-scale environments. \hyperref[sec:results]{\Sec{sec:results}} presents the key differences between \lsb\ and \hsb\ systems, along with their evolution over cosmic time. A detailed discussion of our findings, including a comparison with previous studies and our conclusions, is provided in \hyperref[sec:discussion]{\Sec{sec:discussion}}. Finally, \hyperref[sec:summary]{\Sec{sec:summary}} summarises our work.

\section{Methodology and sample selection} \label{sec:sample}

In this section, we describe the methodology used to formulate these criteria and explain how we combine and cross-match various catalogues from the \eagle\ hydro-dynamical simulation to create a comprehensive and unique dataset for studying LSB galaxies and their evolution. 

\subsection{Data and simulation overview}\label{sec:data}

Our primary catalogue consists of the public data from simulation the \eagle-project \citep{Schaye15_EAGLE,Crain15_EAGLE,McAlpine16_EAGLE_release}, a suite of cosmological hydro-dynamical simulation code \texttt{P-Gadget 3} a modified version of the $N$-body smooth particle hydrodynamics code \texttt{Gadget-2} \citep{Springel05_Gadget2}. We make use of the largest simulation box available in the \eagle\ project, \texttt{Ref-L100N1504}, with a side-length of 100 $h^{-1}$ comoving Megaparsec (cMpc\footnote{Throughout this paper, we refer to comoving distances by preceding a ``c'' in ckpc. Otherwise we mean physical lengths such as \kpc\ or \Mpc.}) and a dark matter (baryon) mass resolution of $9.7\times10^6$ ($1.81\times10^6$) \hMsun. The adopted cosmology consists of a Lambda Cold Dark Matter ($\Lambda$CDM) model with the following parameters: $\Omega_\mathrm{m}=0.307, \Omega_\mathrm{b}=0.048, \Omega_\Lambda=0.693, \sigma_\mathrm{8}=0.823, n_\mathrm{s}=0.96$, $Y=0248$ \citep[see Table 1 from ][for details]{Schaye15_EAGLE}, and a dimensionless Hubble parameter $h=0.6778$ \citep{Planck13}. Hereafter, $h$ is absorbed in the numerical value of its property throughout the text and in all tables and figures.

The simulation was post-processed using the \texttt{SUBFIND} algorithm \citep{Springel01_Subfind,Dolag09_Subfind}, which employs the "Friend-of-Friends" method to identify haloes. Descendant subhaloes, and consequently galaxies, were identified using the \texttt{D-Trees} algorithm \citep{Jiang14_D-Trees}, with a detailed adaptation of this method to the \eagle\ simulations provided in \citet{Qu17_EAGLE_merger_trees}. The definition of the dark matter halo mass is given by $M_{\rm ref}(<R_{\rm ref}) = \Delta_{\rm ref} \rho_{\rm c} \frac{4\pi}{3} R_{\rm ref}^3$ where $\Delta_{\rm ref}=\Delta_{\rm 200c}$ for \Mc\ with $\Delta_{\rm 200c}$ being 200 times the critical density of the universe, $\rho_{\rm c}$. $R_{\rm ref}$ being the corresponding halo radius for which the interior mean density matches the desired value.

The model features element-by-element radiative cooling and heating \citep{Wiersma09_cooling_ref}, an effective temperature pressure floor preventing the gas from artificial fragmentation \citep{Schaye+DallaVecchia08}, time-dependent stellar mass loss due to winds from massive stars and AGB stars, core collapse and type Ia supernovae \citep{Wiersma09_chemical_ref}. Star formation is implemented as a stochastic, pressure-based \textit{Kennicutt-Schmidt} law with a metal-dependent threshold, assuming a \citet{Chabrier03} initial stellar mass function. Black holes are seeded in haloes above $10^{10}$ \Msun, with gas accretion based on a modified \textit{Bondi-Hoyle} rate \citep{Rosas-Guevara15_angMom_bh} and active galactic nucleus (AGN) feedback via stochastic heating \citep{Schaye15_EAGLE,Booth+Schaye09_AGN_ref}.

\subsection{Large-scale environment and further methodological assumptions}\label{sec:sample_envr}

Another goal of this study is to analyse the abundance of LSBGs in various large-scale environments of the cosmic web, such as dense clusters and low-density field environments. To achieve this, we crossmatch our catalogue with the sample selection from \citep[][hereafter RG22]{Rosas-Guevara22_EAGLE_voids} who studied the properties and evolution of central galaxies in the \eagle\ simulation as a function of their large-scale environmental affiliation. RG22 utilised the void-finder algorithm of \citet{Paillas17_cosmic_web_ref} at $z=0$, where galaxies with stellar mass of \Mstar\ $>10^8$ \Msun\ are used as void tracers based on the spherical under-density finder algorithm developed by \citet{Padilla05_void_finder}. We adopt their classification on the large-scale environment for all central galaxies in our dataset. For more information, we refer readers to Section 2.2 of RG22, where all technical details are described comprehensively.

Following R22 we categorising our galaxies into three large-scale environments ranging from very over-dense to less dense: \textit{skeleton} (S), \textit{walls} (W), and \textit{inner} and \textit{outer voids}. We note that we do not differentiate between \textit{inner} and \textit{outer} voids as RG22 do, but instead combine them into a single category: \textit{voids} (V).

About half of the properties studied in this work are obtained by us or one of our collaborators using an aperture of \onefiveropt\ to estimates properties such as stellar mass (\Mstar) or star formation rates (SFRs), while the rest are taken from the public \eagle\ database assuming a 30 kpc aperture. It is important to note that \Mstar\ enclosed within \onefiveropt\ and 30 kpc is equivalent. Additionally, for redshift evolution calculations, we rely solely on data from the public \eagle\ database, which generally assumes a 30 kpc aperture.

We restrict our study to central galaxies -- to avoid the effects of environment which strongly impact satellite galaxies -- above a threshold of \Mstar$>10^9$ \Msun\ and with a well defined half-mass radius. For our statistical analysis, we use the Python packages \texttt{Seaborn} and \texttt{Statsmodels}\footnote{\url{https://www.statsmodels.org/stable/index.html}}, employing the latter for robust regression, contour plots, $p$-value estimation, and bootstrap errors. Unless stated otherwise, bootstrap errors are estimated using at least 10,000 repetitions, corresponding to the 90\% and 10\% confidence levels. \textit{Probability density function} (PDF) plots use kernel density estimation from \texttt{Statsmodels}. Due to the larger scatter in certain properties, such as radii, we favour median estimates for their greater robustness, when noted. Regression lines include bootstrap-derived error bars, and marginal axes display box-and-whisker plots with PDFs, where boxes represent the interquartile range (IQR), medians are marked in red, and outliers as “$\times$”. In contour plots, we adopt levels of [0.25, 0.5, 0.75, 0.9, 0.99].

\subsection{Estimation of the surface brightness density}\label{sec:sample_corr}

The surface brightness density (\SBB) of a galaxy is defined as:

\begin{equation}
   \SBB = \MB + 2.5\log_{10}(2\pi~R^2_{\mathrm{opt}}).
	\label{eq:SB}
\end{equation}

Where \MB\ is the absolute magnitude in the $B$-band of the \textit{Johnson-Morgan} photometric system \citep{Johnson+Morgan53_UBV} and \ropt\ is the optical radius, defined as the radius that encloses 83\% of the stellar mass \citep{Tissera+Dominguez98_ropt_ref,Saiz01_83mass}. We use the conversion algorithm from Lupton (2005)\footnote{\url{www.sdss3.org/dr10/algorithms/sdssUBVRITransform.php}} on dust-corrected $r$- and $g$-band magnitudes in the AB photometric system \citep{Fukugita96_ugriz} -- as provided in the \eagle\ database table \texttt{RefL0100N1504\_Stars} \citep[see][for further details]{Trayford15_EAGLE_mags_ref,Trayford17_EAGLE_DustyMags} -- to obtain \MB. 

\begin{figure}
   	\includegraphics[width=0.665\columnwidth,angle=0]{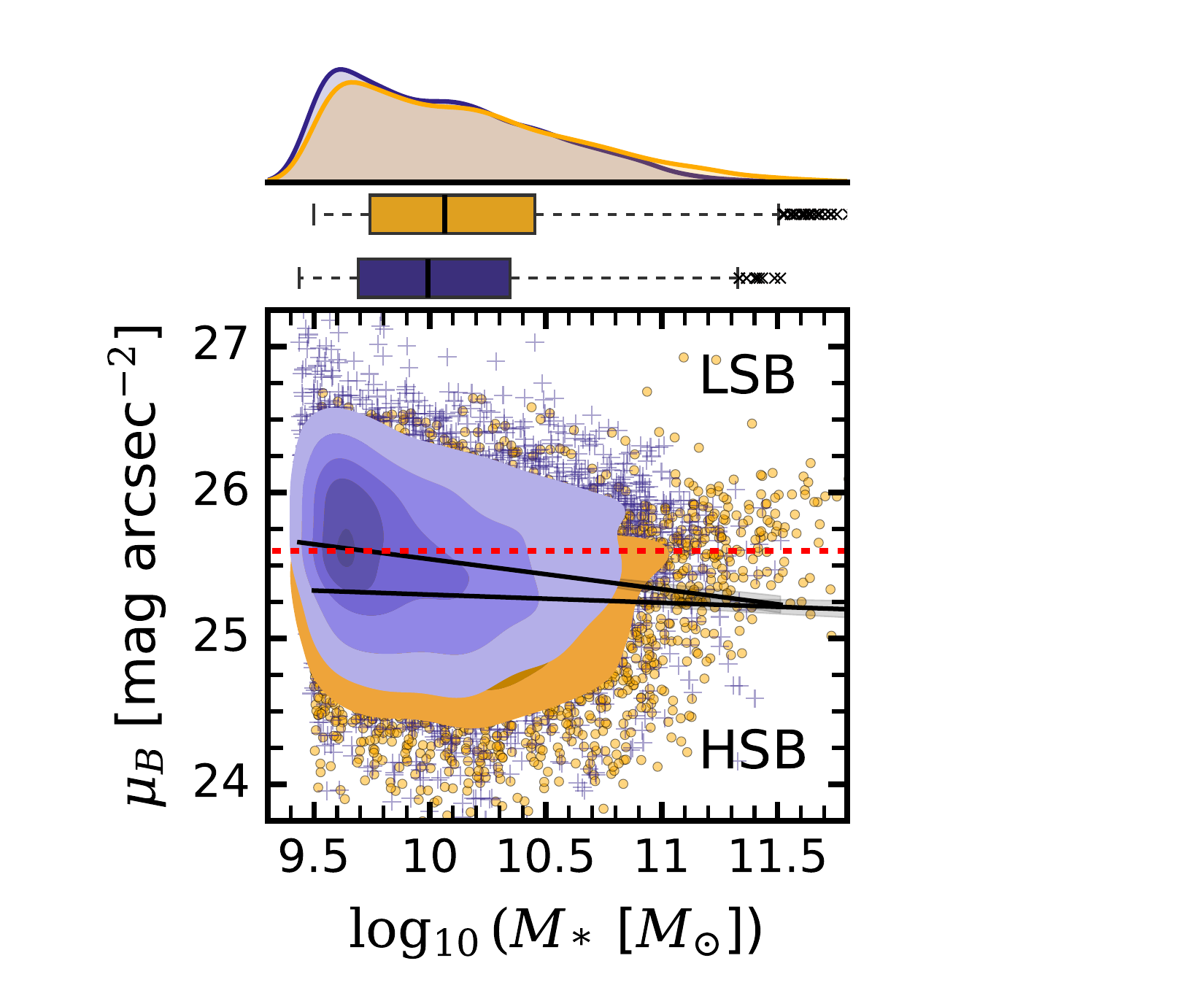}\hspace{-3.1cm}
 	\includegraphics[width=0.665\columnwidth,angle=0]{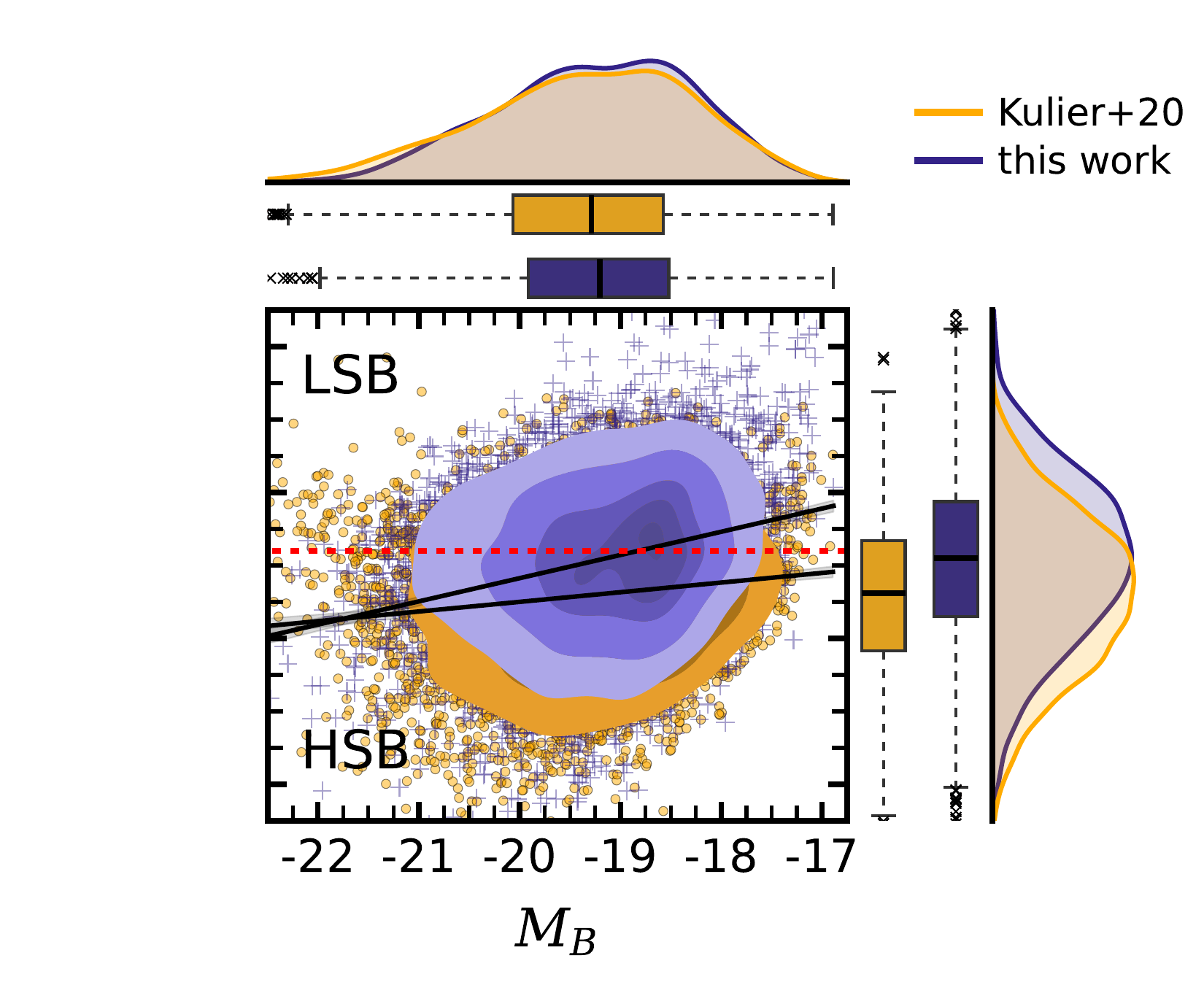}
    \caption{\SBB\ as a function of \Mstar\ (left panel) and \MB\ (right panel) at $z=0$. We used \hyperref[eq:SB]{\Eq{eq:SB}} to estimate the surface brightness density and corrected optical radii \roptc\ as described in \hyperref[sec:correct_ropt]{\Sec{sec:correct_ropt}}. The horizontal red dashed line marks the standard cut-off threshold of $\mu_{\mathrm{cut}}=25.6$ \magarcs\ to divide the dataset into \lsb\ and \hsb\ populations. Refer to \hyperref[sec:sample_envr]{\Sec{sec:sample_envr}} for statistical definitions.}\label{fig:SB_vs_Kulier}
\end{figure}

Our comparison with K20 -- who used the same simulation for their study -- estimates on radius, \Mstar, and \MB\ shows broadly consistent results (see the upper panel of \hyperref[fig:ropt_vs_Kulier_corrected]{\Fig{fig:ropt_vs_Kulier_corrected}} in \hyperref[sec:correct_ropt]{\Sec{sec:correct_ropt}}). However, we tend to underestimate larger radii ($\gtrsim30$ kpc), which may result in an overestimation of LSBG numbers in our sample. K20 estimations of radii is more directly comparable to observations, but computationally more demanding, requiring individual particle data from large simulations. Therefore, we introduce a correction scheme which is more efficient and feasible, detailed in \hyperref[sec:correct_ropt]{\Sec{sec:correct_ropt}}, to align our \ropt\ values with K20’s estimates. The corrected optical radius (\roptc) is used in \hyperref[eq:SB]{\Eq{eq:SB}} to calculate \SBB. \hyperref[fig:SB_vs_Kulier]{\Fig{fig:SB_vs_Kulier}} shows the comparison between our and K20’s \SBB\ distributions in terms of \Mstar\ (left) and \MB\footnote{The small difference between our and K20’s magnitudes is explained by the dust correction applied directly to the \eagle\ database magnitudes in our work, while K20 performed an independent correction using the same model from \citet{Trayford17_EAGLE_DustyMags}.} (right). We use cyan contours and crosses for our data and yellow contour and dots for K20’s. A threshold of $\mu_{\mathrm{cut}}=25.6$ \magarcs\ (red dashed line) divides our sample into \lsb\ (with $\SBB \geq 25.6$, $\sim2,300$ galaxies) and \hsb\ galaxies ($\SBB < 25.6$, $\sim2,000$ galaxies).

\subsection{Halo and stellar mass from an unbiased sample selection}\label{sec:sample_dens}

As shown by various authors in the literature, galaxy properties are closely linked to their stellar and halo masses \citep[e.g.][and therein]{Wechsler+Tinker18_rev}. Therefore, comparing galaxies across a broad mass range can introduce bias, as the properties of the dataset may be dictated by their mass range and environmental affiliation, rather than their intrinsic properties. Such datasets cannot be compared objectively as long as a bias in stellar and halo mass persists. Since the primary goal of this study is to investigate how LSB galaxies form and evolve in comparison to their HSB counterparts, we must ensure that our galaxy sample is as minimally biased by mass as possible.

\begin{figure}
    \centering
 	\includegraphics[width=0.9\columnwidth,angle=0]{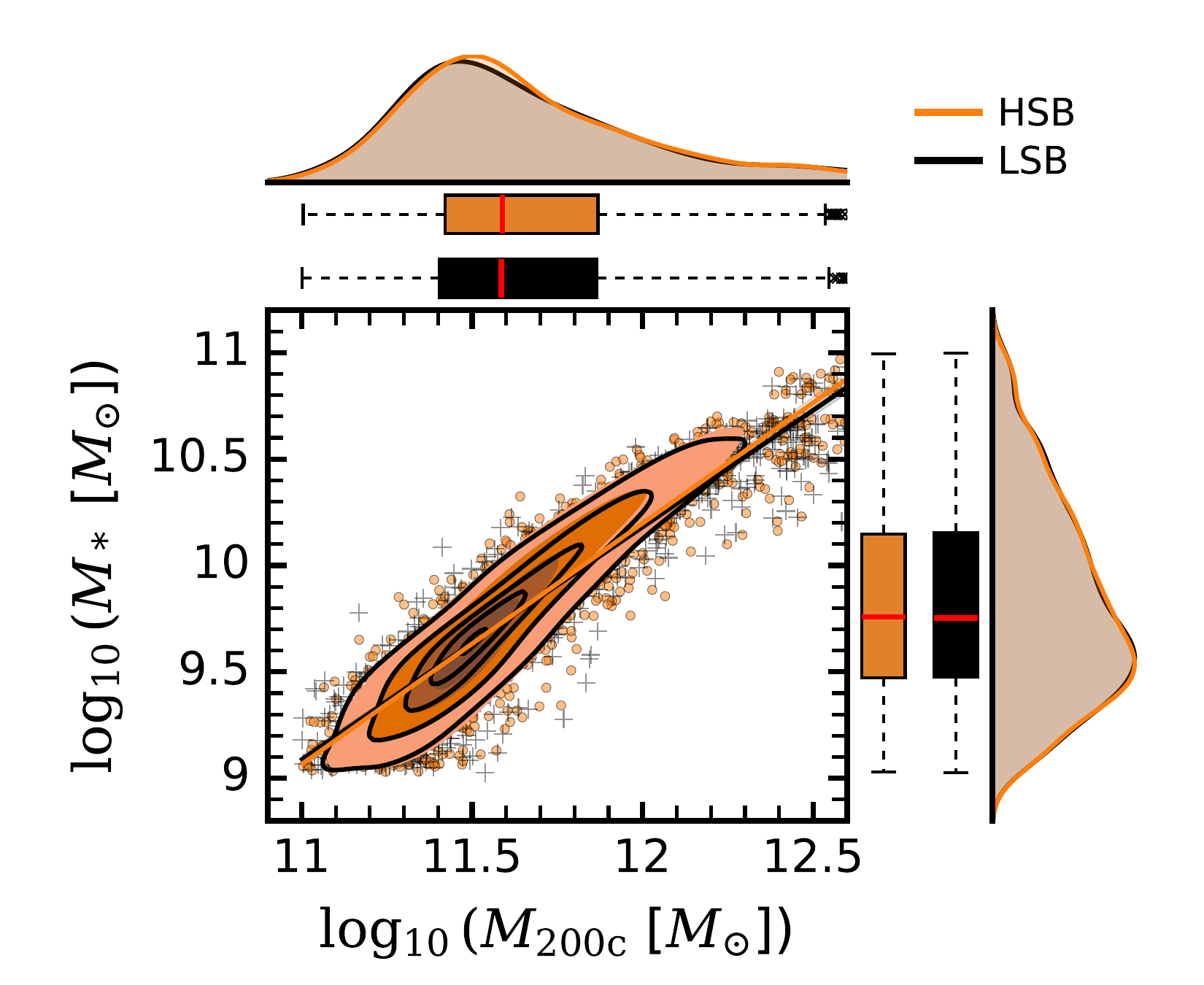}
    \caption{\Mstar\ as a function of \Mc\ at $z=0$ for the in stellar and halo mass unbiased populations of \lsbs\ (black solid contours and black crosses) and \hsbs\ (filled contours in shades of orange and orange dots). Refer to \hyperref[sec:sample_envr]{\Sec{sec:sample_envr}} for statistical definitions.}\label{fig:mstar2mhalo_unbiased}
\end{figure}

To construct an unbiased dataset, we use the \eagle\ parent sample at $z=0$ with realistic merger trees as described in \hyperref[sec:trees]{\Sec{sec:trees}}. Then we generate histograms of halo mass, \Mc, for samples of \lsbs\ and \hsbs\ using 25 linear bins of $11 < \log_{10}(\Mc~[\Msun]) < 13.5$ and randomly downsample the objects in each bin to match the number density distribution of \lsbs\ and \hsbs\ simultaneously. We repeat the process for the stellar mass, \Mstar, using 20 linear bins within the range of $9 < \log_{10}(\Mstar~[\Msun]) < 11$. The resulting dataset consists of 3,190 central galaxies, evenly split between \lsbs\ and \hsbs\ allowing for a fair comparison between the two populations by eliminating any dependency on halo and stellar mass as shown in \hyperref[fig:mstar2mhalo_unbiased]{\Fig{fig:mstar2mhalo_unbiased}}. As expected, both populations occupy the same stellar and halo mass parameter space. Statistical means and median values as well as uncertainty estimations and confidence levels on \Mstar\ and \Mc\ can be found in \hyperref[tab:dens6b_new_SB]{\Tab{tab:dens6b_new_SB}}.

\section{Results}\label{sec:results}

This section is divided into two parts. Firstly, we present results on the general halo and stellar properties of galaxies classified as \lsb\ and \hsb\ at $z=0$, using the unbiased dataset described earlier. This sample was generated by aligning both the stellar and halo mass as shown in \hyperref[fig:mstar2mhalo_unbiased]{\Fig{fig:mstar2mhalo_unbiased}}. In \hyperref[tab:dens6b_new_SB]{\Tab{tab:dens6b_new_SB}}, we list selected galaxy and halo properties studied in this work for the parent sample (\eagle), along with the corresponding values for our \lsb\ and \hsb\ populations. Secondly, we investigate the redshift evolution of our dataset in order to determine when low-surface brightness features emerge in \lsb\ galaxies. 

\subsection{Results on statistically highest-ranked properties}\label{sec:res_mstar}

\begin{figure}
    \includegraphics[width=0.87\columnwidth,angle=0]{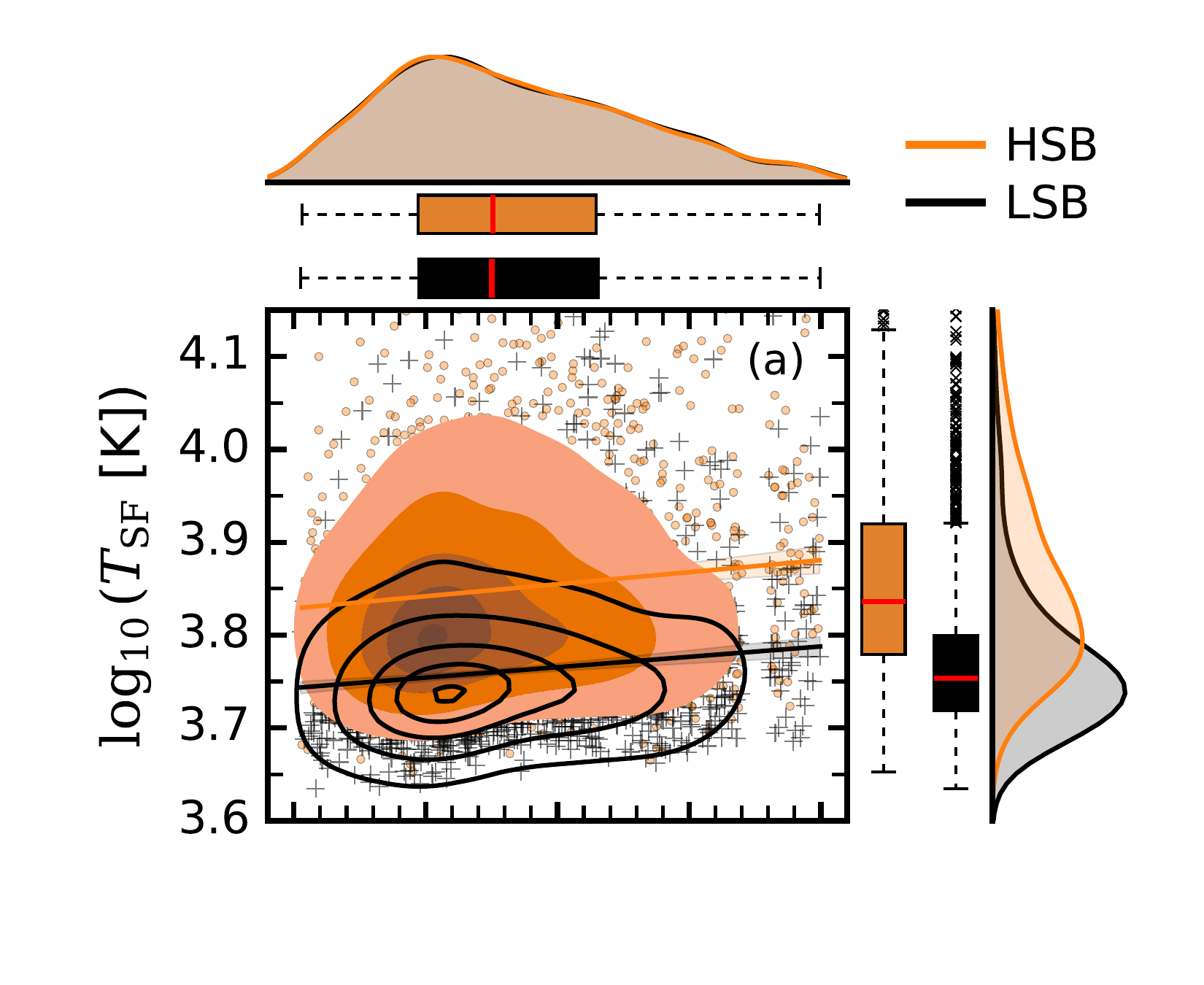}\vspace{-3.05cm}
    \includegraphics[width=0.87\columnwidth,angle=0]{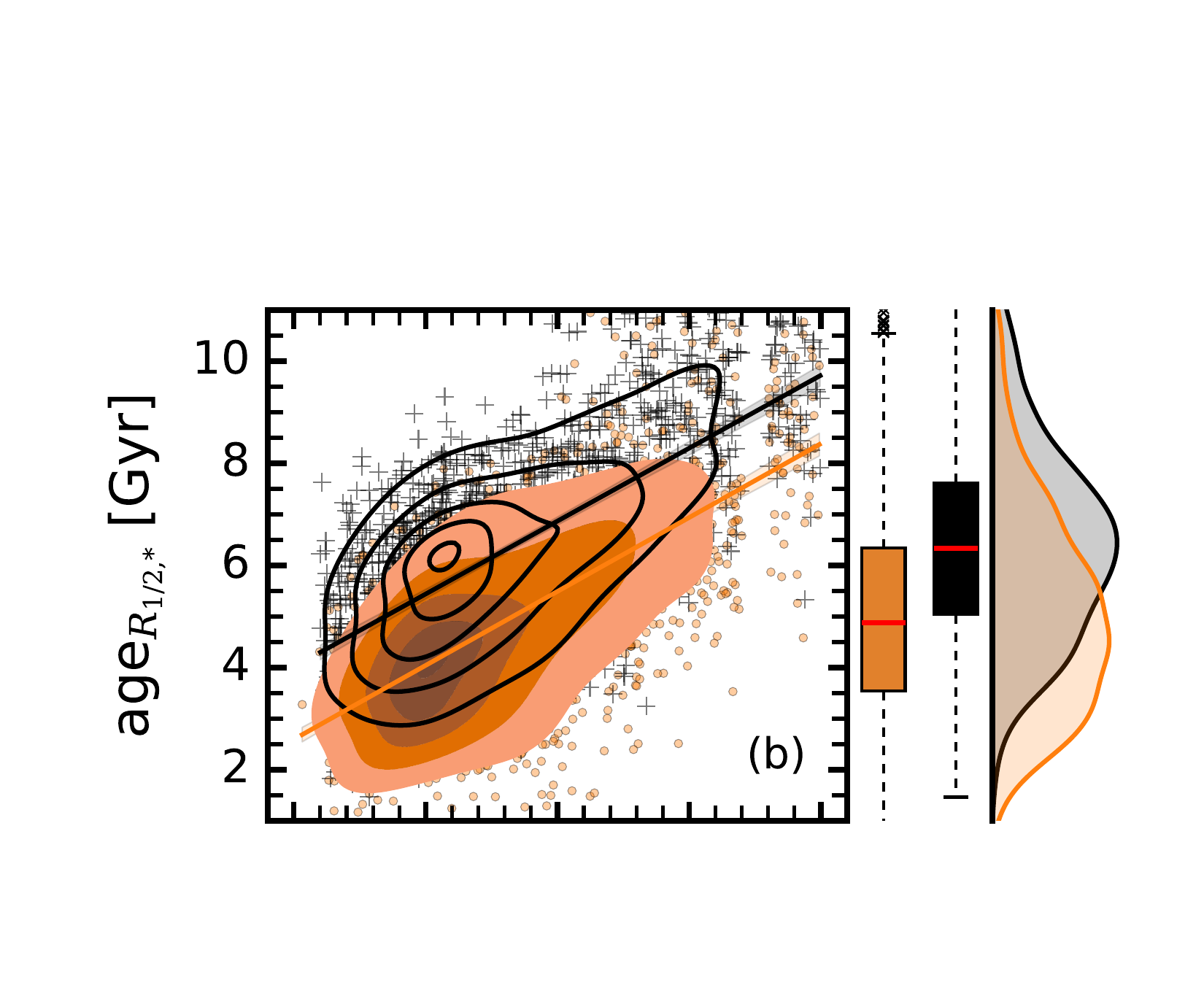}\vspace{-3.05cm}
    \includegraphics[width=0.87\columnwidth,angle=0]{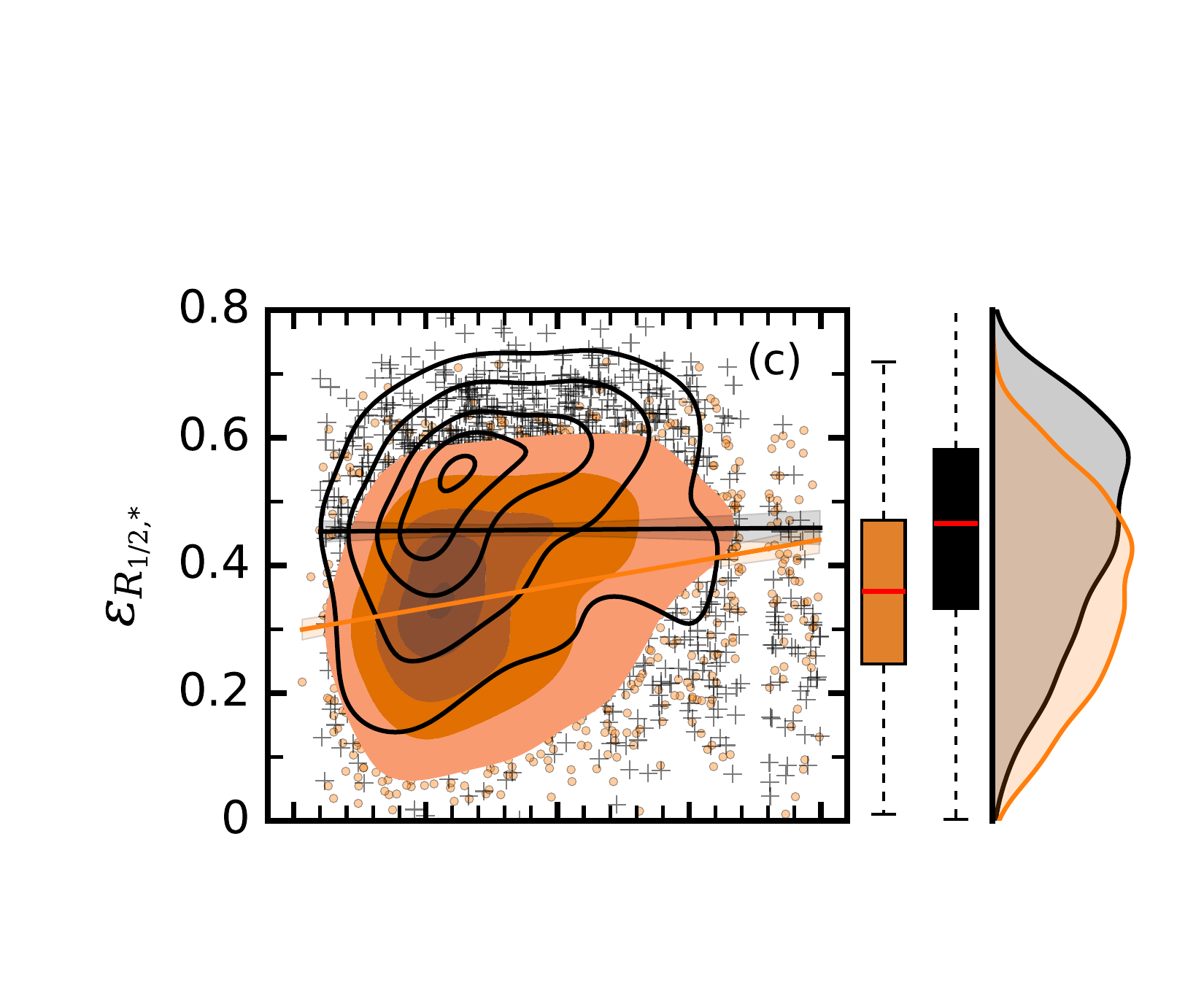}\vspace{-3.05cm}
    \includegraphics[width=0.87\columnwidth,angle=0]{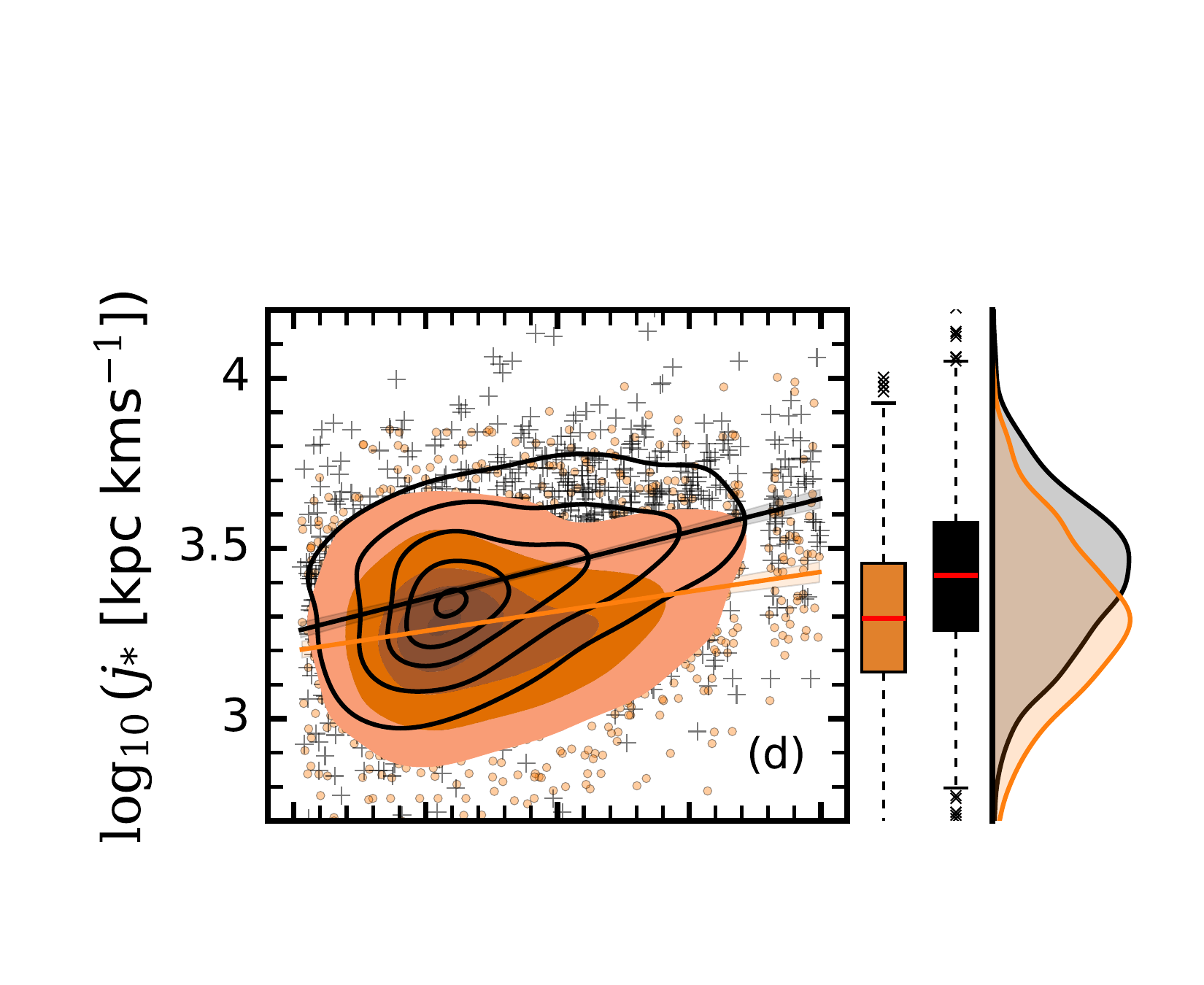}\vspace{-3.05cm}
  	\includegraphics[width=0.87\columnwidth,angle=0]{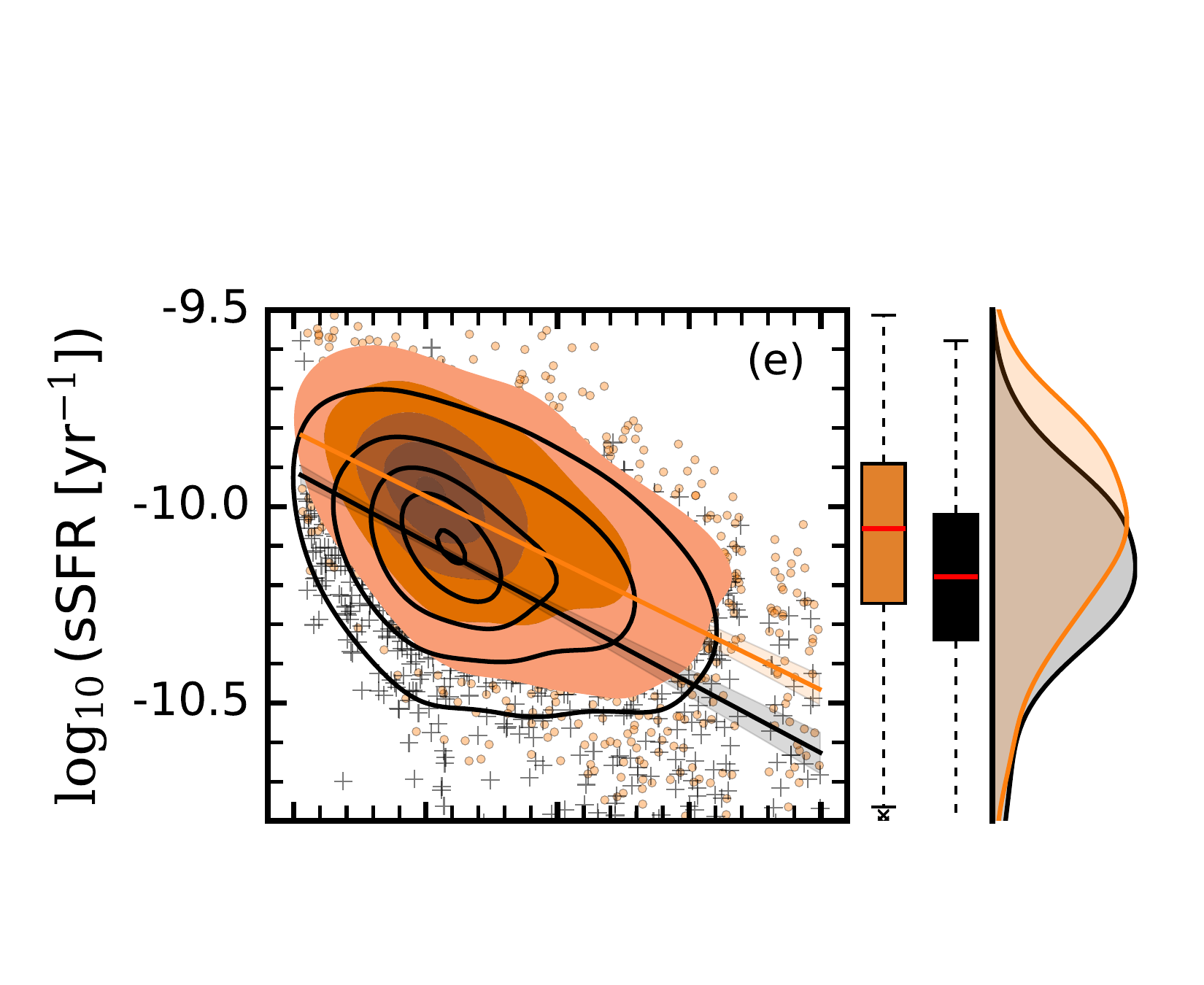}\vspace{-3.05cm}
    \includegraphics[width=0.87\columnwidth,angle=0]{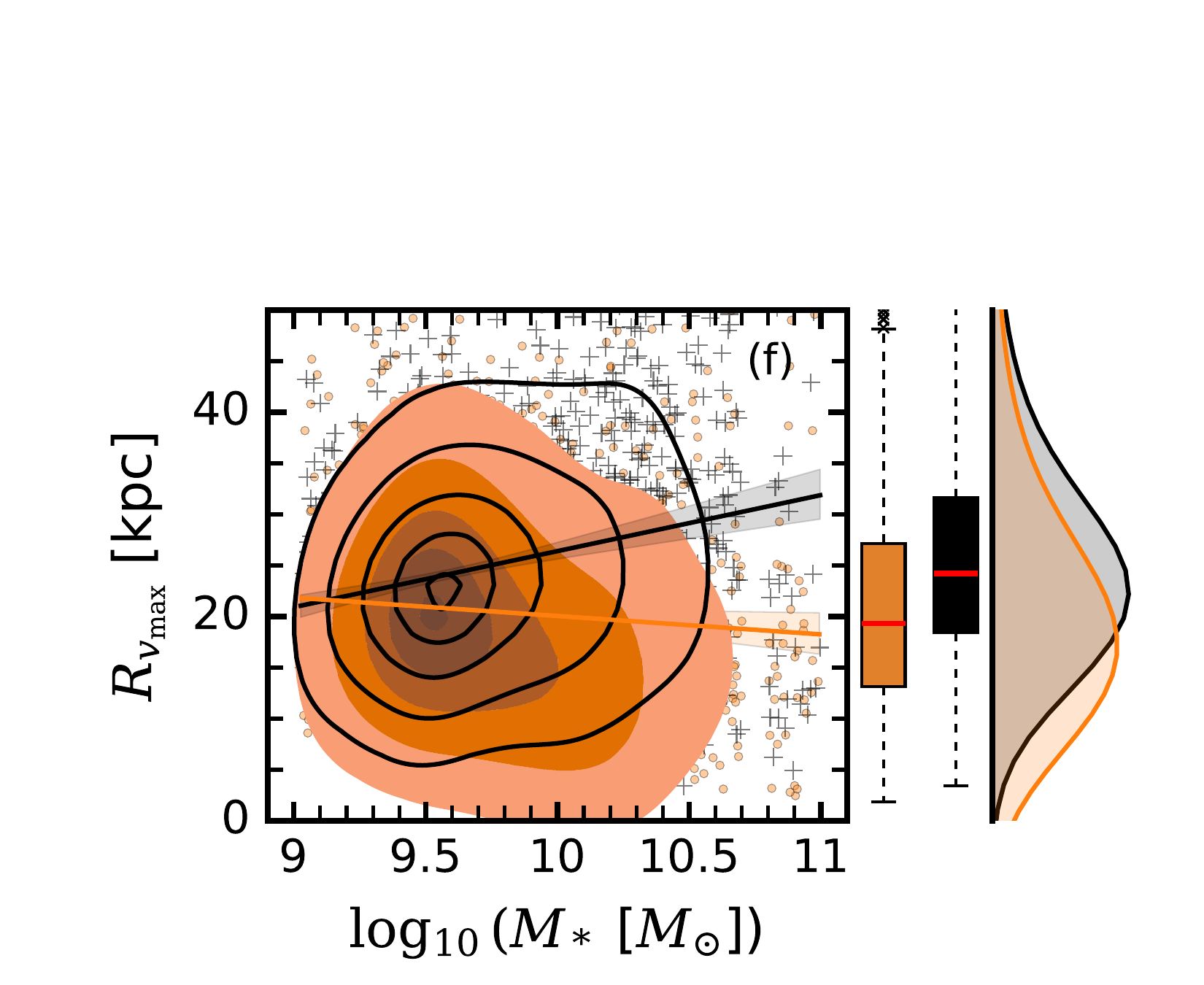}
    \caption{Statistically highest ranked properties: (a) \tempSF, (b) \agest, (c) \ellTwost, (d) $j_{*}$, (e) sSFR, and (f) \rvmax, as a function of \Mstar\ at $z=0$. See \hyperref[fig:mstar2mhalo_unbiased]{\Fig{fig:mstar2mhalo_unbiased}} for colour and symbol coding, and refer to \hyperref[sec:sample_envr]{\Sec{sec:sample_envr}} for statistical definitions.}\label{fig:results_mstar_unbiased}
\end{figure}

To identify which of the $\sim150$ properties warrant further exploration, we conducted various statistical tests comparing the \lsb\ and \hsb\ samples, as detailed in \hyperref[sec:sample_envr_test]{\Sec{sec:sample_envr_test}}. The results, ranked by the highest scores in the \textit{Kolmogorov-Smirnov test} (KS-test), are shown in \hyperref[fig:stats_test]{\Fig{fig:stats_test}}. The KS-test quantifies differences between two samples by measuring the maximum deviation between their cumulative distribution functions; thus, higher scores indicate greater dissimilarity. All statistical tests shown in \hyperref[fig:stats_test]{\Fig{fig:stats_test}} confirm that the \lsb\ and \hsb\ galaxy populations are statistically distinct in terms of their distributions as well as differences in mean values and variances. Furthermore, the two samples exhibit statistical robustness, as confirmed by non-parametric resampling and rank-based assessments.

Focusing on properties unrelated to our target selection criteria, we identified six key galaxy and halo properties that best encapsulate these differences and present them from top to bottom as a function \Mstar\ at $z=0$ in \hyperref[fig:results_mstar_unbiased]{\Fig{fig:results_mstar_unbiased}}:

\begin{itemize}
    \item[(a)] \tempSF\ as the mass-weighted temperature of the starforming gas e.g.\ in the interstellar medium (ISM) retrieved from the \eagle\ database,
    \item[(b)] \agest\ as the $r$-band luminosity-weighted line-of-sight stellar age measured within the projected half-mass radius adopted from \citet{Lagos18_dynamics} (see also \citet{Bender94,Varela-Lavin22_Eagle}),
    \item[(c)] \ellst\ as the edge-on ellipticity of star particle orbits within the projected half-mass radius from \citet[][see their Eqs.\ 1 \& 2]{Lagos18_dynamics} following the methodology outlined by \citet{Cappellari07_ell},
    \item[(d)] $j_*$ as the specific stellar angular momentum within \onefiveropt,
    \item[(e)] sSFR as the specific star formation rate within \onefiveropt, and
    \item[(f)] \rvmax\ as the radius of the maximum velocity of the total mass of the galaxy retrieved from the \eagle\ database.
\end{itemize}

Except for sSFR, all y-axis properties have significantly lower median values for \lsbs, as highlighted by the thick red lines. Notably, \lsb\ values often lie at or just beyond the IQR edges, indicating that HSB medians differ by about half the IQR, confirming statistical significance.

The \tempSF\ in \hyperref[fig:results_mstar_unbiased]{\Fig{fig:results_mstar_unbiased}}a yields the highest KS-test score, identifying it as the most significant differentiator between \lsbs\ and \hsbs. On average, \lsbs\ exhibit temperatures approximately 1,300 Kelvin lower than \hsbs\ across the entire stellar mass range shown in the figure. In the \eagle\ simulations, a temperature floor of 8,000 K is imposed, slightly exceeding the values found in our samples. Consequently, the \tempSF\ should not be interpreted literally but rather as a reflection of the effective pressure imposed on the unresolved, multiphase ISM \citep[see Sec.\ 4.3 in][]{Schaye15_EAGLE}. This suggests that \lsbs\ have lower ISM densities than \hsbs\ across the full stellar mass range.

In \hyperref[fig:results_mstar_unbiased]{\Fig{fig:results_mstar_unbiased}}b we show \agest\ and find similar distributions as for \tempSF, that the stellar population of \lsbs\ are approximately 1-2 Gyr older than the of \hsbs\ within the whole stellar mass range shown in the figure. In panel (c), we plot \ellst\ which displays a constant function of stellar mass for \lsbs\ while for \hsbs, the ellipticity is decreasing moderately from lower to higher stellar masses. In other words, this indicates that \lsbs\ are more discy ($\ellst\gtrsim0.3$) across all stellar masses, whereas \hsbs\ tend to become more discy only at higher \Mstar. Furthermore, \lsbs\ have slightly higher $j_*$ shown in panel (d) of the same figure. At lower stellar masses ($\Mstar>10^{9.5}$) \lsbs\ and \hsbs\ hold the same amount of angular momenta, while with \lsbs\ exhibit higher amounts of $j_*$ (0.25 dex) with increasing stellar mass ($\Mstar>10^{10.5}$). \hsbs\ exhibit higher star formation activity than \lsbs\ at a given stellar mass, as shown by sSFR in \hyperref[fig:results_mstar_unbiased]{\Fig{fig:results_mstar_unbiased}}e. While the differences are small, they remain statistically significant.

Notably, we find significantly larger radii of \rvmax\ for \lsb\ galaxies, with $\rvmax=24.22_{-5.77}^{+7.39}$ kpc, compared to $\rvmax=19.34_{-6.20}^{+7.81}$ kpc for \hsb\ galaxies, within the $25^\mathrm{th}$ and $75^\mathrm{th}$ percentiles, as shown in  \hyperref[fig:results_mstar_unbiased]{\Fig{fig:results_mstar_unbiased}}f. \rvmax\ is the radius at which the rotational velocity ($\vrot$) of a galaxy or halo reaches its maximum value \citep{Klypin11,Trujillo-Gomez11_HAM}. It provides insight into the mass distribution and concentration of the dark matter halo, with larger \rvmax\ indicating a more extended mass profile \citep{Diemand07_halos,Dutton+Maccio14_profiles}. \lsbs\ exhibit a strong trend of increasing their radii with halo mass which is in agreement with other studies \cite[e.g.\ see][]{McGaugh01_LSBGs}. Hence, more massive \lsbs\ reach higher \vmax\ at larger radii, even though both populations share a similar halo mass range and exhibit aligned \vmax\ distributions.

\subsection{Results on dynamical, kinematic, and morphological properties}\label{sec:res_gas_dyn}

We examine the dynamical and kinematic differences between the two populations, as ellipticity ranks among the highest in statistical significance (see \hyperref[fig:stats_test]{\Fig{fig:stats_test}}). In kinematic studies, ellipticity and the stellar spin parameter are closely tied to the observed velocity field, stellar orbits, and angular momentum \citep{Binney1978_ell,Kormendy+Illingworth82_rot_ref,Emsellem07_lambdaR_ref,Cappellari07_ell}. These properties help differentiate between dynamically hot, dispersion-supported systems (e.g., ellipticals and bulges) and cold, rotation-supported systems (e.g., discs) \citep{deVaucouleurs1948_ref,Cappellari06_morph_ref,Ho2007_kin_morph}. We investigate these properties within this framework to provide insights into the observational characteristics of LSBGs \citep{Buyle06_kin_LSBGs,Pahwa+Saha18_LSBGs_OBS,Cardona-Barrero20_NIHAO_dyn}.

We adopt the following properties from \citet{Lagos18_dynamics} (see also \citet{Varela-Lavin22_Eagle}) to characterise the degree of stellar rotational support in galaxies. These properties are measured along the line of sight at twice the projected stellar half-mass radius ($2\rhalfstars$) and are weighted by the $r$-band luminosity. We study

\begin{figure}
    \centering
 	\includegraphics[width=0.75\columnwidth,angle=0]{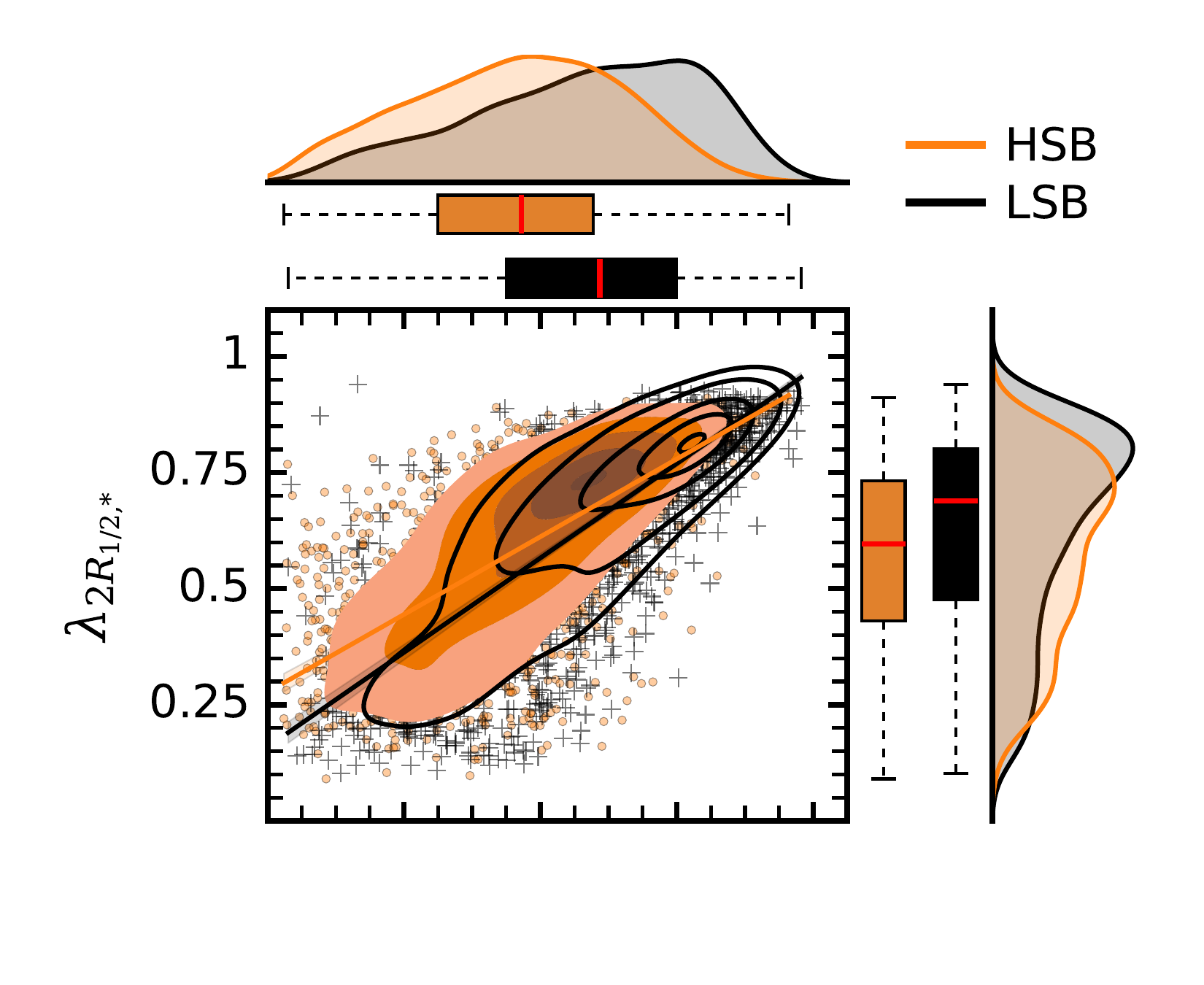}\vspace{-2.65cm}
     \includegraphics[width=0.75\columnwidth,angle=0]{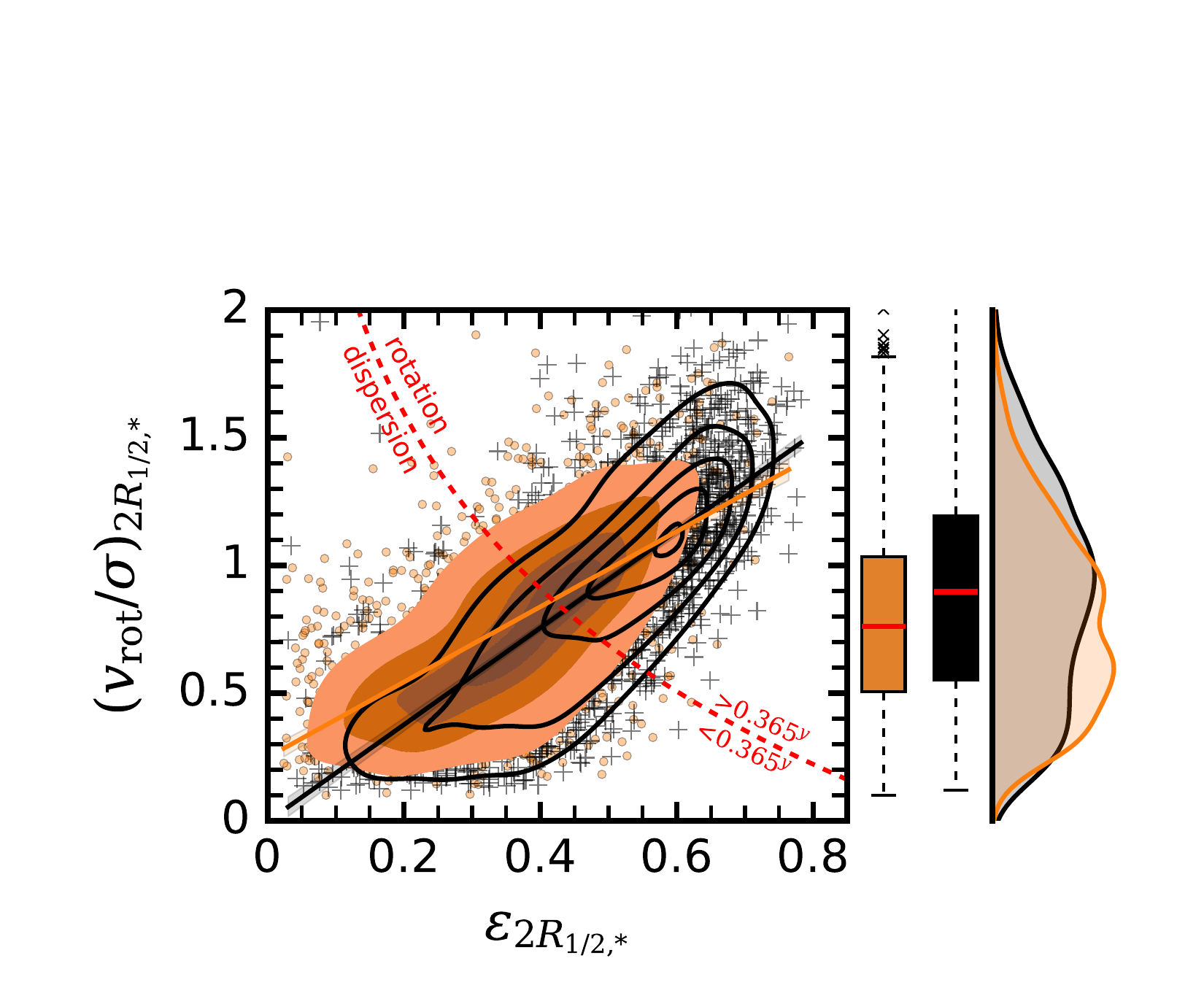}
    \caption{\lamTwost\ on the top and \VvStwoR\ on the bottom as a function of \ellTwost\ at $z=0$. See \hyperref[fig:mstar2mhalo_unbiased]{\Fig{fig:mstar2mhalo_unbiased}} for the colour and symbol coding, and refer to \hyperref[sec:sample_envr]{\Sec{sec:sample_envr}} for statistical definitions. The red dashed line represents the empirical boundary separating the rotation-supported sub-population from the dispersion-supported one.}\label{fig:results_dyn_unbiased}
\end{figure}

\begin{itemize}
    \item \lamTwost\ as the edge-on stellar spin parameter as defined by \citet{Emsellem07_lambdaR_ref,Emsellem11_lambdaR_ref},
    \item \VvStwoR\ as the ratio of \vrot\ to the velocity dispersion (\vdisp), where \vrot\ is defined as the velocity at which the Gaussian peaks, and \vdisp\ as the square root of the variance \citep[see][for details]{Lagos18_dynamics} 
\end{itemize}

as a function of \ellTwost\ measured within $2\rhalfstars$ in \hyperref[fig:results_dyn_unbiased]{\Fig{fig:results_dyn_unbiased}}.\footnote{For the same property measured within one \rhalfstars\ and as a function of \Mstar\ refer to \hyperref[fig:results_mstar_unbiased]{\Fig{fig:results_mstar_unbiased}}c} We observe that a significant number of \lsbs\ are skewed towards higher values of \lamTwost\ (top panel) and \VvStwoR\ (bottom panel) in comparison to their \hsb\ counterparts. Examining the \VvStwoR~--~\ellTwost\ plane in the lower panel of \hyperref[fig:results_dyn_unbiased]{\Fig{fig:results_dyn_unbiased}} in more detail, we observe that both \lsbs\ and \hsbs\ populations can be further divided into a more flattened, \textit{rotation-supported} (upper-right region) sub-population and a more circular, \textit{dispersion-supported} (lower-left region) sub-population as marked by the dashed red line following the empirical function $x=0.365^y$. The parameter of the function where obtained by fitting a power law through the median values of the distributions on the $x$ and $y$ axes. 

Consequently, the galaxy populations of \lsbs\ and \hsbs\ represent a mix of morphological types typical of the stellar mass range adopted. We observe that a significant number of \lsb\ galaxies are more rotational-supported than their \hsbs, however, most massive population of \lsbs\ ($\logT(\Mstar\ [\Msun])>10.25$) has a significant dispersion-supported population which is less pronounced for their \hsb\ counterparts. Furthermore, \lsbs\ and \hsbs\ exhibit similar trends in dynamical and structural properties (e.g. ellipticity and spin parameters) across both rotation- and dispersion-supported systems. However, the variation within each category is greater among \lsbs\ than among \hsbs.

Additionally, \lsbs\ generally have \rhalfstars\ that are 1-2 kpc larger than their \hsb\ counterparts at a fixed \Mstar, with mean values of $\rhalfstars = 5.10_{-0.80}^{+1.08}$ kpc for \lsbs\ and $\rhalfstars = 3.83_{-0.51}^{+0.78}$ kpc for \hsbs. Their \textit{S\'ersic} indices ($n_{\text{S\'ersic}}$) are also higher ($n_{\text{S\'ersic}} = 1.87_{-0.81}^{+0.99}$ for \lsbs\ vs.\ $n_{\text{S\'ersic}} = 1.55_{-0.44}^{+0.49}$ for \hsbs), indicating that \lsbs\ deviate more from an exponential profile ($n_{\text{S\'ersic}}=1$) and lean slightly more towards bulge-like structures.

So far, we have discussed properties where \lsbs\ and \hsbs\ differ significantly at $z=0$. However, several properties exhibit remarkable similarities between the two populations (see \hyperref[tab:dens6b_new_SB]{\Tab{tab:dens6b_new_SB}}). Notably, both have comparable iron abundance (Fe/H) and oxygen abundance ($12+\logT(\mathrm{O/H})$), despite previous studies reporting lower metallicities in LSBGs \citep[e.g.][]{Das07_gLSBGs_AGN,Puzia08_LSBGs_GCs,Kulier20_LSBGs_Eagle,Cao23_bELSBGs,Tang24_LSBGs_TNG_metals}. Other properties that show similar values at $z=0$, aside from stellar and halo masses, include black hole mass (\Mbh), baryon and molecular gas fractions, baryon-to-dark matter particle fraction, stellar maximum velocity (\vmax), velocity dispersion (\vdisp), dark matter halo radius (\rc), halo concentration parameter assuming a \textit{Navarro-Frenk-White} profile \citep{Navarro97}, and disc-to-total mass ratio (D/T)\footnote{D/T is defined using the AM-E method, which decomposes galaxies into discs and spheroids based on the angular momentum (AM) content and binding energy (E) \citep[see][for description and method]{Tissera12_ref}.} for gas and stars. 

While the median velocity dispersion shows no significant differences between our \lsbs\ and \hsbs, their gradients ($\Delta_{\vdisp}$) -- defined as the difference of \vdisp\ measured at \rhalfstars\ and at 2\rhalfstars -- do. \lsbs\ exhibit twice the \lsbs\ values of \hsbs\ ($\Delta_{\vdisp} = 1.54_{-1.80}^{+1.96}$ for \lsbs\ and $\Delta_{\vdisp} = 0.69_{-1.72}^{+1.85}$ for \hsbs) consistent across both rotation- and dispersion-supported systems. A similar trend is observed for the age gradient ($\Delta_{\mathrm{age}}$) between 1–2\rhalfstars, though the difference appears only in rotation-supported systems. Notably, the most massive 35\% of rotation-supported \lsbs\ exhibit negative age gradients of approximately $\Delta_{\mathrm{age}}=~-0.5$ Gyr. These differences in \vdisp\ and age could serve as observational markers for identifying LSBGs.

\subsection{Results on the large-scale environment}\label{sec:res_envr}

In this section, we examine the differences between \lsbs\ and \hsbs\ in relation to their large-scale environment. Our selected galaxies are distributed across three environments: \textit{skeleton} (S, high density), \textit{walls} (W, intermediate density), and \textit{voids} (V, low density), as defined in RG22 and \hyperref[sec:sample_envr]{\Sec{sec:sample_envr}}. Both populations exhibit similar abundances in each environment, with 11\% in S, $\sim$33\% in W, and $\sim$7\% in V, indicating no clear environmental preference for \lsbs. Additionally, galaxy and halo properties of \lsbs\ and \hsbs\ show comparable results within a given large-scale environment. Statistical tests, as applied previously in \hyperref[sec:sample_envr_test]{\Sec{sec:sample_envr_test}}, yield no significant signal, suggesting that any environmental influence on \lsbs\ and \hsbs\ must be mild.

We remind the reader that our sample selection aligns stellar and halo masses, which may account for the comparable abundances and properties observed. This suggests that stellar and halo mass are more strongly linked to environmental trends than intrinsic properties associated with LSB features. However, a more complex entanglement between LSB characteristics and mass dependence might explain the absence of a clear environmental trend -- an aspect we aim to explore in future studies.

\subsection{Results on the assembly histories and redshift evolution}\label{sec:res_zevol}

\begin{figure}
    \centering
    \includegraphics[width=0.75\columnwidth,angle=0]{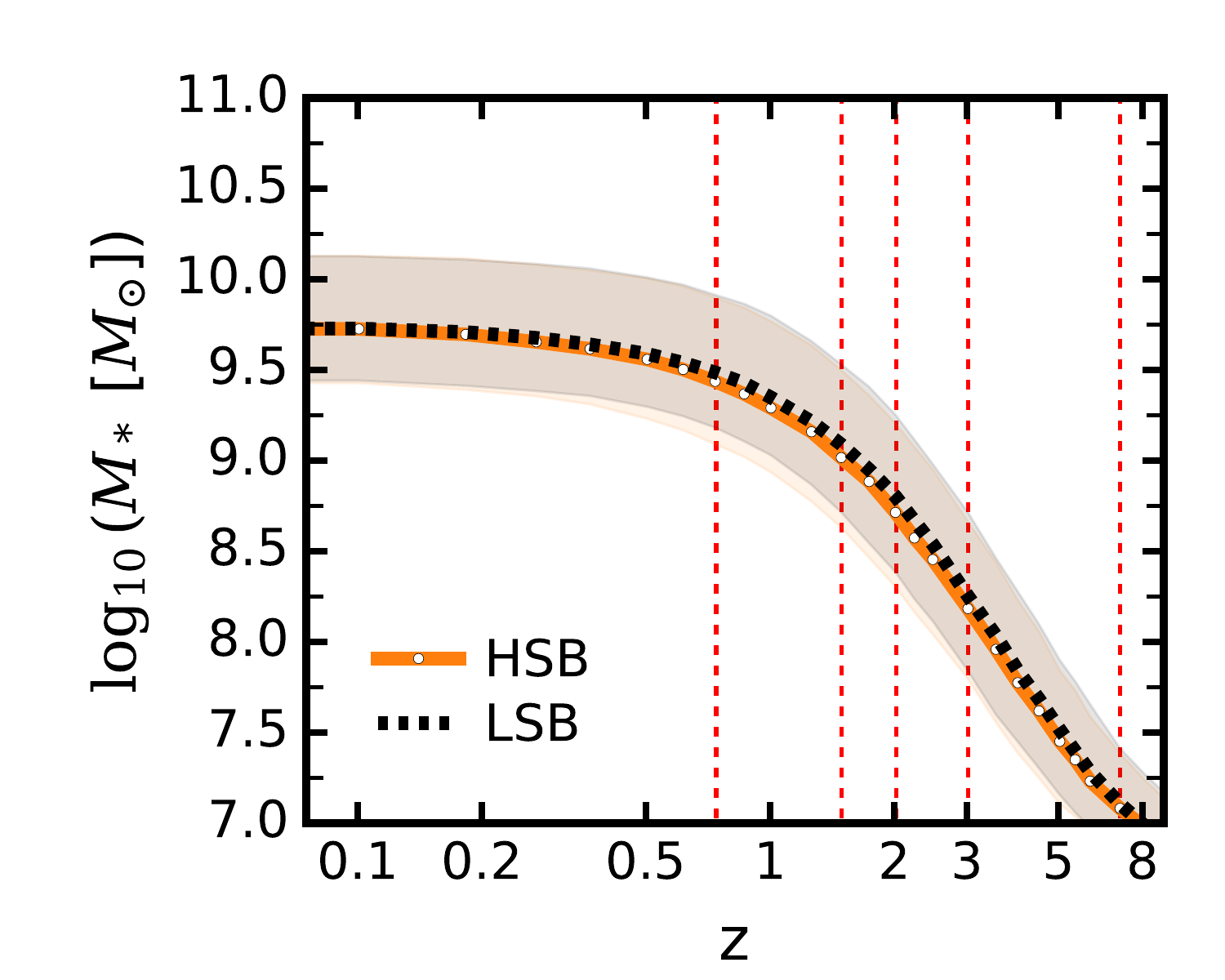}
    \caption{Median values for the \Mstar\ assembly history within $25^{th}$ and $75^{th}$ percentiles for \lsbs\ (dashed black line) and \hsbs\ (solid orange line with with dots), respectively.}\label{fig:mstar_assembly}
\end{figure}

We trace each galaxy’s evolution using the IDs of their host dark matter haloes to follow their main progenitors through merger trees. A detailed method and quality assessment of these trees is provided in \hyperref[sec:trees]{\Sec{sec:trees}}, including examples across halo, stellar, and black hole components. Around 10\% of galaxies are excluded for not meeting quality criteria.

In \hyperref[fig:mstar_assembly]{\Fig{fig:mstar_assembly}}, we examine the evolution of mass assembly through the median values for \Mstar, for \lsbs\ (dashed black line) and \hsbs\ (solid orange line with white dots), within the $25^{\text{th}}$ to $75^{\text{th}}$ percentiles. The evolution of \Mstar\ for both populations is highly consistent. For this reason, we omit the assembly histories for the \Mc\ as they are very similar for both populations.

\begin{figure}
    \centering
    \includegraphics[width=0.67\columnwidth,angle=0]{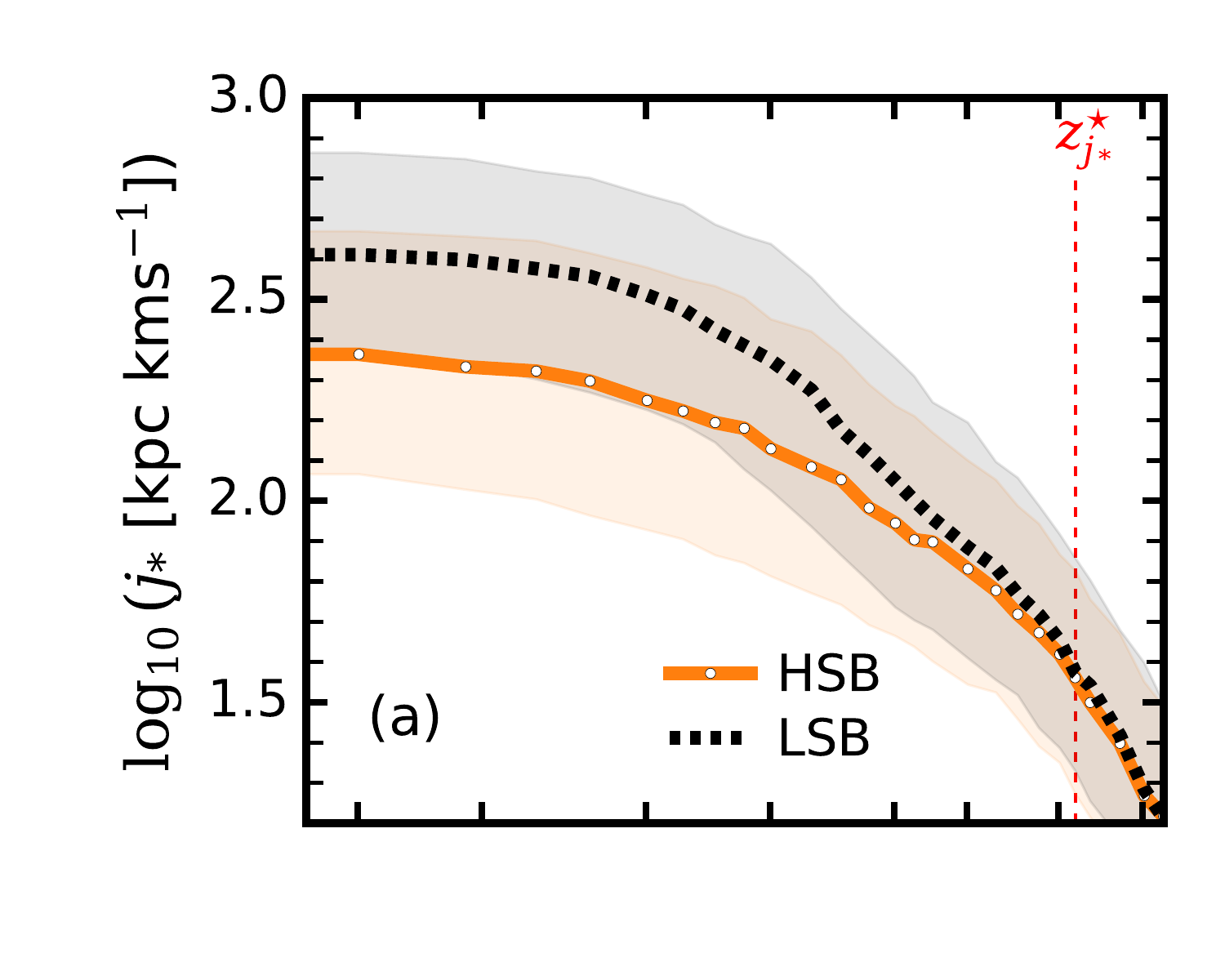}\vspace{-1.22cm}
    \includegraphics[width=0.67\columnwidth,angle=0]{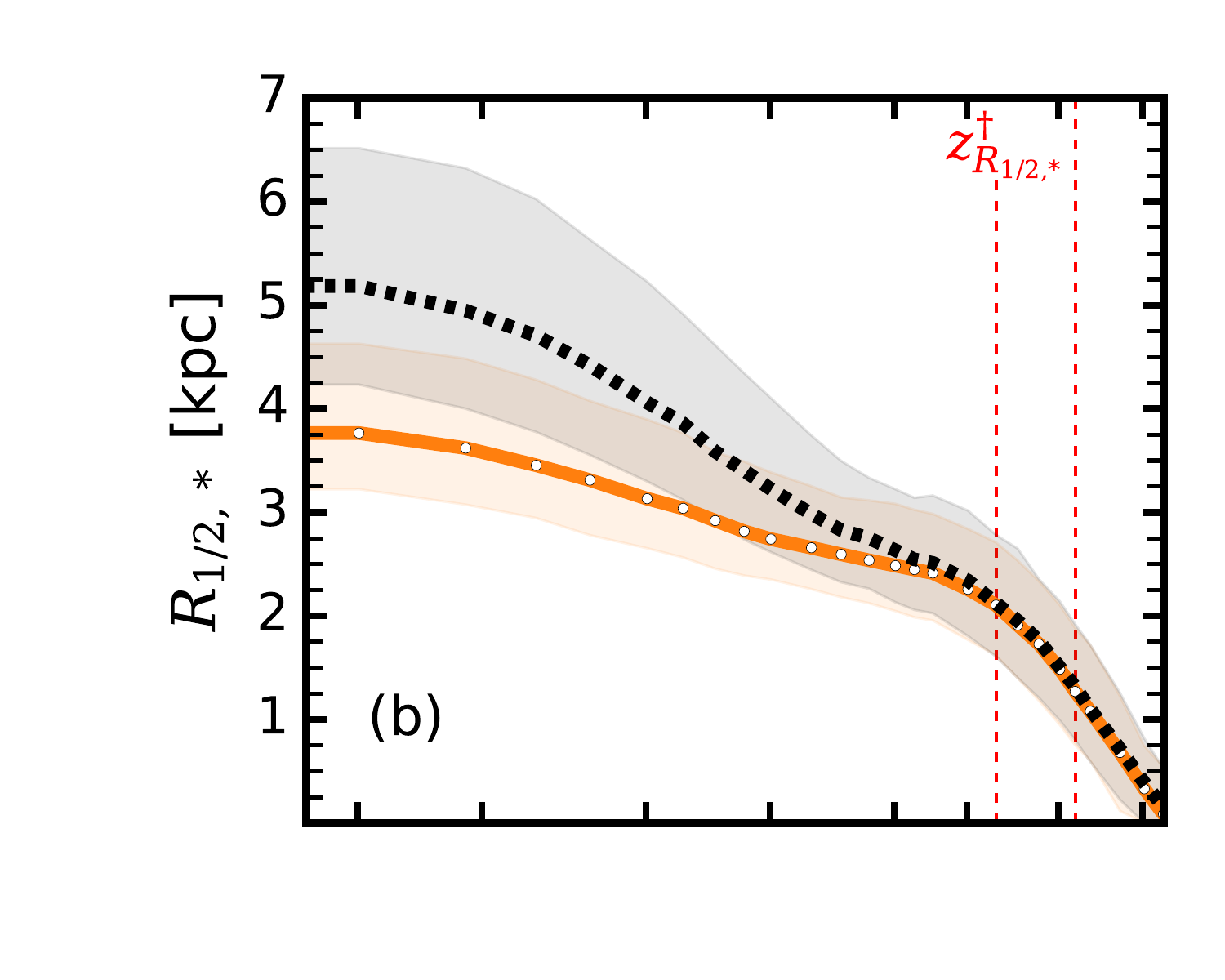}\vspace{-1.22cm}
    \includegraphics[width=0.67\columnwidth,angle=0]{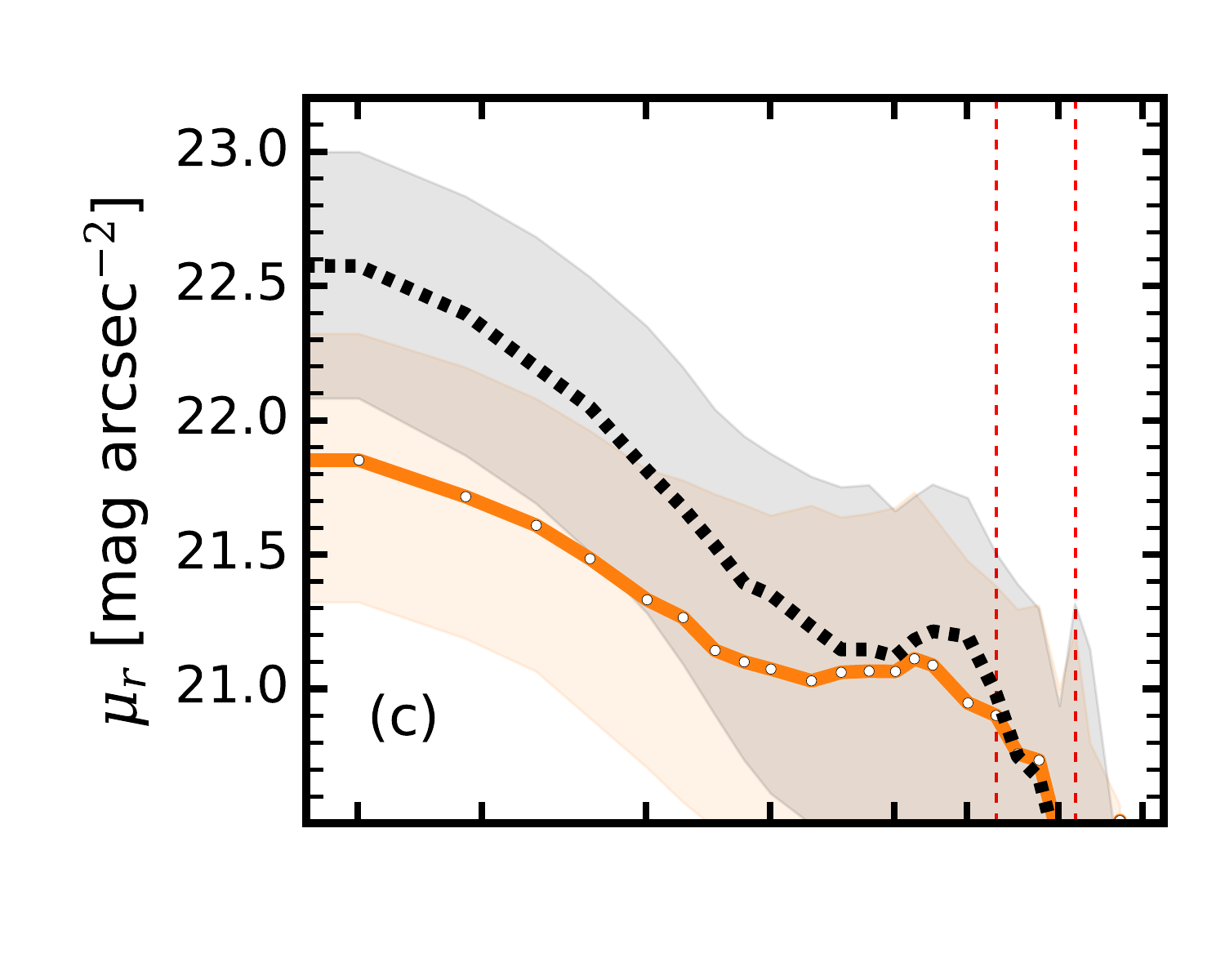}\vspace{-1.22cm}
    \includegraphics[width=0.67\columnwidth,angle=0]{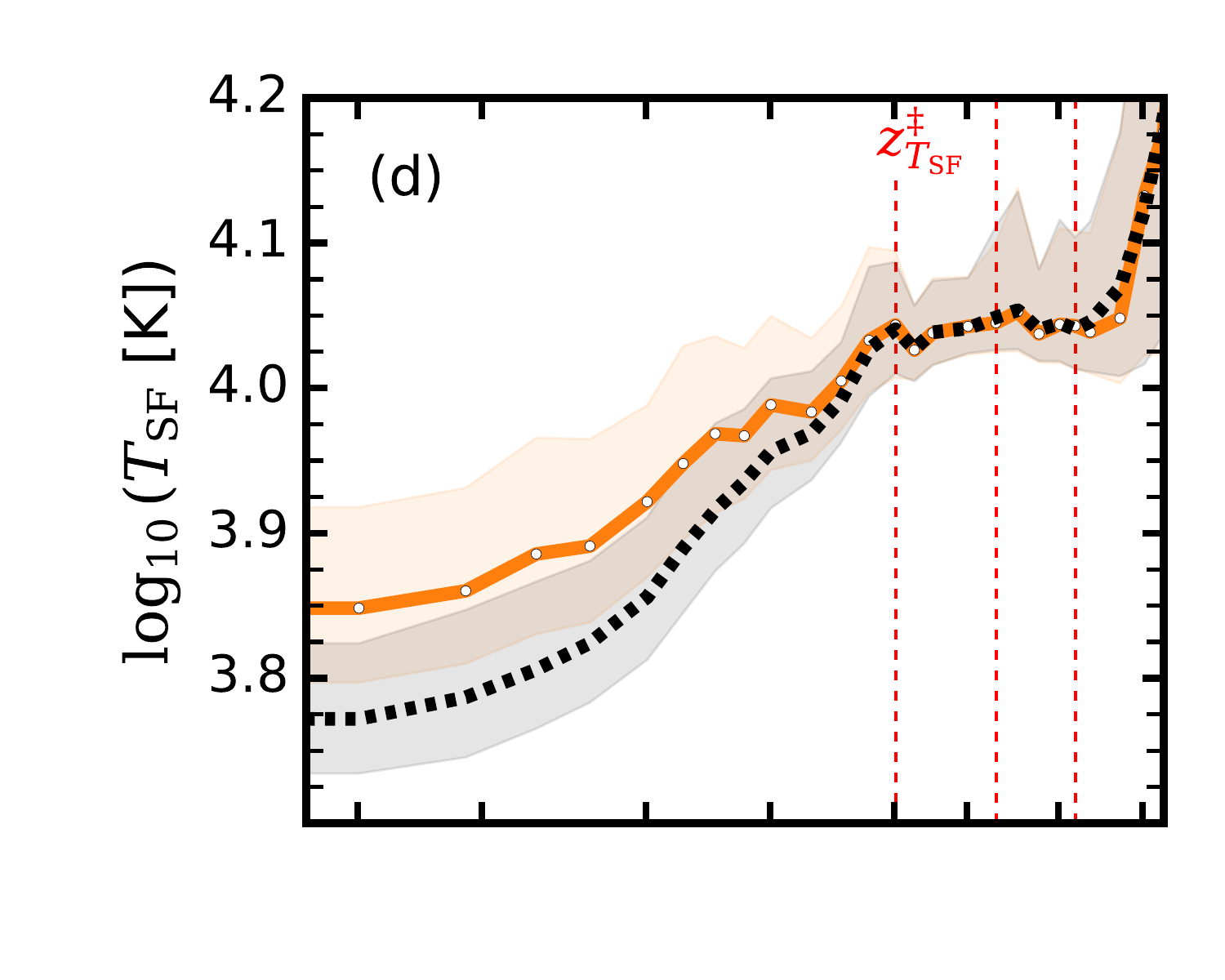}\vspace{-1.22cm}
    \includegraphics[width=0.67\columnwidth,angle=0]{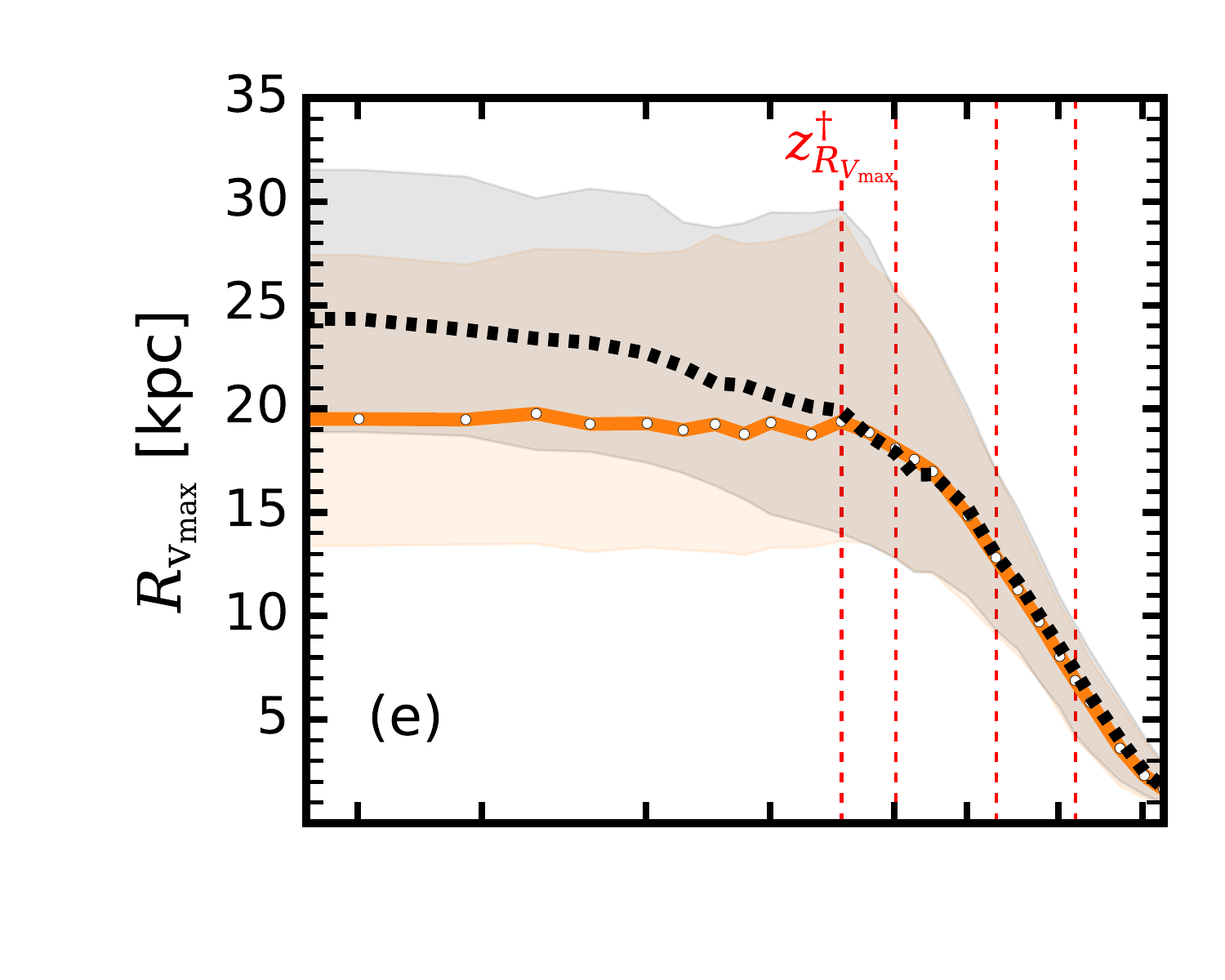}\vspace{-1.22cm}
    \includegraphics[width=0.67\columnwidth,angle=0]{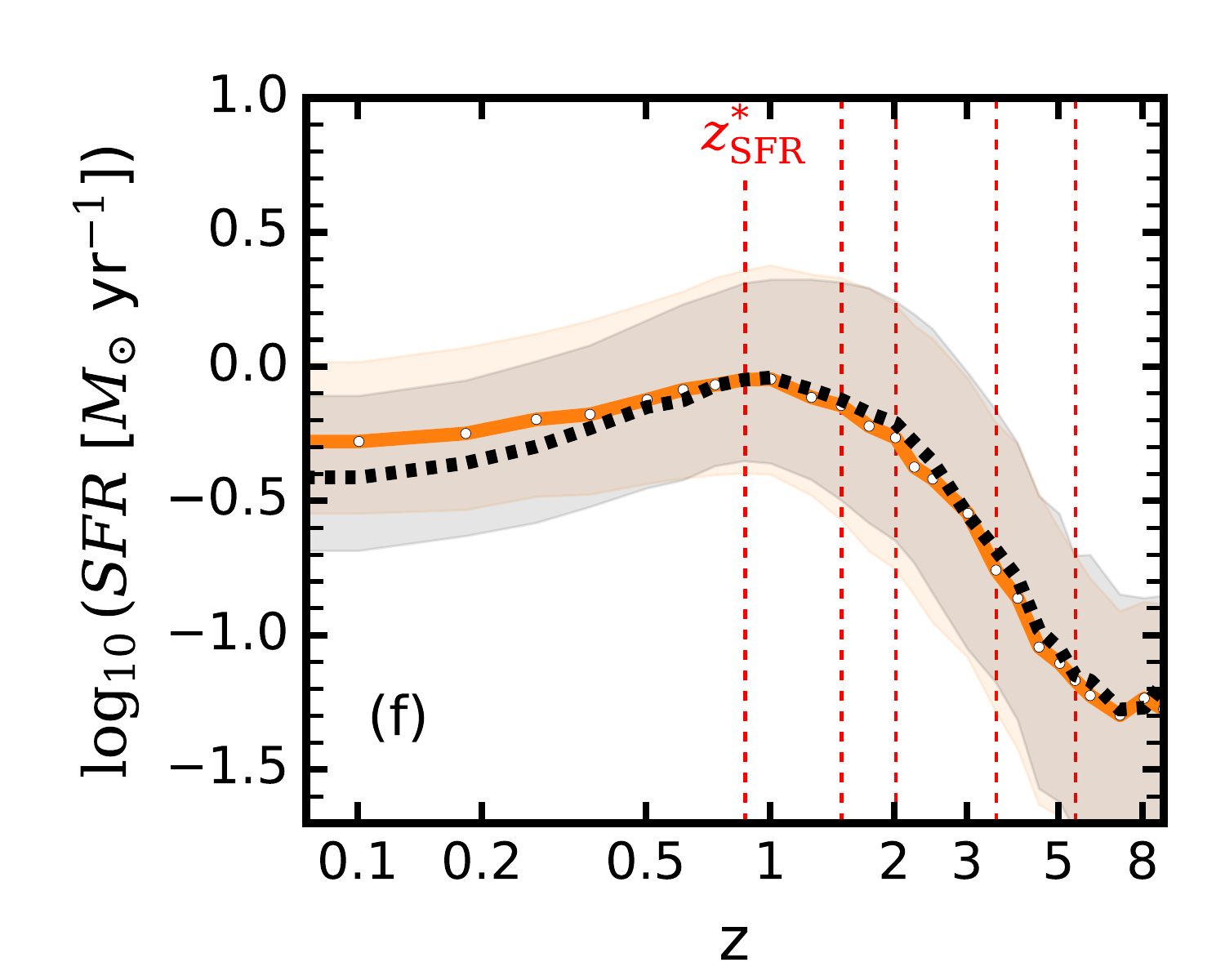}
    \caption{Redshift evolution of the median values for \lsbs\ and \hsbs\ displayed from top to bottom for $j_*$, \rhalfstars, \SBr, \tempSF, \rvmax, and SFR. The order of the plots corresponds to the sequence of divergence events in their histories, as marked by vertical dashed red lines with annotations (see \hyperref[tab:events]{\Tab{tab:events}}). See \hyperref[fig:mstar_assembly]{\Fig{fig:mstar_assembly}} for colour and symbol coding, and statistical definitions.}\label{fig:zevol_props}
\end{figure}

In the following, we show the redshift evolution of properties that exhibit the most significant divergence at some point in cosmic history ordered by their appearance starting at high redshift. From top to bottom in \hyperref[fig:zevol_props]{\Fig{fig:zevol_props}}, we plot the evolution of $j_*$, \rhalfstars, the surface brightness density in the $r$-band (\SBr),\footnote{Here, we use the $r$-band magnitude and stellar radius instead of the $B$-band magnitude and optical radius, as these properties are available across all snapshots studied, while the properties used to define our \lsb\ and \hsb\ samples can only be obtained at $z=0$.} \tempSF, \rvmax, and SFR. The redshifts at which the two populations diverge are indicated by vertical dashed lines with annotations, as described in \hyperref[tab:events]{\Tab{tab:events}}. To determine the redshift of divergence events, we first narrow down a plausible range through visual inspection. We then calculate the percentage difference between \lsbs\ and \hsbs\ at each redshift, defined as $(x_{\lsbs} - x_{\hsbs}) / [0.5(x_{\lsbs} + x_{\hsbs})] \times 100$, where $x$ denotes the median value of the property in question. The divergence redshift is identified as the redshift at which this percentage difference reaches a local maximum between two consecutive snapshots.

In \hyperref[fig:zevol_props]{\Fig{fig:zevol_props}}a, we examine the evolution of $j_*$. The figure shows that both populations gain constantly angular momenta during their evolution. Notable, \lsbs\ subtly diverge from \hsbs\ starting from $z\sim6-7$ -- marked as \zjstars\ and described in \hyperref[tab:events]{\Tab{tab:events}} -- but more progressively gaining momenta at later cosmic times around $z\sim2$.

From $z=3$ onwards, \lsbs\ exhibit significantly faster growth of \rhalfstars\ compared to their \hsbs\ counterparts as shown in \hyperref[fig:zevol_props]{\Fig{fig:zevol_props}}b. This divergence is marked as \zrhalf\ by the vertical dashed line. This divergence is also evident in \SBr\ shown in panel (c) of the same figure because \rhalfstars\ is the principal component in our estimation of surface brightness (see \hyperref[eq:SB]{\Eq{eq:SB}}) and therefore strongly correlated with \SBr. The figure clearly illustrates that \lsb\ galaxies have maintained their nature since $z \sim 2-3$. Panel (d) of the same figure shows \tempSF. We identify a divergence event around $z\sim3$, where the median temperature of \lsbs\ begins to decrease more rapidly than that of their \hsb\ counterparts. 

The divergence of \rvmax at $\zrvmax\sim1.5$ is highlighted in \hyperref[fig:zevol_props]{\Fig{fig:zevol_props}}e. It represents the most distinct and prominent divergence observed in our analysis. While \hsbs\ exhibit a nearly constant \rvmax\ from \zrvmax\ onwards, the \rvmax\ of \lsbs\ continue to grow, coinciding with their \rhalfstars\ expanding over time. It is important to note that stellar and halo masses were homogenised for this analysis, and \rvmax\ serves as a proxy for the concentration of the dark matter halo. A growing \rvmax\ in \lsbs\ suggests that these systems undergo significantly more dynamical and structural evolution compared to their \hsb\ counterparts. This observation aligns with our findings on the dynamical properties at $z=0$ discussed in \hyperref[sec:res_gas_dyn]{\Sec{sec:res_gas_dyn}}.

Although the distinctive features of \lsbs\ are established at higher redshifts, SFR begins to diverge only around $\zSFR\sim0.9$ as shown in \hyperref[fig:zevol_props]{\Fig{fig:zevol_props}}f. This suggests that star formation activity does only play a minor role in the development of LSB characteristics. Decreasing values of SFR and lower star formation activity is instead a low-redshift outcome of processes initiated during the early stages of galaxy evolution. 

In \hyperref[eq:zdiv]{\Eq{eq:zdiv}}, we summarise the sequence of divergence redshifts, starting with \zjstars\ as the earliest, followed by the subsequent divergence redshifts of other properties, as detailed in \hyperref[tab:events]{\Tab{tab:events}} and illustrated in \hyperref[fig:zevol_props]{\Fig{fig:zevol_props}}. Our results show that $j_*$ is the first property where \lsbs\ and \hsbs\ diverge in their redshift evolution, occurring at $z\sim5-7$. This early divergence likely drives the development of LSB features. Similar trends are observed for the angular momentum of both the starforming and non-starforming gas, though their divergences are significantly less pronounced than in $j_*$.

\begin{table}
    \begin{center}
    \caption{Redshifts at which the properties of \lsbs\ and \hsbs\ begin to diverge.}\vspace{-0.4cm}
    \setlength{\tabcolsep}{3pt}
        \begin{tabular}{M{0.12\columnwidth}||M{0.12\columnwidth}|M{0.18\columnwidth}|M{0.22\columnwidth}}
        \hline
        symbol & z & $t_{\mathrm{LB}}$ [Gyr] & property \\\hline
            \hline
            \zjstars    & $\sim5-7$ & $\sim12.6-13$ & $j_*$ \\\hline
            \zrhalf    & $\sim3.5$ & $\sim12$      & \rhalfstars\ and \SBr\\\hline   
            \ztempSF    & $\sim2$   & $\sim10.5$    & \tempSF  \\\hline
            \zrvmax     & $\sim1.5$ & $\sim9.5$     & \rvmax   \\\hline
            \zSFR       & $\sim0.9$ & $\sim7.4$     & SFR     \\\hline
            \hline
            (i)                         & (ii)      &  (iii)     & (iv) \\
        \end{tabular}
    \tablefoot{Column (i) lists the symbols representing each divergence event. The columns (ii) and (iii) provide the redshift and corresponding lookback time in Gyr, whereas the column (iv) the property which diverges.}\label{tab:events}
    \end{center}
\end{table}

\large
\begin{equation}
     \zjstars \gtrapprox\  \zrhalf\ \gtrapprox\ \ztempSF \gtrapprox\ \zrvmax\ \gtrapprox\ \zSFR\ \label{eq:zdiv}
\end{equation}
\normalsize

\subsection{Results on the evolution and assembly history in narrow stellar mass bins}\label{sec:res_bins}

\begin{figure}
    \centering
    \includegraphics[width=0.75\columnwidth,angle=0]{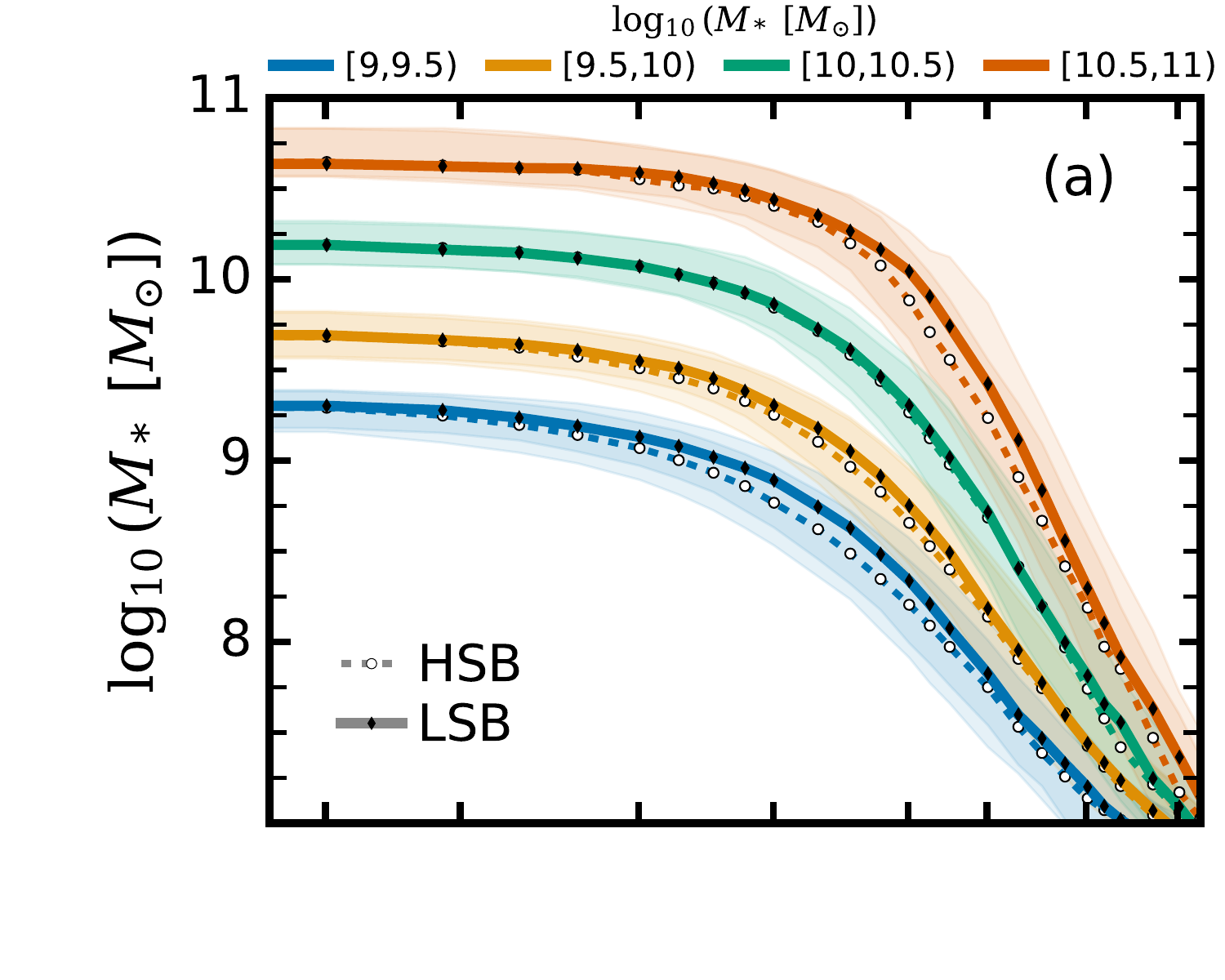}\vspace{-1.36cm}
    \includegraphics[width=0.75\columnwidth,angle=0]{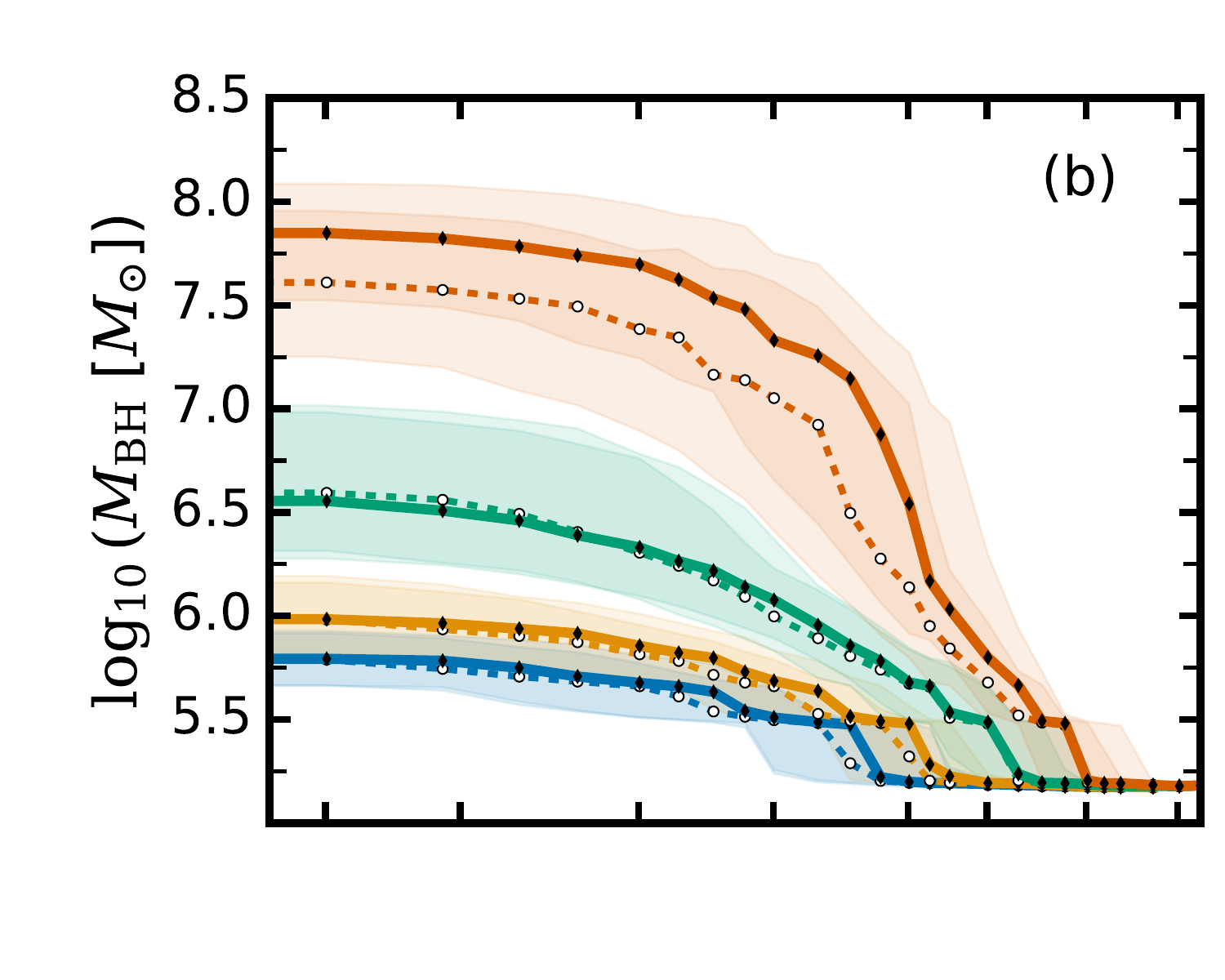}\vspace{-1.36cm}
    \includegraphics[width=0.75\columnwidth,angle=0]{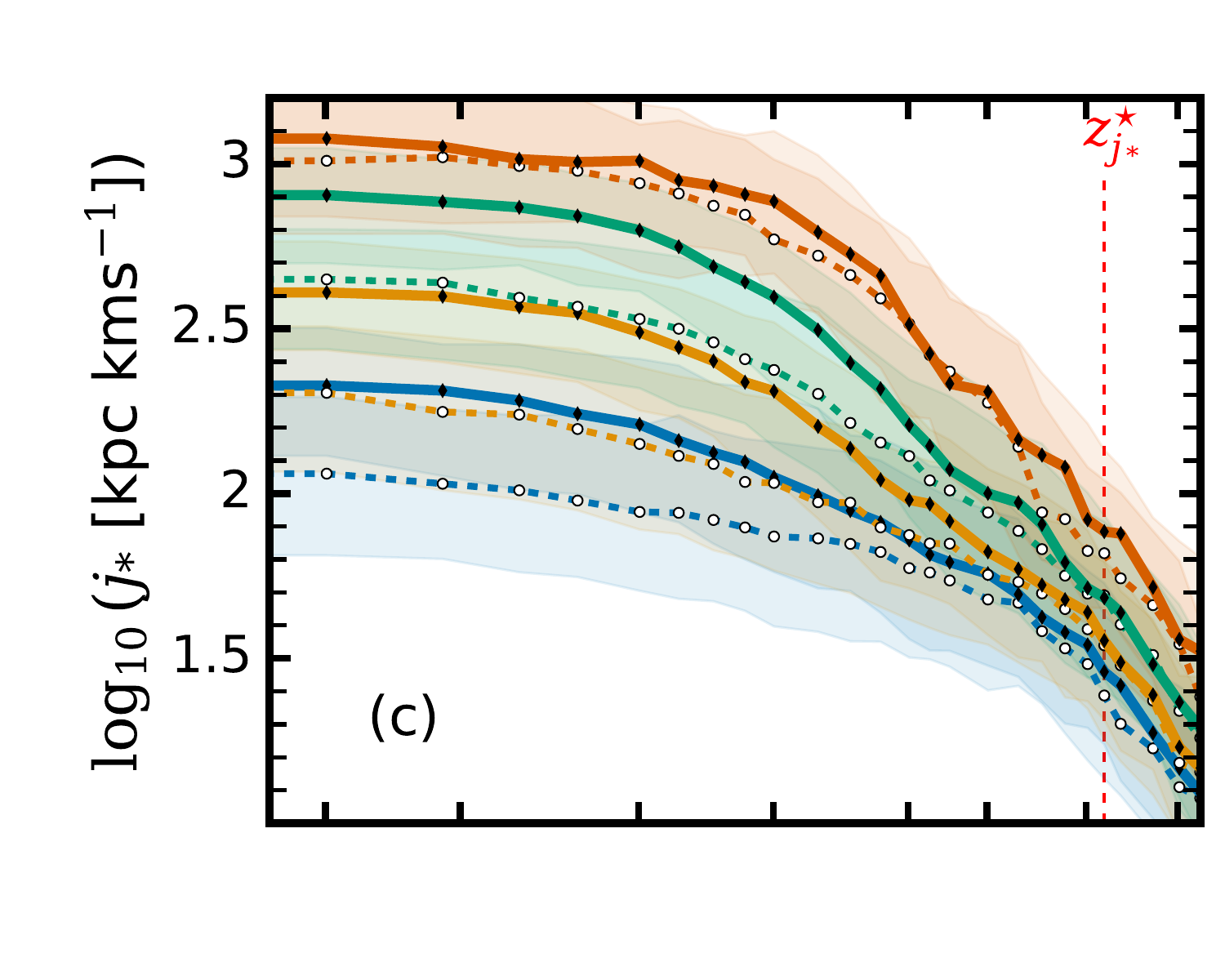}\vspace{-1.36cm}
    \includegraphics[width=0.75\columnwidth,angle=0]{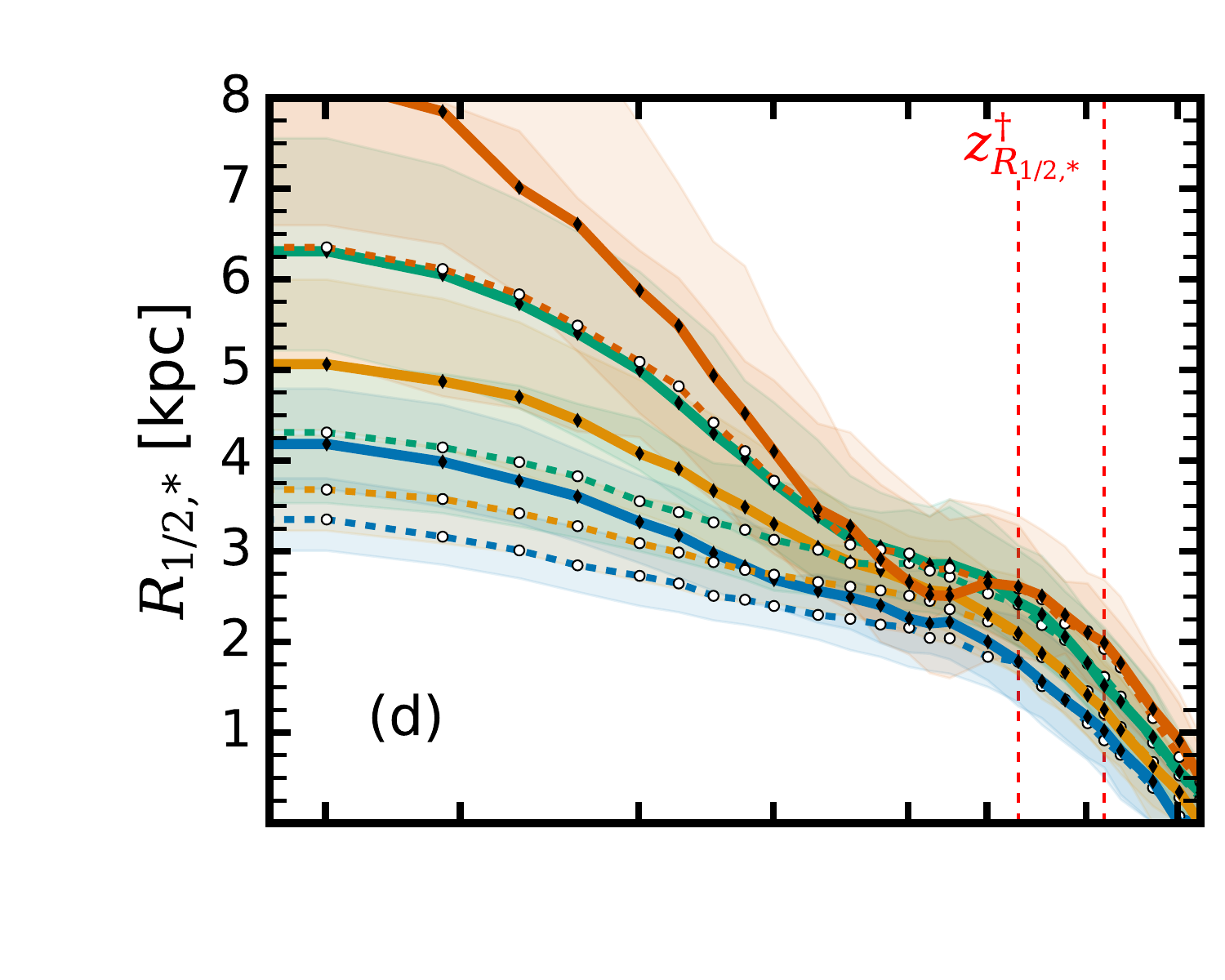}\vspace{-1.36cm}
     \includegraphics[width=0.75\columnwidth,angle=0]{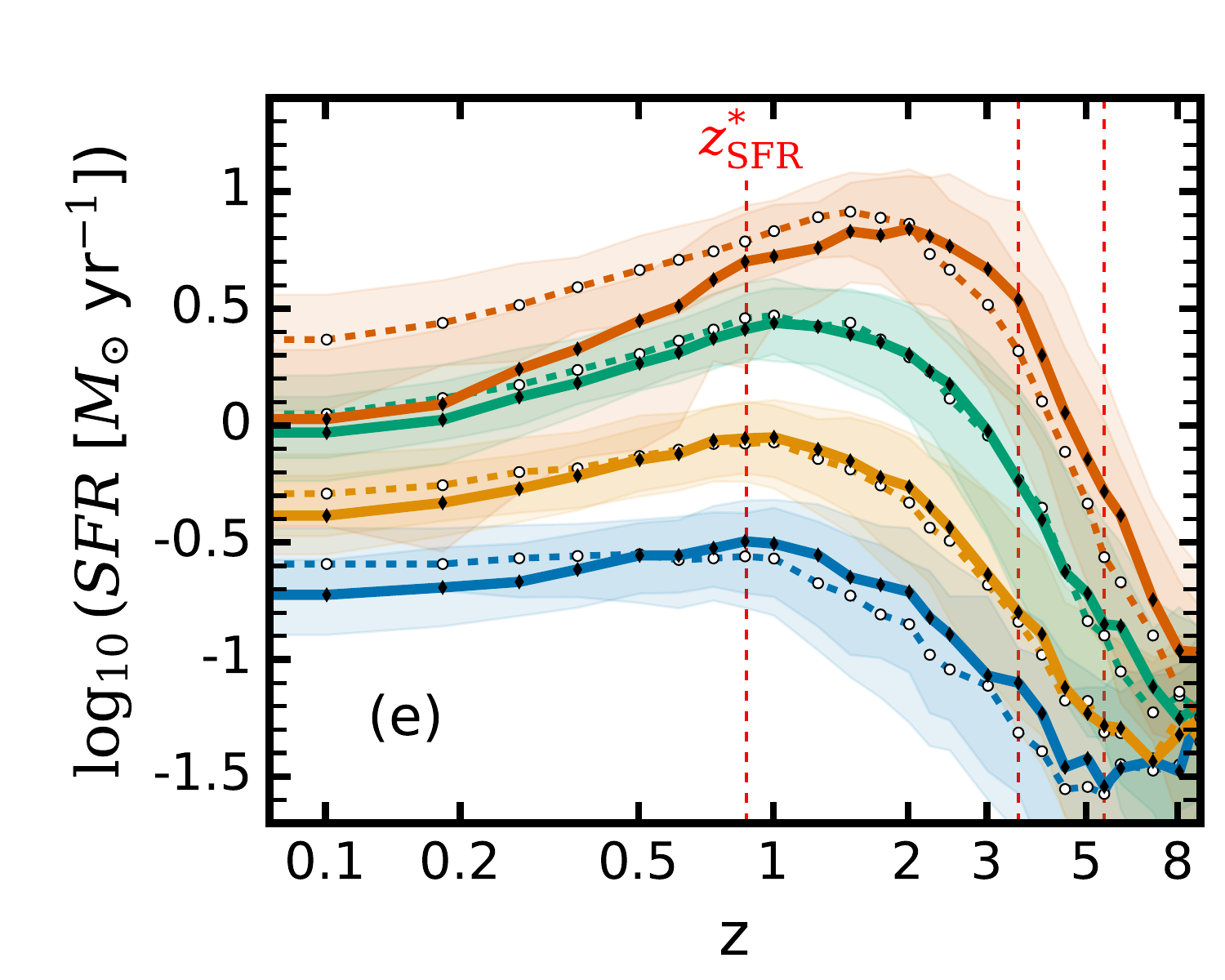}
    \caption{Redshift evolution of the median values alongside the $25^\text{th}$ and $75^\text{th}$ percentiles of \Mstar, \Mbh, $j_*$, \rhalfstars, and SFR (from top to bottom), binned by stellar mass at $z=0$. \lsbs\ are shown as solid thick coloured lines with black diamond markers, while \hsbs\ are represented by dashed thin coloured lines with white circular markers. Vertical dashed red lines with annotations represent the divergence events as described in \hyperref[tab:events]{\Tab{tab:events}}.}\label{fig:zevol_histo2D_M2}
\end{figure}

In \hyperref[fig:mstar_assembly]{\Fig{fig:mstar_assembly}} we demonstrated that \lsbs\ and \hsbs\ share basically identical median stellar mass assembly tracks using the entire mass range of our homogenised sample ($10^9<\Mstar\ [\Msun]<10^{11}$). However, this is a fairly brought mass range, therefore we repeat our analysis and examining the redshift evolution within four narrow bins at $z=0$: $10^{9} < \Mstar\ [\Msun] < 10^{9.5}$ (436 galaxies, 27\%), $10^{9.5} < \Mstar\ [\Msun] < 10^{10}$ (622 galaxies, 39\%), $10^{10} < \Mstar\ [\Msun] < 10^{10.5}$ (378 galaxies, 24\%), and $10^{10.5} < \Mstar\ [\Msun] < 10^{11}$ (159 galaxies, 10\%). It is important to note that the galaxy number counts in each bin are fixed at $z=0$ for both \lsbs\ and \hsbs\ and remain consistent across higher redshifts. For example, in the most massive bin, each population constitutes 10\% of the total sample.

In \hyperref[fig:zevol_histo2D_M2]{\Fig{fig:zevol_histo2D_M2}}, we show from top to bottom the assembly/redshift evolution of \Mstar, \Mbh, $j_*$, and \rhalfstars, and SFR within the redshift range $0<z<8$. We plot \lsbs\ as solid lines with diamond-shaped markers and \hsbs\ as dashed lines with circular markers. The colours correspond to the mass bins defined in the legend of the figure. As before, we plot median values and the shaded regions represent the $25^\text{th}$ and $75^\text{th}$ percentiles. Additionally, data points are shown only if there are more than 25 objects in each histogram bin.

The overall trends remain consistent with our findings for the full mass range of \lsbs\ and \hsbs\ as shown for the stellar mass assembly in \hyperref[fig:zevol_histo2D_M2]{\Fig{fig:zevol_histo2D_M2}}a. However, analysing different mass bins separately  reveals additional trends that remain hidden when considering the entire population, allowing for a more detailed analysis. The most massive \lsbs\ ($\Mstar>10^{10.5}~\Msun$, brown solid lines) exhibit significantly higher \Mbh\ at $z=0$ and \Mbh\ more rapidly than their \hsb\ counterparts (see \hyperref[fig:zevol_histo2D_M2]{\Fig{fig:zevol_histo2D_M2}}b). \lsbs\ in this mass bin have \Mbh\ values about 0.5 dex higher than \hsbs\ and show increased black hole accretion rates between $3<z<5$. In contrast, intermediate-mass galaxies ($10^{9.5} < \Mstar\ [\Msun] < 10^{10}$ and $10^{10} < \Mstar\ [\Msun] < 10^{10.5}$, blue and yellow lines) display the most similar evolutionary tracks between \lsbs\ and \hsbs, however small differences are visible although not as pronounced as for \lsbs\ in the highest mass bins.

Another sign of the faster assembly of \lsbs\ is their earlier presence in higher stellar mass bins when studying 2D histograms, whereas \hsbs\ either appear later or in lower numbers, indicating a general "delay" in their evolution. This is reflected in their half-mass assembly times (\thalf), which mark when 50\% of each mass component is assembled relative to $z=0$. \lsbs\ build up their stellar, halo, and black hole masses earlier and more rapidly. While the differences in stellar and halo mass assembly are modest (0.25–0.5 Gyr), \lsbs\ reach their black hole half-mass assembly approximately 0.6 Gyr earlier than \hsbs\ (see  \hyperref[tab:dens6b_new_SB]{\Tab{tab:dens6b_new_SB}}).

In \hyperref[fig:zevol_histo2D_M2]{\Fig{fig:zevol_histo2D_M2}}c,d, we show the evolution of $j_*$ and \rhalfstars\ across stellar mass bins. For $j_*$ the divergence starts at \zjstars\ for all mass bins expect the most massive one mass bins, consistent with the full sample (\hyperref[fig:zevol_props]{\Fig{fig:zevol_props}}a), however, the divergence seems not to exist for galaxies with $\Mstar > 10^{10.5}~\Msun$. The separation in \rhalfstars\ is most pronounced in the intermediate-mass bin ($10^{10} < \Mstar\ [\Msun] < 10^{10.5}$, green lines), followed by the most massive bin, while the lowest mass bin shows a weaker divergence. 

For the most massive \lsbs, we find that \rhalfstars\ contracts around $z\sim2$ (dip in the solid brown line in \hyperref[fig:zevol_histo2D_M2]{\Fig{fig:zevol_histo2D_M2}}d), a behaviour absent in their \hsb\ counterparts. We propose that a disruptive event at this redshift alters the baryonic and dark matter mass distribution, with \lsbs\ being more susceptible due to their intrinsic properties. Notably, in this highest mass bin, the evolution of $j_*$ and \rhalfstars\ appears decoupled: while $j_*$ follows a similar trajectory in both \lsbs\ and \hsbs\ (\hyperref[fig:zevol_histo2D_M2]{\Fig{fig:zevol_histo2D_M2}}c), \rhalfstars\ exhibits a distinct contraction, suggesting that angular momentum alone does not dictate size evolution in these massive systems.

In \hyperref[fig:zevol_histo2D_M2]{\Fig{fig:zevol_histo2D_M2}}e, we present the evolution of the SFR. All \lsb\ mass bins exhibit a continuous decline in SFR towards lower redshifts as shown for the overall population in \hyperref[fig:zevol_props]{\Fig{fig:zevol_props}}f. More massive systems quench earlier and more rapidly, in line with the classical picture of galaxy star formation histories. For both the lowest-mass ($10^{9} < \Mstar\ [\Msun] < 10^{9.5}$) and the most massive ($\Mstar\ [\Msun] > 10^{10.5}$) systems, \lsbs\ show higher SFRs at earlier cosmic times, reaching their turnover points around $z\sim2$ for the most massive, and $z \sim 0.5$ for the lowest-mass systems, respectively.

\subsection{Results on dispersion- and rotation-supported systems}\label{sec:res_bins_dyn}

\begin{figure}
    \centering
    \includegraphics[width=0.75\columnwidth,angle=0]{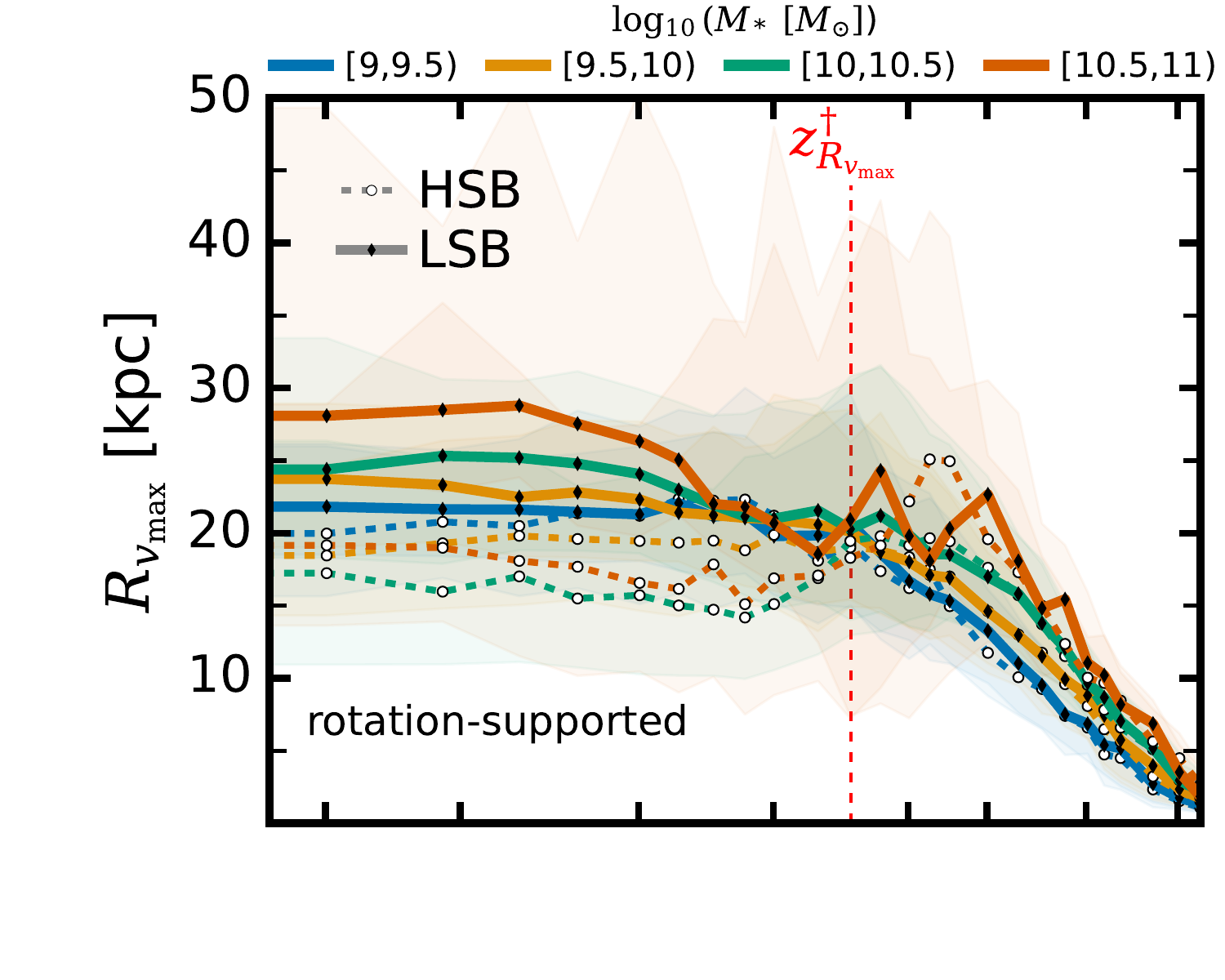}\vspace{-1.36cm}
    \includegraphics[width=0.75\columnwidth,angle=0]{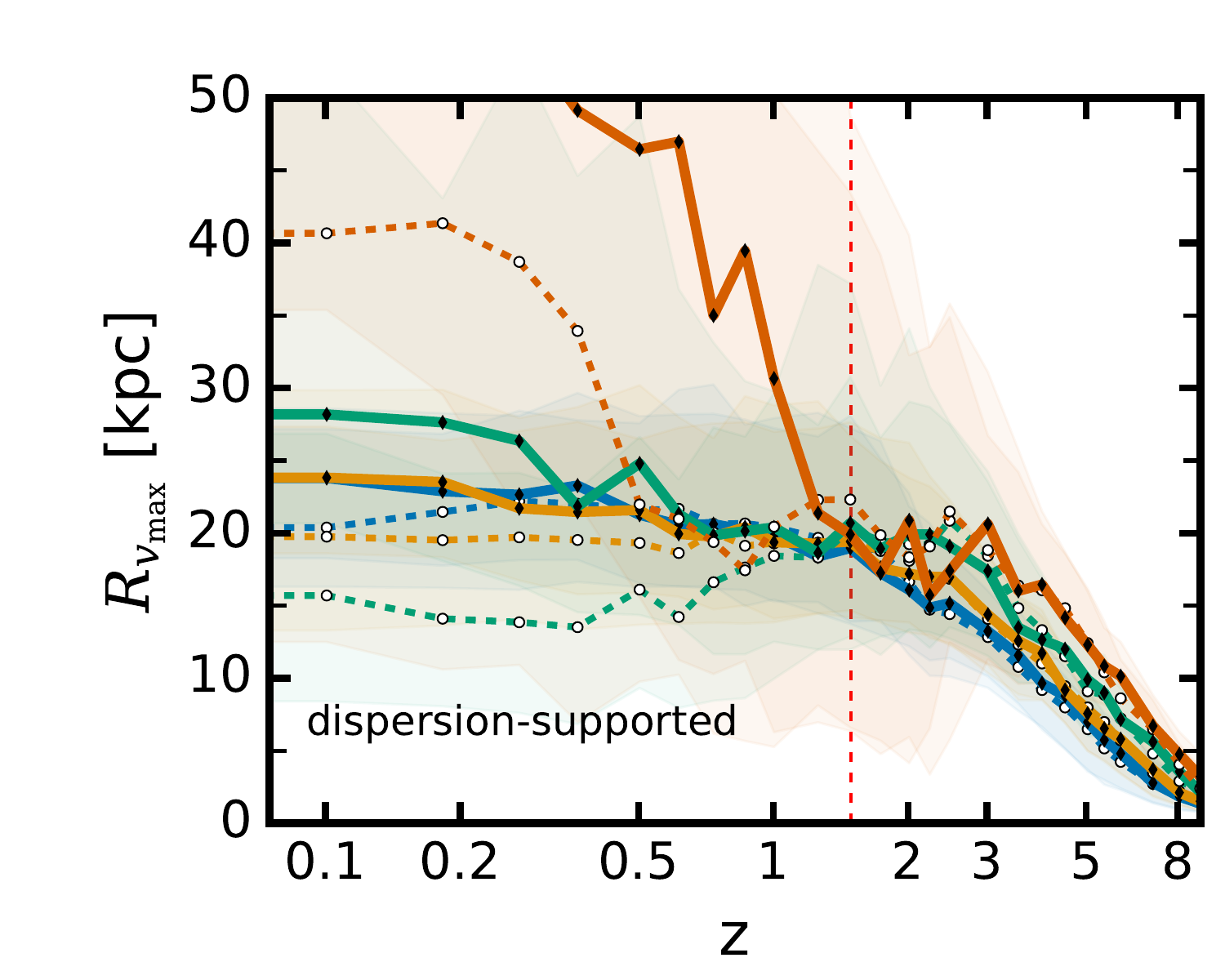}
    \caption{Redshift evolution of the median values of \rvmax, binned by stellar mass at redshift $z=0$, is shown for dispersion-  (top) and rotation-supported systems (bottom). See \hyperref[fig:zevol_histo2D_M2]{\Fig{fig:zevol_histo2D_M2}} for colour and symbol coding, and refer to \hyperref[sec:sample_envr]{\Sec{sec:sample_envr}} for statistical definitions.}\label{fig:zevol_rvmax_dyn_M2}
\end{figure}

In this section we study the evolution of \rvmax\ in bins of stellar mass as introduced in \hyperref[sec:res_bins]{\Sec{sec:res_bins}} together with their dynamical state. The dynamical state is obtained from the \VvStwoR -- \ellTwost\ plane shown in \hyperref[fig:results_dyn_unbiased]{\Fig{fig:results_dyn_unbiased}} and sub-categories \lsbs\ and \hsbs\ in dispersion- and rotation-supported system as discussed in \hyperref[sec:res_gas_dyn]{\Sec{sec:res_gas_dyn}}. We find that 28\% of \lsbs\ in our dataset are rotation-supported and 22\% dispersion-supported, while only 19\% of \hsbs\ are rotation-supported, but 31\% dispersion-supported. That is aligned with the general picture that \lsbs\ tend to me more disc-oriented \citep{Pohlen03_LSBGs_bar,Pahwa+Saha18_LSBGs_OBS,Saburova19_UGC1378}. 

In general the differences in the evolution of properties between rotation-supported than dispersion-supported systems with similar stellar masses is more pronounced in \lsbs. For example the star formation activity is basically the same for dispersion- and rotation-supported \hsbs\ at fixed \Mstar, however, the most massive ($\Mstar>10^{10.5}~\Msun$) dispersion-supported \lsbs\ show a significantly decrease in SFR at $z<1.5$ in comparison to their rotation-supported counterparts. A similar trend can be found for other properties such as gas masses and gas fractions, baryon-to-dark matter particle fractions, \Mbh, and \rvmax.

\rvmax\ highlights the most pronounced dynamical differences, as shown in \hyperref[fig:zevol_rvmax_dyn_M2]{\Fig{fig:zevol_rvmax_dyn_M2}}. We present its evolution separately for rotation-supported (top) and dispersion-supported (bottom) systems. Our general result, that \lsbs\ on average have larger \rvmax, still holds. However, the most massive \lsbs\ and \hsbs\ ($\Mstar>10^{10.5}$ \Msun) are actually those which exhibit the largest \rvmax\ at $z=0$ and are predominantly dispersion-supported, whereas their rotation-supported counterparts do not show this extreme behaviour. Low-mass \lsbs\ and \hsbs\ ($10^{9} < \Mstar\ [\Msun] < 10^{9.5}$) with dispersion- and rotation-supported systems follow similar \rvmax\ evolution with only mild divergence found at low redshift.

High-mass dispersion-supported \lsbs\ appear to be most sensitive to disruption events which could affect their baryonic and dark matter distribution and expand their \rvmax. To explore this further, we measured the offset between the centre of mass and the centre of potential, finding that \lsbs\ exhibit larger displacements than \hsbs, particularly in the redshift range $1.5<z<3$, possibly indicating the aftermath of a significant merger event. At lower redshifts, \lsbs\ realign with their centres of mass, and by $z=0$, they are generally closer to them than \hsbs. Additionally, \lsbs\ show a higher baryon-to-dark matter fraction between $z\sim2-5$, but this trend reverses at later times: while \hsbs\ retain or gain baryons up to $z=0$, \lsbs\ experience a gradual decline in their baryonic fraction.

\subsection{Results on occurrences of last merger events}\label{sec:res_merger}

\begin{figure*}
    \centering
    \includegraphics[width=0.8\textwidth,angle=0]{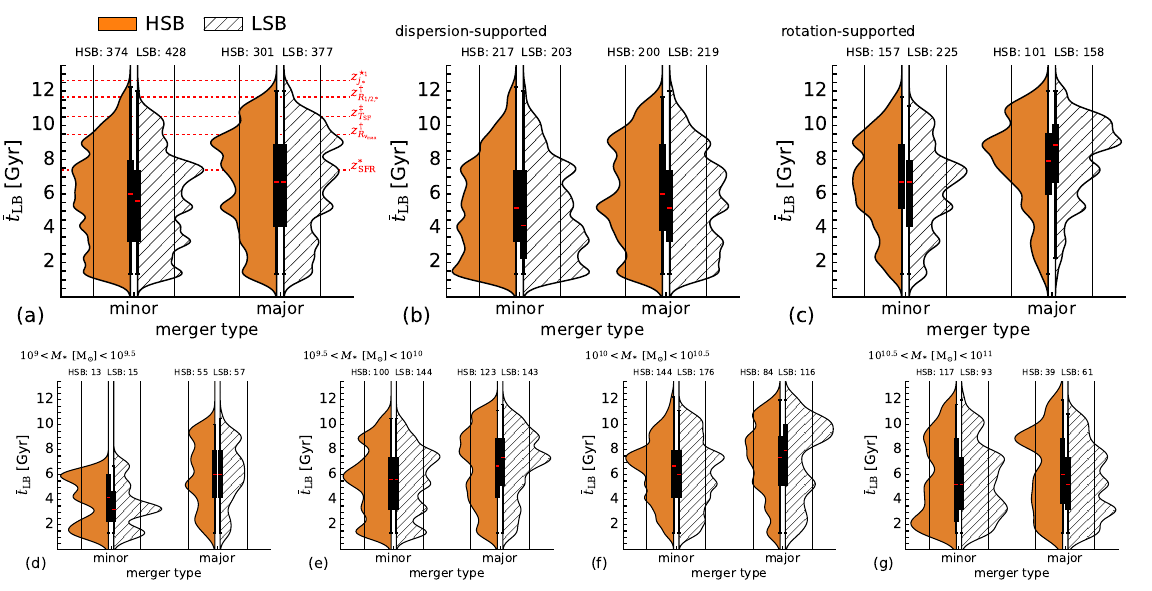}
    \caption{PDFs of last \tLB\ for minor and major mergers, showing distributions for the full \lsb\ and \hsb\ populations (panel a), dispersion-supported (panel b) and rotation-supported galaxies (panel c), as well as by stellar mass bin (panels d–g). \hsbs\ are represented by solid orange bars and \lsbs\ are as white bars with diagonal hatching. To aid visual interpretation of number densities, we include mirrored vertical black lines along the boxes. In panel (a), the horizontal red dashed lines with annotations mark the lookback times of the divergence events listed in \hyperref[tab:events]{\Tab{tab:events}}. For statistical definitions, refer to \hyperref[sec:sample_envr]{\Sec{sec:sample_envr}}.}\label{fig:merger_history}
\end{figure*}

We obtain median lookback times (\tLB) in Gyrs for the last minor and major mergers from R22 for approximately 50\% of our \lsb\ and \hsb\ samples. Mergers are classified as \textit{major} when the stellar mass ratio of the secondary to the primary galaxy exceeds 0.25, while \textit{minor} mergers have mass ratios between 0.1 and 0.25.

In \hyperref[fig:merger_history]{\Fig{fig:merger_history}}, we present PDFs of \tLB\ for minor (on the left side of each panel) and major mergers (on the right side of each panel) alongside the median values (box plot with whiskers), showing distributions for the full \lsb\ and \hsb\ populations (panel a), dispersion-supported (panel b) and rotation-supported galaxies (panel c), as well as by stellar mass bin (panels d–g). \hsbs\ are shown as solid orange bars, while \lsbs\ as white bars with diagonal hatches. To aid visual interpretation, vertical thin black lines are mirrored along the boxes, and the number count of galaxies is displayed as PDFs in each half-violin on the upper halves of the panels. Additionally, in \hyperref[fig:merger_history]{\Fig{fig:merger_history}}a, horizontal red dashed lines with annotations mark the lookback times of divergence events from \hyperref[tab:events]{\Tab{tab:events}}.

\hyperref[fig:merger_history]{\Fig{fig:merger_history}}a presents \tLB\ for all \lsbs\ and \hsbs. Both populations show a peak in last major mergers around $\tLB\sim10-11$ Gyr, coinciding with the divergence of \tempSF\ and \rvmax, while \lsbs\ experienced fewer last minor mergers in the same period. The last minor merger generally occurred at slightly lower \tLB\ in \lsbs, whereas the timing of the last major merger is similar for both populations. Additionally, \lsbs\ exhibit a more varied last minor merger occurrence, particularly between $\tLB\sim8-4$ Gyr, with distinct local peaks aligning with the \zSFR\ divergence event, whereas \hsbs\ follow a smoother, more continuous distribution.

We find that \lsbs\ generally experience last major mergers earlier, but last minor mergers later compared to \hsbs. This pattern emerges when comparing median \tLB\ values (red bars in the box plots in \hyperref[fig:merger_history]{\Fig{fig:merger_history}}). In six cases (panels b–g), \lsbs\ exhibit higher median \tLB\ for last major mergers three times (panels c, e, and f), while \hsbs\ do so twice (panels b and g). For last minor mergers, the trend reverses: \hsbs\ show higher \tLB\ three times (panels b, d, and f), while in three cases (panels c, e, and g), both populations share similar median values.

The merger occurrences of dispersion-supported systems are largely similar for \lsbs\ and \hsbs\ (\hyperref[fig:merger_history]{\Fig{fig:merger_history}}b), except that \lsbs\ exhibit lower \tLB\ values for both last minor and major mergers. In contrast, rotation-supported systems (\hyperref[fig:merger_history]{\Fig{fig:merger_history}}c) show that the earliest major merger in \lsbs\ occurred around $\tLB\sim10$ Gyr, roughly 0.5 Gyr earlier than in \hsbs, though their overall last major merger occurrence remain comparable. The minor merger histories of \lsbs\ and \hsbs\ reveal notable differences in distribution, yet their median lookback times are similar ($\tLB\sim7$ Gyr).

In the lower panels of \hyperref[fig:merger_history]{\Fig{fig:merger_history}}d-g, we examine the PDFs of \lsbs\ and \hsbs\ across the stellar mass bins introduced in \hyperref[sec:res_bins]{\Sec{sec:res_bins}}, ranging from the lowest to the highest bin (left to right). The merger occurrences of galaxies in the lowest mass bin exhibit similar PDF shapes, with the key difference that \lsbs\ experience peaks in last minor merger activity significantly later in cosmic history ($\tLB\sim3$ Gyr) compared to their \hsb\ counterparts. However, the distribution of the occurrences of the last major merger of both populations are well aligned. In the intermediate mass bins, shown in \hyperref[fig:merger_history]{\Fig{fig:merger_history}}e, the PDFs for last minor and major mergers are also largely similar, except for a pronounced peak in \lsb\ last major mergers around $\tLB\sim8$ Gyr.

The most distinct differences emerge in the higher mass bins in \hyperref[fig:merger_history]{\Fig{fig:merger_history}}f,g. We find a notable accumulation of \lsb\ last major mergers at $\tLB\sim8-10$ Gyr in panel (f), as well as generally higher late-time merger activity ($\tLB>4$ Gyr) for \lsbs\ compared to \hsbs\ in both mass bins. Additionally, in the highest mass bin (panel g), \lsbs\ exhibit last merger events later than their \hsb\ counterparts, as indicated by the whiskers in the box plot.

\section{Discussion and conclusion}\label{sec:discussion}

\subsection{What processes led to the development of the observed LSB features in galaxies at low redshifts?}

Our findings suggest that events before $z\sim5$ are crucial in shaping LSB features observed at $z=0$. In particular, the divergence in $j_*$ initiates a critical transition phase that foster the emergence of those features (see \hyperref[sec:res_zevol]{\Sec{sec:res_zevol}}). Furthermore, subsequent evolution is mass-dependent and with minor influences of the large-scale environment (see \hyperref[sec:res_bins]{\Sec{sec:res_bins}} and \hyperref[sec:res_envr]{\Sec{sec:res_envr}}). Furhtermore, we propose that early gas accretion and quicker hierarchical growth, where global angular momentum could have been conserved, drove the formation of LSB features, generally in agreement with previous studies \citep[see][]{DiCintio19_LSBGs_NIHAO,Martin19_LSBGs,Perez-Montano+Cervantes-Sodi19_LSBGs_env,Tremmel20_RomulusC_UDGs,Benavides21_TNG_UDGs,Wright21_Romulus25_UDGs,Perez-Montano22_LSBGs_TNG,Perez-Montero24_LSBGs_envr_angM}.

We gather evidence of more rapid evolution of \lsbs\ in comparison to \hsbs\ at $z>3$ for the most massive galaxies in our sample (see \hyperref[fig:zevol_histo2D_M2]{\Fig{fig:zevol_histo2D_M2}}), similar to what was reported by \citet{Ferre-Mateu25_dLSBs_ass}, who studied dwarf-LSBGs. In contrast, other studies describe LSBGs as a slowly evolving population \citep[e.g.][]{Bothun90_LSBGs_Malin2,Bothun93_LSBGs_env,Mo94_LSBGs_clustering,deBlok+McGaugh97_LSBGs_halo,Gerritsen+deBlok99_LSBGs_env,vanDenHoek00_LSBGs_SFH}. We suggest that LSBGs form quickly but transition to a slower evolutionary phase later on where their early assembly within immature dark matter haloes helps in setting the foundation for their LSB nature. This process appears mass-dependent, with early stellar feedback potentially playing a role by regulating supermassive black hole growth and feedback \citep{Martin19_LSBGs,Tillman+23_AGN}.

\subsection{Is there a uniform channel for LSB evolution?}

Our results indicate that there is no single evolutionary pathway leading to low-surface brightness; instead, multiple channels exist, shaped by both stellar mass and the galaxy’s dynamical state. For the most massive \lsbs\ ($\Mstar>10^{10.5} \Msun$), we identify at least two distinct formation channels, likely linked to the galaxy's dynamical state: either dispersion- or rotation-supported. In these cases, early mass assembly, mergers, gas accretion, and angular momentum all appear to play crucial roles (see \hyperref[sec:res_bins]{\Sec{sec:res_bins}} and \hyperref[sec:res_bins_dyn]{\Sec{sec:res_bins_dyn}}). A third channel may contribute to the development of LSB features in lower-mass galaxies ($10^9<\Mstar [\Msun]<10^{9.5}$), where angular momentum appears to be the key driver. This is likely supported by ongoing minor merger activity, as these galaxies tend to have accumulated their last minor merger at $\tLB<10$ Gyr (see \hyperref[fig:merger_history]{\Fig{fig:merger_history}}d).

The key difference between \lsbs\ and \hsbs\ in our dataset lies in their response to external events such as mergers and gas accretion. The most massive 10\% of \lsbs, which are predominantly dispersion-supported, undergo significant structural transformations around $z\sim3$ ($t_{\mathrm{LB}}\sim11$ Gyr). These events initially cause a contraction in \rhalfstars, followed by an expansion (see \hyperref[fig:zevol_histo2D_M2]{\Fig{fig:zevol_histo2D_M2}}d and \hyperref[fig:zevol_histo2D_M2]{\Fig{fig:zevol_histo2D_M2}}c), supporting our hypothesis that \lsbs\ are dynamically and structurally distinct from their \hsb\ counterparts at a given mass. This disruption is also reflected in the evolution of \rvmax, which exhibits oscillatory behaviour within the redshift range $1.5<z<4$ for both dispersion- and rotation-supported galaxies across various mass bins (see \hyperref[fig:zevol_rvmax_dyn_M2]{\Fig{fig:zevol_rvmax_dyn_M2}}). These oscillations likely indicate the occurrence of major merger events during this epoch.

Additionally, the assembly of \Mbh\ appears to play as well a key role in the development of LSB features, particularly in dispersion-supported systems with $\Mstar\ [\Msun]>10^{10}$. Although not shown in a separate figure (but available upon request), we find that central black holes in these galaxies tend to form earlier and more rapidly, resulting in higher black hole efficiencies (\Mbh/\Mc) and larger final black hole masses at $z=0$. Furthermore, they also exhibit lower baryon to dark matter particle fraction. These trends are not observed in rotation-supported galaxies within the same mass range.

Massive, rotation-supported \lsbs\ such as \texttt{Malin 1} \citep{Zhu18_IllTNG_Malin1,Galaz22_Malin1} on the other hand, may have formed through a major merger or accretion event at high to intermediate redshift, as these galaxies typically show a accumulation of last major merger events at earlier cosmic times (see \hyperref[fig:zevol_histo2D_M2]{\Fig{fig:zevol_histo2D_M2}}c,g) than their dispersion-supported counterparts. In combination with disc instabilities, as proposed by \citet[][]{Noguchi01_gLSBGs_disc}, such events may have driven material outward, promoting the formation of an extended, low-density disc consistent with the observed properties of systems like \texttt{Malin 1}.

\subsection{Comparison to previous studies}

In the previous section we proposed and discussed various channels of how LSB features could have been developed in our simulation. A Similar scenario was previously proposed by \citet{DiCintio19_LSBGs_NIHAO} using zoom-in hydrodynamical simulations. They identified two formation pathways for LSBGs: a feedback-dominated process for low-mass galaxies ($\Mstar\sim10^{7-9}$ \Msun) and an angular momentum-driven mechanism for more massive systems ($\Mstar\sim10^{9.5-10}$ \Msun), where co-planar mergers contribute angular momentum to the disc in LSBGs, whereas perpendicular mergers in HSBGs might instead remove angular momentum. While our study does not cover the same stellar mass range, our findings similarly suggest a dichotomy in the evolutionary pathways of low- and high-mass LSBGs, as well as a distinction between those that are more rotation- or dispersion-supported (see  \hyperref[fig:zevol_histo2D_M2]{\Fig{fig:zevol_histo2D_M2}} and \hyperref[fig:zevol_rvmax_dyn_M2]{\Fig{fig:zevol_rvmax_dyn_M2}}).

Our findings using the \eagle\ simulation align with studies such as \citet{Martin19_LSBGs}, who examined so-called \textit{ultra-diffuse galaxies} (UDGs) -- the extreme diffuse end of LSBG population -- in the \texttt{HorizonAGN} simulation \citep{Dubois14_HorizonAGN}. They found that UDGs and HSBGs share common progenitor and show identical distribution of gas fraction and effective radii until $z\sim2$. Furthermore, they also formed stars more rapidly at early epochs. Similarly, our results support the findings of \citet{Perez-Montano22_LSBGs_TNG,Perez-Montero24_LSBGs_envr_angM} using \IllTNG, especially in that LSBGs are more extended and exhibit higher angular momentum at $z=0$. These studies also report a similar "contraction" in \rhalfstars\ for the most massive galaxies around $z\sim2-3$ (see \hyperref[sec:res_bins]{\Sec{sec:res_bins}}). However, their sample is built similar as ours using the effective radii in the $r$-band the threshold for surface brightness density as we in \hyperref[sec:res_zevol]{\Sec{sec:res_zevol}}, their stellar mass range $10^{9}<\Mstar~[\Msun]<10^{12}$ is slightly broader and not homogenised across the distribution of stellar and halo mass (see \hyperref[sec:sample_dens]{\Sec{sec:sample_dens}}).

Furthermore, \citet{Perez-Montano22_LSBGs_TNG,Perez-Montero24_LSBGs_envr_angM} report that their LSB population is less massive, more gas-rich, and exhibits later-type morphologies with younger stellar populations. In contrast, we find only slightly higher gas masses in \lsbs\ (see \hyperref[tab:dens6b_new_SB]{\Tab{tab:dens6b_new_SB}}), with KS-tests showing no significant differences in neutral or molecular hydrogen, challenging the assumption that LSBGs are gas-rich \citep{Burton99_LSBGs_LV,Schombert01_dLSBGs_SF-scenario,ONeil04_massive_LSBGs}. Additionally, \lsbs\ in our dataset host stellar populations at least 1 Gyr older (see \hyperref[tab:dens6b_new_SB]{\Tab{tab:dens6b_new_SB}} and \hyperref[fig:results_mstar_unbiased]{\Fig{fig:results_mstar_unbiased}}b).

Furthermore, \citet{Perez-Montano22_LSBGs_TNG} reported lower \Mbh\ in LSBGs at $z=0$, whereas we find no such difference for the majority of our galaxies; both populations show comparable \Mbh\ and assembly histories with no statistically significant discrepancy (see \hyperref[fig:zevol_histo2D_M2]{\Fig{fig:zevol_histo2D_M2}}b). This contrast may arise from a subtle stellar mass dependence at $z=0$ (see their Fig.\ 3), which we account for through our strict mass-matching procedure.

Compared to other studies \citep[i.e.,][]{Rosenbaum09_LSBGs_env,Galaz11_LSBGs_SDSS,Du15_LSBGs_obs,Alabi20_LSBGs_COMA,Zhu23_TNG100_gLSBGs,Kulier20_LSBGs_Eagle}, we find no strong evidence that \lsbs\ favour a particular large-scale environment such as dense regions like cluster environments or less dense such as filaments and walls. Any detected trend appears mild, consistent with previous findings \citep[e.g.,][]{Shao15_LSBGs_env,Perez-Montano+Cervantes-Sodi19_LSBGs_env}. However, properties such as SFR may be influenced by a more complex interplay -- such as a superposition of mass and environmental effects -- that is challenging to disentangle when analysing the full sample. In future works, we plan to investigate this aspect in more detail by exploring the star formation histories and the \textit{Kennicutt–Schmidt relation} across different environments, and within narrow stellar mass bins. Cutting-edge observational experiments, such as \citet{Garcia-Benito24_CAVITY}, will provide excellent data for this endeavour.

Differences in LSBG properties across studies may also arise from variations in the physical models employed by different simulations. For instance, \IllTNG\ and \eagle\ adopt distinct feedback mechanisms: the former implements stronger AGN feedback, redistributing material to larger radii \citep[see e.g.][]{Zinger20_TNG_feedback}. Additionally, black hole seeding occurs later (at higher masses) in \IllTNG\ than in \eagle\ \citep{Habouzit21_BH}. These fundamental differences in physical implementations likely contribute to the divergent black hole properties reported in different works \citep{Crain23_Hydro_comparion_rev,Wright24_Hydro_comparison}.

We do not compare our results to observational data in this work. Our primary goal was to contrast \lsbs\ and \hsbs\ under controlled conditions, specifically eliminating stellar and halo mass biases. Constructing an observational sample with comparable selection criteria is challenging due to inherent selection effects and limitations in detecting LSB systems. Therefore, we leave a meaningful comparison to observations to future works.

Our study provides a comprehensive view of LSBG evolution in the \eagle\ simulation, aligning with key trends found in other models. We also reveal new insights, such as the divergence of \rvmax\ at $z\sim1.5$ for \lsbs\ and \hsbs, a feature that has not been shown before. Though the overall picture remains consistent across different simulations, the discrepancies highlighted in this section underscore the impact of model choices and sample selection, since LSBG evolution is mass-dependent and sensitive to the dynamical state of the galaxies (i.e.; dispersion- or rotation-supported). By carefully matching stellar and halo mass, we have minimised some of these biases. We reinforce the importance of using unbiased samples in future studies to achieve a clearer and more robust understanding of LSBG evolution.

\section{Summary} \label{sec:summary}

Through a series of statistical tests, we analysed the structural, kinematical, and environmental properties of galaxies classified as low-surface brightness (\lsbs) and high-surface brightness (\hsbs) in our homogenised samples, ensuring fixed number densities in stellar and halo mass bins for both populations. Using the cosmological hydrodynamical simulation \eagle\ \citep{Schaye15_EAGLE}, we demonstrate that \lsbs\ and \hsbs\ exhibit statistically significant differences at $z=0$. By tracing their redshift evolution, we identified key transitional phases in cosmic history where their properties begin to diverge. Our main findings are summarised as follows:

\begin{itemize}
    \item \textbf{Properties of \lsbs\ compared to \hsbs\ at $z=0$:}
    At fixed stellar and halo mass, \lsbs\ are predominantly extended, rotation-supported, low-density galaxies with lower star formation activity and older stellar populations. They exhibit higher angular momenta in both their stellar and gas components, as well as larger \rvmax\ than their \hsb\ counterparts at $z=0$ (see \hyperref[fig:results_mstar_unbiased]{\Fig{fig:results_mstar_unbiased}}). \lsbs\ formed their stellar, halo, and black hole components at least 0.25-0.6 Gyr earlier than \hsbs. See \hyperref[sec:res_mstar]{\Sec{sec:res_mstar}} and \hyperref[tab:dens6b_new_SB]{\Tab{tab:dens6b_new_SB}}.

    \item \textbf{Manifestation of LSB features:}
    We identified five key transition events in their evolution and assembly history, where the properties of \lsbs\ and \hsbs\ begin to diverge after an initial phase of co-evolution (see \hyperref[tab:events]{\Tab{tab:events}}). These transitions are linked to the development of LSB features observed at $z=0$. $j_*$ is the first property to diverge at $z\sim5-7$, likely marking a critical phase that sets the foundation for future LSB characteristics (see \hyperref[sec:res_zevol]{\Sec{sec:res_zevol}}). \lsb\ galaxies have maintained their low-surface brightness nature for most of their lifetimes (see \hyperref[fig:zevol_props]{\Fig{fig:zevol_props}}). 
    
    \item \textbf{Star formation activity and large-scale environment play a minor role on the development of LSB features:}
    The lower star formation activity observed in \lsbs\ at $z<0.9$ appears to be a consequence of early evolutionary processes such as gas accretion linked with angular momenta evolution rather than a direct driver for low-surface brightness (see \hyperref[fig:zevol_props]{\Fig{fig:zevol_props}}f). Additionally, we find no strong preference for \lsbs\ to reside in specific large-scale environments; their distribution across under- and over-dense regions closely resembles that of \hsbs\ (see \hyperref[sec:res_envr]{\Sec{sec:res_envr}}). A more complex interplay -- such as a superposition of mass and environmental effects -- might underlie these trends, making them difficult to disentangle when analysing the full sample.
    
    \item \textbf{Evolutionary pathways that foster low-surface brightness:}
    We propose that \lsbs\ do not originate from a single formation channel but instead follow mass-dependent pathways, where their dynamical and structural states play a significant role (see \hyperref[sec:res_bins]{\Sec{sec:res_bins}} and \hyperref[sec:res_bins_dyn]{\Sec{sec:res_bins_dyn}}). Our study suggests that, unlike \hsbs, \lsbs\ form more rapidly in their early stages but experience a slowdown over time. These differences, along with their distinct structural and dynamical properties, likely influence their ability to respond to external factors such as mergers or gas accretion (see \hyperref[sec:res_merger]{\Sec{sec:res_merger}}). Ultimately, these processes effect their radii and baryon distribution in their systems, leading to the emergence of LSB characteristics (see \hyperref[sec:discussion]{\Sec{sec:discussion}}).
\end{itemize}

\begin{acknowledgements}
DS is funded by the Spanish Ministry of Universities and the European Next Generation Fond under the \textit{Margarita Salas Fellowship} CA1/RSUE/2021-00720 and wants to thank the \textit{JUPP Family Weiz} for their creative spirits. PBT acknowledges partial funding by Fondecyt-ANID 1240465/2024, and ANID Basal Project FB210003. YRG and DS acknowledge support from the European Union’s HORIZON-MSCA-2021-SE-01 Research and Innovation Programme under the \textit{Marie Sklodowska-Curie} grant agreement number 101086388 (LACEGAL). This work has benefited from the publicly available software tools: \textsc{Matplotlib} \citep{Matplotlib}; Python Software Foundation 1990-2025 and Anaconda\footnote{\url{https://www.python.org}, \url{https://www.anaconda.com}}; \textsc{pyenv}\footnote{\url{https://github.com/pyenv/pyenv}}; The CentOS and Fedora Projects\footnote{\url{https://www.centos.org}, \url{https://fedoraproject.org/}}, \textsc{Topcat} \citep{Taylor13_Topcat}. We used OpenAI's \texttt{GPT-4} language model for assistance with language refinement and drafting parts of the manuscript.
\end{acknowledgements}

\bibliographystyle{aa}
\bibliography{archive}

\appendix

\section{Evaluating the mass assembly histories of our dataset}\label{sec:trees}

This section explains how we trace merger trees across cosmic history and assess the quality of these trees for central galaxies in the \eagle\ simulation. Tracing the merger tree is straightforward, but estimating the redshift at which half of the mass components formed is more challenging due to discrete snapshots and possible disruptions in the assembly histories caused by physical or numerical effects. To address this, we developed a strategy to identify realistic trees and exclude those affected by numerical artifacts.

We define the "half-mass assembly times" (\thalf) as the lookback time at which 50\%  or \tseventy\ where 70\% of the halo (\Mc), stellar (\Mstar), or black hole (\Mbh) mass was formed. These times are estimated using the following methods:

\begin{enumerate}
    \item \textbf{Standard Interpolation:}  using the \texttt{numpy.interp} function, which assumes the data is monotonically ascending. However, due to numerical uncertainties and other factors, this criterion is not always satisfied for a particular merger tree.

    \item \textbf{Fitting Approaches:} Curve fitting and cubic spline fitting from the \texttt{SciPy} package are employed to correct the data when needed. After fitting the data, we interpolate the results using the \texttt{interpolate.interp1d} function from the same package.
\end{enumerate}

That provides us with three values for \thalf\ and \tseventy\ for each mass component. The following criteria are applied to evaluate the quality of the merger trees:

\begin{enumerate}
    \item[$\blacksquare$] Starting from $z=0$, the tree must be detectable until at least $z \sim 3.5$ (excluding two merger trees).
    
    \item[$\blacksquare$] The variation in measurement between redshift $z=0$ and the following two snapshots ($z=0.1$ and $z=0.18$, respectively) should not exceed 50\% of the value at $z=0$ (in 266 trees detected).
\end{enumerate}

\begin{figure}
    \centering
    \includegraphics[width=0.7\columnwidth,angle=0]{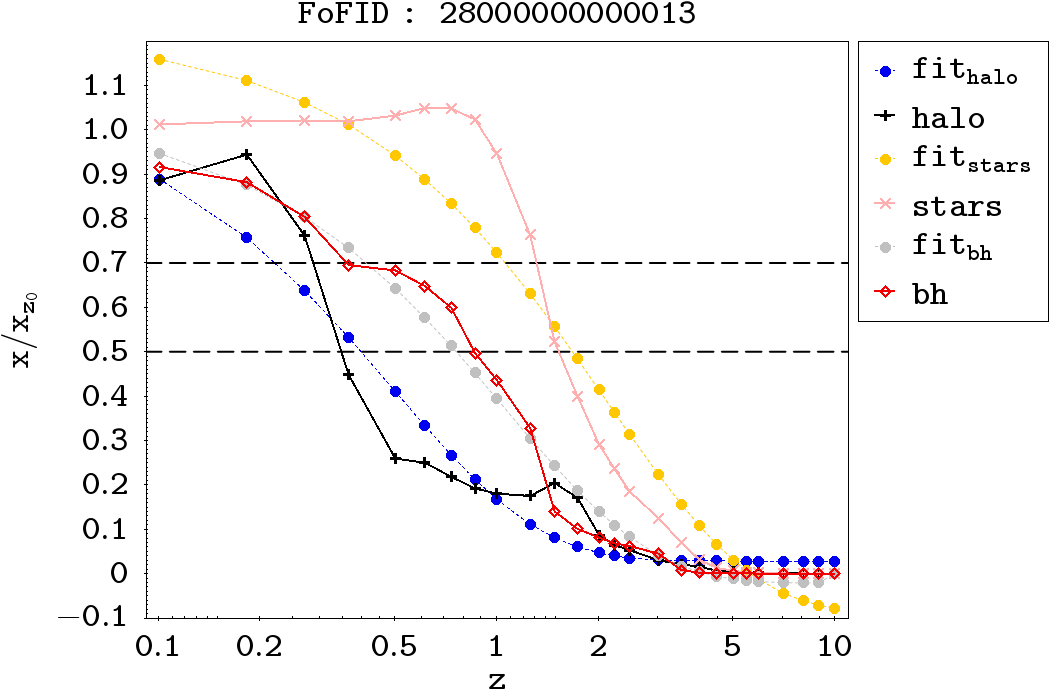}
    \caption{Well-behaved assembly histories for all mass components of the halo identified as \texttt{FoFID 28000000000013}. We plot \Mc\ as a solid black line with crosses, \Mstar\ as a solid salmon line with "$\times$"-shaped markers, and \Mbh\ as a solid red line with small diamonds. The corresponding curve fit functions are shown as dashed lines with coloured dots (blue for \Mc, yellow for \Mstar, and light grey for \Mbh). The horizontal black dashed lines represent  \thalf\ and \tseventy. We observe that these histories display discrete behaviour with some degree of discontinuity. Using methods like curve fitting could result in significantly different estimates for the half-mass assembly time, which we aim to avoid. This example demonstrates that the interpolation approach, without any fit, provides a more accurate result.}\label{fig:inter}
\end{figure}

\begin{figure}
    \centering
    \includegraphics[width=0.7\columnwidth,angle=0]{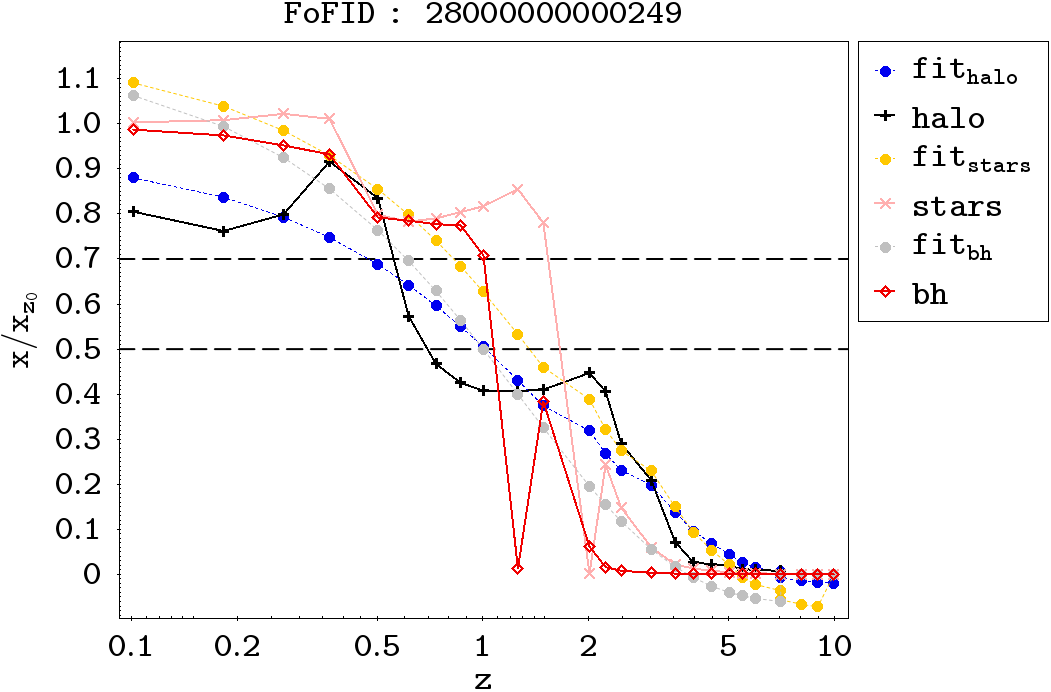}
    \caption{Discounted assembly histories for the halo mass components for the halo identified as \texttt{FoFID 28000000000249}. We use the same colour scheme, marker types, and line style keys as in \hyperref[fig:inter]{\Fig{fig:inter}}. The figure shows that the assembly time \tseventy\ for the halo mass component (black solid line with crosses) cannot be accurately using interpolation alone; therefore, the curve fit procedure is employed, which in this case provides a more realistic result on both half assembly times.}\label{fig:fit}
\end{figure}

We consider a tree "realistic" if the standard deviation between the three approaches is $\sigma < 0.05$\footnote{We chose the threshold for $\sigma$ to be 0.05 because this value corresponds to half the difference in redshift between two consecutive snapshots for at least the first ten snapshots when tracing the merger tree from $z=0$ backwards.}. For such trees, we use the standard interpolation approach, which provides more accurate results, as demonstrated in \hyperref[fig:inter]{\Fig{fig:inter}}. If $\sigma > 0.05$, we evaluate further:

\begin{enumerate}
    \item[$\blacksquare$] If the difference between interpolation and cubic spline fit is smaller than 0.01, we prefer the interpolation method (379 cases).
    
    \item[$\blacksquare$] If the difference exceeds 0.01, we use the curve fit method (68 cases), as it better handles unstable trees (see \hyperref[fig:fit]{\Fig{fig:fit}}.
\end{enumerate}

\begin{figure}
    \centering
    \includegraphics[width=0.7\columnwidth,angle=0]{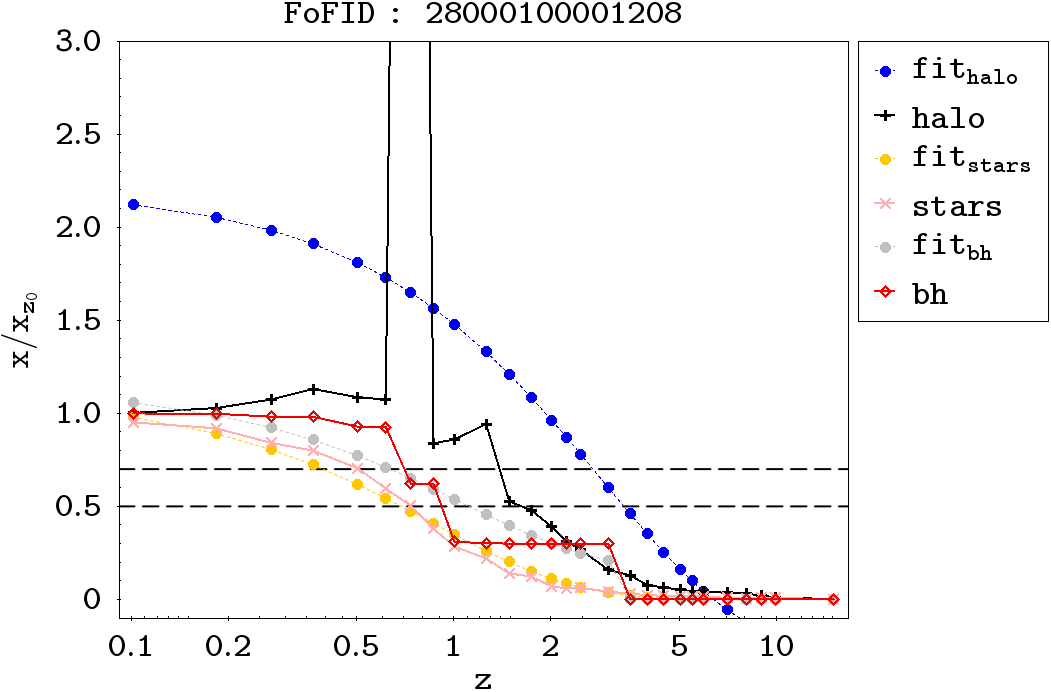}
    \caption{Discarded merger tree with \texttt{FoFID 28000100001208} because of the significant discontinuity of the halo mass assembly history (black line with a prominent peak). We detect an unrealistic curve fit function for the halo mass at $z=0$, where the value exceeds 200\% ($x/x_{z_0} > 2$). Although the stellar mass (salmon "$\times$"-shaped marker) and the central black hole mass (small red diamonds), along with their corresponding curve fit functions, are realistic, we still exclude this merger tree from our study.}\label{fig:dis}
\end{figure}

Severe discontinuities in the assembly history are identified when \thalf\ is lower than \tseventy, with 561 cases detected \hyperref[fig:dis]{\Fig{fig:dis}}. If the mass components’ values at $z=0$ exceed 200\%, the tree is classified as unrealistic, and we exclude these cases (91 galaxies). From a total of 4,466 central galaxies in the \eagle\ simulation, 503 merger trees (about 10\%) were excluded based on the criteria described in this section.

\section{Correction of the optical radius}\label{sec:correct_ropt}

\begin{figure}
    \centering
    \includegraphics[width=0.665\columnwidth,angle=0]{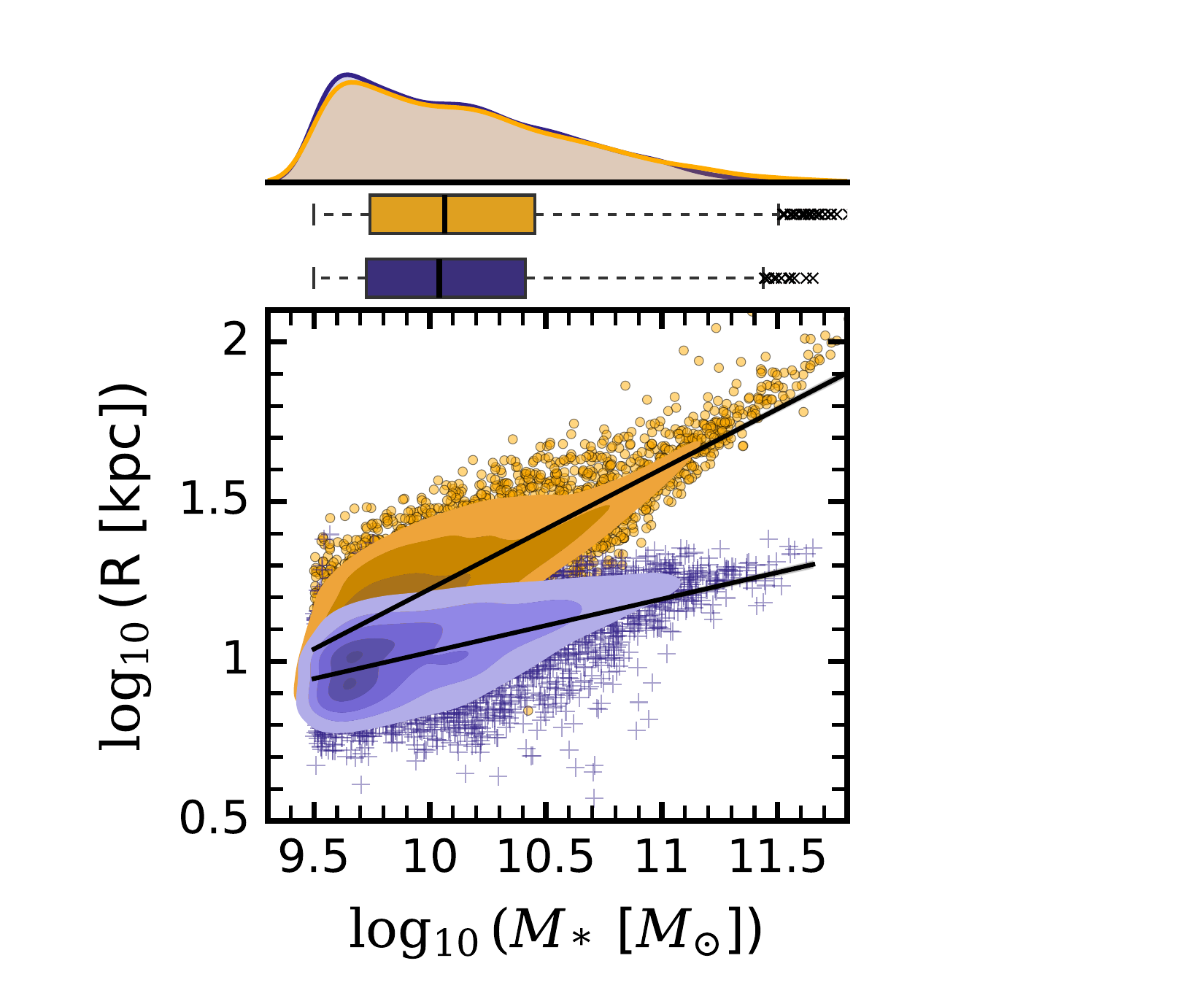}\hspace{-3.1cm}
 	\includegraphics[width=0.665\columnwidth,angle=0]{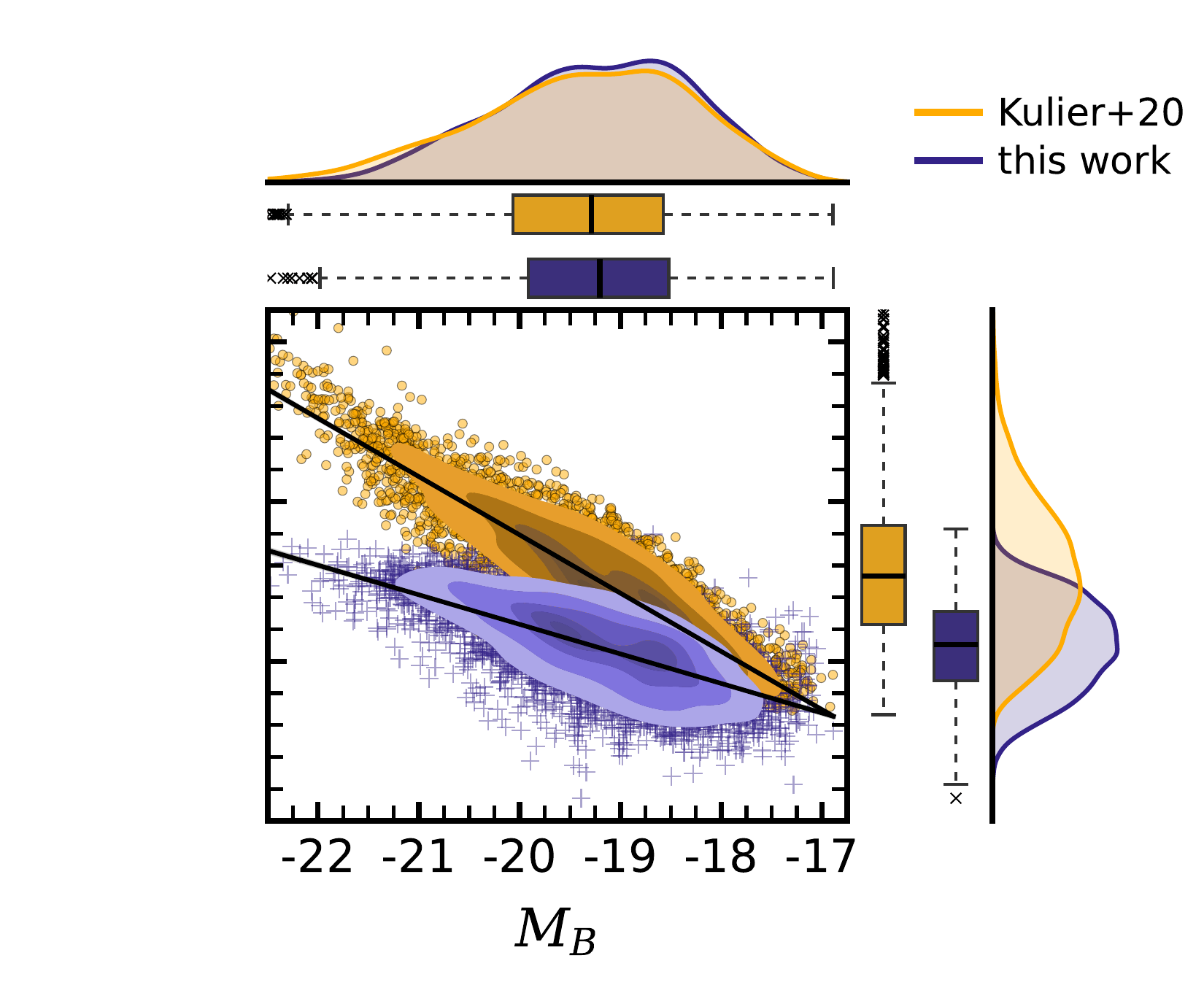}\vspace{-0.2cm}\\
 	\includegraphics[width=0.665\columnwidth,angle=0]{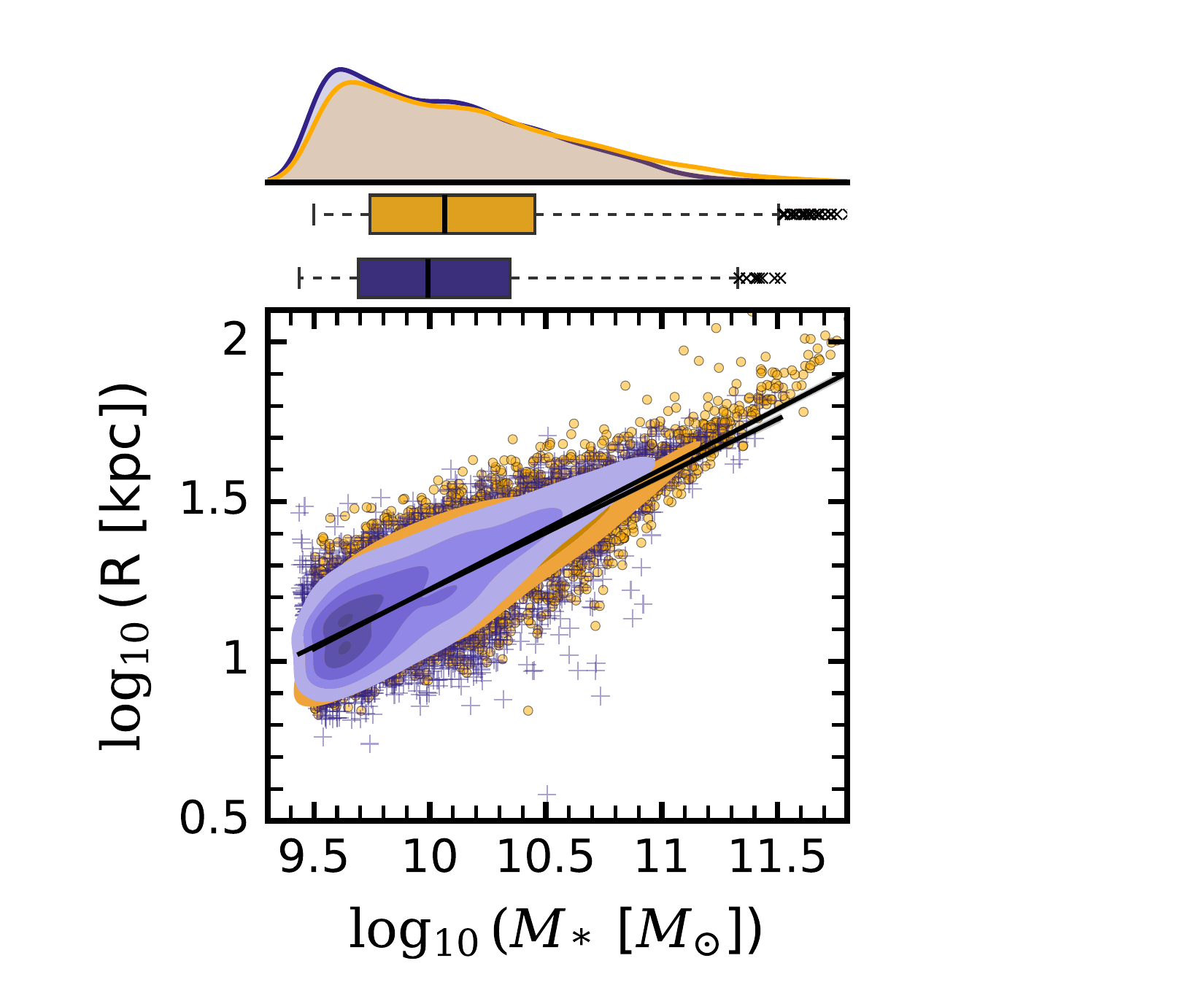 }\hspace{-3.1cm}
 	\includegraphics[width=0.665\columnwidth,angle=0]{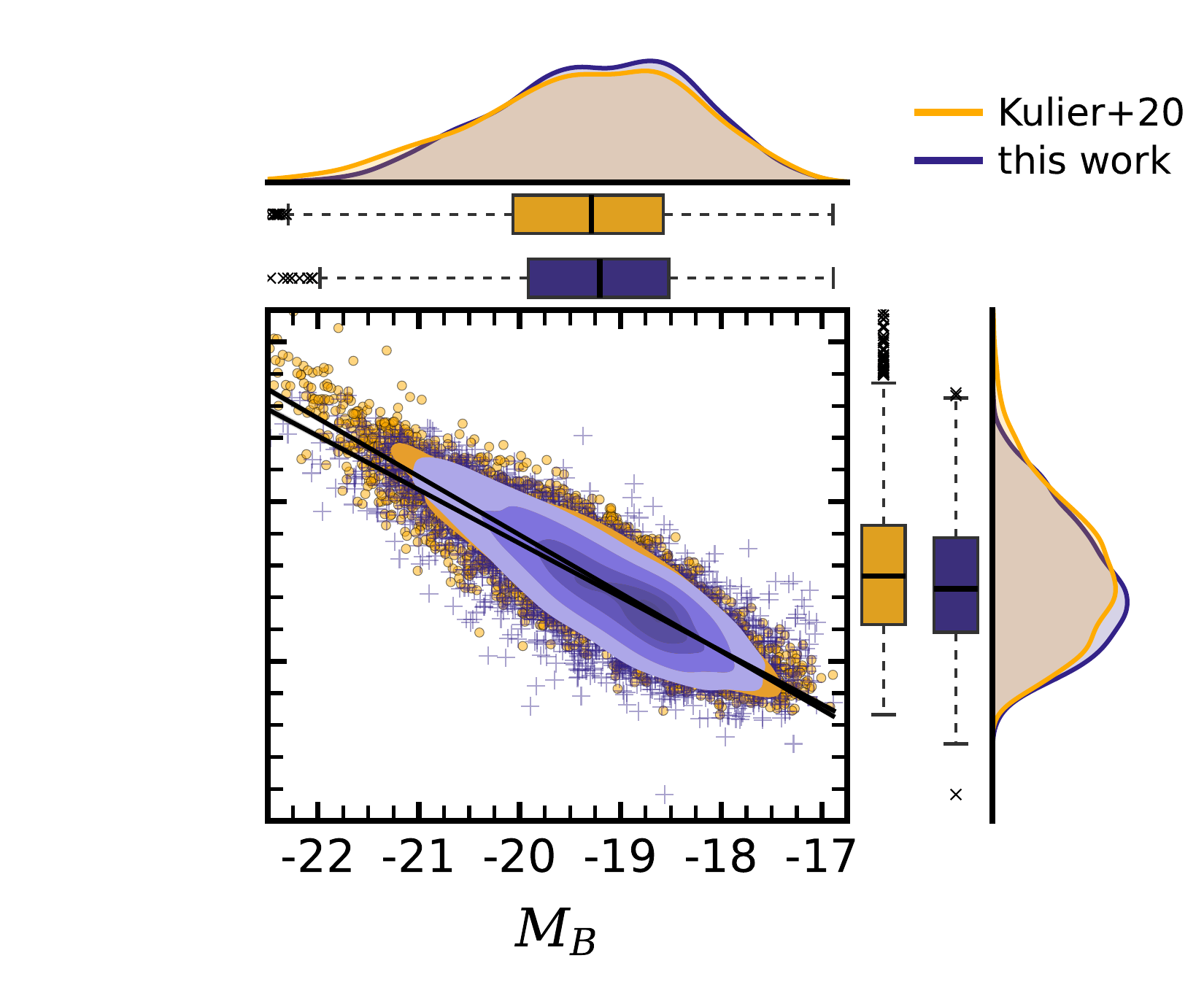}
    \caption{Uncorrected radii (top) and corrected radii (bottom) as a function of \Mstar\ (left) and \MB\ (right) both at $z=0$. Note that, "$R$" denotes for this work the optical radius (\ropt), while for K20, "$R$" denotes the circularised 28 mag/arcsec$^2$ isophotal radius in the $B$-band. The figure utilises the same colour and symbol coding as in \hyperref[fig:SB_vs_Kulier]{\Fig{fig:SB_vs_Kulier}}, and the statistical definitions are as described in \hyperref[sec:sample_envr]{\Sec{sec:sample_envr}}.}\label{fig:ropt_vs_Kulier_corrected}
\end{figure}

We compare our surface brightness density (\SBB) estimates from \hyperref[eq:SB]{\Eq{eq:SB}} with those of K20, who computed \SBB\ for the same \eagle\ simulation at a fixed isophote radius of 28 \magarcs\ at $z=0$ (described in detail in their Sec.\ 2.3). K20's method is computationally intensive, requiring individual particle data, so we aim to develop a faster, computationally simpler method for estimating surface brightness using only radius and \MB, applicable to models without detailed particle information (e.g., semi-analytical models).

In \hyperref[fig:ropt_vs_Kulier_corrected]{\Fig{fig:ropt_vs_Kulier_corrected}}, the upper panels show the distribution of our optical radius, \ropt, as a function of \Mstar\ (left) and \MB\ (right). To correct \ropt\ in our dataset, we align the regression lines in the \ropt-\Mstar\ plane (left panel). This alignment is essentially a rotation of the data by the angle $\theta$ between the regression lines. Using \textit{Ordinary Least Squares} from the Python \texttt{SciPy-statsmodels} package, we calculate the components of the linear equations ($y~=~kx+d$) for the regression lines. The intersection ($x_{\mathrm{inter}}$) and $\theta$ are then determined by expressing the regression lines as vectors, \vec{a} and \vec{b}, and computing $\theta$ with the following formula:

\begin{equation}\label{eq:calc_theta}
      \theta = \arcsin \left (\frac{ |\vec{ab}| }{|\vec{a}||\vec{b}|} 180/\pi \right ), \quad \vec{a} = \begin{pmatrix}  x_{\mathrm{inter}}\\ y_1 \end{pmatrix}, \quad \vec{b} = \begin{pmatrix}  x_{\mathrm{inter}}\\ y_2 \end{pmatrix} 
\end{equation}

We find $\theta = 11^\circ$. To align with the K20 regression line we use the following equation:

\begin{equation}\label{eq:ropt_corr}
    \ropt^{\dagger} = o_{\mathrm{y}} + \sin\theta (p_{\mathrm{x}} - o_{\mathrm{x}}) + \cos\theta (p_{\mathrm{y}} - o_{\mathrm{y}}) 
\end{equation}

Here, \roptc\ is the corrected optical radius, and $p_{\mathrm{x}}$, $p_{\mathrm{y}}$ are the coordinates of our data points in the \ropt--\Mstar\ plane. $o_{\mathrm{x}}$ and $o_{\mathrm{y}}$ serve as the coordinates of the origin of the rotation (point of interception). The results are shown in the bottom panels of \hyperref[fig:ropt_vs_Kulier_corrected]{\Fig{fig:ropt_vs_Kulier_corrected}}.

\section{Statistical tests and tables}\label{sec:stats_tables}

\subsection{Statistical tests}\label{sec:sample_envr_test}

\begin{figure*}
    \centering
    \includegraphics[width=0.75\textwidth,angle=0]{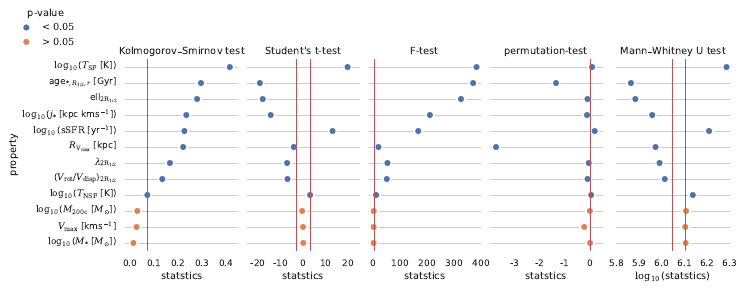}
    \caption{Results of several statistical tests comparing \lsb\ and \hsb\ galaxy samples for selected galaxy and halo properties. Test statistics with $p<0.05$ are marked in blue, indicating statistical significance, while those with $p>0.05$ are shown in orange. The vertical red lines denote the critical thresholds for each test beyond which the null hypothesis is rejected.}\label{fig:stats_test}
\end{figure*}

We performed a series of statistical tests to demonstrate that \lsb\ and \hsb\ galaxies exhibit distinct distributions of properties, and that these differences are statistically significant. In \hyperref[fig:stats_test]{\Fig{fig:stats_test}}, we present the results of these statistical analyses, conducted using the \texttt{Python} package \texttt{SciPy.stats}\footnote{\url{https://docs.scipy.org/doc/scipy/reference/stats.html}} for selected galaxy and halo properties. To ensure a robust comparison, we employed various statistical tests that examine different aspects of the two samples, \lsbs\ and \hsbs. The results, shown from left to right in the figure, correspond to the following tests: the KS-test, the \textit{Student's t-test}, the \textit{F-test}, the \textit{Permutation test}, and the \textit{Mann–Whitney U-test}. Each statistical test conducted in this analysis focuses on specific properties, such as the distributions, variances, medians, or the exchangeability of the two samples being tested against each other.

The properties are ordered based on the results of the KS-test, highlighting the six highest-ranked properties that are not related to our target selection criteria\footnote{Properties tied to our target selection, such as \roptc\ or \SBB, exhibit maximal diversity by design, as these criteria were used to define the sample. Consequently, these properties are excluded.} The x-axis shows the results of the statistical tests as coloured dots, with the colour representing the significance level of the test result ($p$-value). We adopt a standard significance threshold of 5\% ($p=0.05$); values with $p<0.05$ are shown as blue dots, indicating significant results, while values with $p>0.05$ are displayed as orange dots, suggesting non-significant results. The vertical red lines denote the critical values for each test. For these thresholds, we use a stricter significance level of $\alpha=0.001$, corresponding to a 99.9\% confidence level.

\subsection{Tables}\label{sec:tables}

\hyperref[tab:dens6b_new_SB]{\Tab{tab:dens6b_new_SB}} provides galaxy and halo properties analysed in this work for the parent sample of the \eagle\ simulation, as well as for the \lsb\ and \hsb\ populations.

\begin{table*}
	\begin{center}
    \caption{Median values (within $25^\mathrm{th}$ and $75^\mathrm{th}$ percentiles) and mean values (with $10^{\mathrm{th}}$–$90^{\mathrm{th}}$ confidence interval and standard deviation) for key galaxy and halo properties of our unbiased parent sample, as well as the \lsb\ and \hsb\ populations at $z=0$.}\vspace{-0.4cm}
	\setlength{\tabcolsep}{2pt}
     \setlength\extrarowheight{-1.5pt}
		\begin{tabular}{M{0.22\textwidth}||M{0.09\textwidth}|M{0.09\textwidth}|M{0.05\textwidth}||M{0.09\textwidth}|M{0.09\textwidth}|M{0.05\textwidth}||M{0.09\textwidth}|M{0.09\textwidth}|M{0.05\textwidth}} 
& \multicolumn{3}{c||}{\multirow{1}{*}{\parbox{0.25\linewidth}{\centering EAGLE $N_{\text{gal}}$: 3190, 100.0\%}}} & \multicolumn{3}{c||}{\multirow{1}{*}{\parbox{0.25\linewidth}{\centering HSB $N_{\text{gal}}$: 1595, 50.0\%}}} & \multicolumn{3}{c}{\multirow{1}{*}{\parbox{0.25\linewidth}{\centering LSB $N_{\text{gal}}$: 1595, 50.0\%}}}\\\hline
property & median  & mean  & std & median & mean & std & median  & mean  & std \\\hline
\hline
$\log_{10}(M_{\mathrm{200c}}$ [$M_{\odot}$])&	11.59$_{-0.18}^{+0.28}$	 & 	11.68$_{-0.38}^{+0.40}$	 & 	0.006	&	11.59$_{-0.17}^{+0.28}$	 & 	11.68$_{-0.36}^{+0.39}$	 & 	0.008	&	11.59$_{-0.18}^{+0.28}$	 & 	11.69$_{-0.38}^{+0.41}$	 & 	0.009	\\\hline
$\log_{10}(M_{\mathrm{*}}$ [$M_{\odot}$])&	9.76$_{-0.28}^{+0.40}$	 & 	9.83$_{-0.45}^{+0.46}$	 & 	0.005	&	9.76$_{-0.29}^{+0.39}$	 & 	9.83$_{-0.45}^{+0.47}$	 & 	0.007	&	9.75$_{-0.28}^{+0.40}$	 & 	9.83$_{-0.45}^{+0.47}$	 & 	0.007	\\\hline
c$_{\mathrm{NFW}}$&	8.72$_{-0.64}^{+0.43}$	 & 	8.54$_{-0.83}^{+0.87}$	 & 	0.012	&	8.71$_{-0.64}^{+0.41}$	 & 	8.54$_{-0.81}^{+0.86}$	 & 	0.016	&	8.72$_{-0.64}^{+0.44}$	 & 	8.53$_{-0.84}^{+0.90}$	 & 	0.017	\\\hline
$R_{\mathrm{200c}}$ [kpc]&	153.59$_{-19.49}^{+36.89}$	 & 	173.69$_{-60.37}^{+65.70}$	 & 	1.609	&	153.67$_{-18.59}^{+36.96}$	 & 	172.62$_{-56.23}^{+62.55}$	 & 	1.927	&	153.32$_{-19.91}^{+36.93}$	 & 	174.75$_{-62.30}^{+70.50}$	 & 	2.490	\\\hline
\rvmax\ [kpc]&	22.09$_{-6.31}^{+7.54}$	 & 	26.91$_{-22.20}^{+28.18}$	 & 	1.817	&	19.34$_{-6.20}^{+7.81}$	 & 	25.04$_{-22.41}^{+30.33}$	 & 	2.446	&	24.22$_{-5.77}^{+7.39}$	 & 	28.79$_{-19.15}^{+28.27}$	 & 	2.776	\\\hline
\vmax\ [kms$^{-1}$]&	124.95$_{-16.34}^{+28.66}$	 & 	136.95$_{-41.13}^{+44.29}$	 & 	0.957	&	124.55$_{-16.66}^{+29.89}$	 & 	136.83$_{-40.03}^{+43.80}$	 & 	1.138	&	125.36$_{-16.16}^{+27.03}$	 & 	137.07$_{-40.90}^{+45.96}$	 & 	1.546	\\\hline
$\sigma$ [kms$^{-1}$]&	60.91$_{-9.57}^{+18.10}$	 & 	68.58$_{-23.71}^{+25.29}$	 & 	0.472	&	60.25$_{-9.36}^{+18.21}$	 & 	68.11$_{-23.24}^{+25.52}$	 & 	0.692	&	61.46$_{-9.70}^{+17.72}$	 & 	69.06$_{-23.48}^{+25.64}$	 & 	0.654	\\\hline
$\log_{10}(M_{\mathrm{BH}}$ [$M_{\odot}$])&	6.09$_{-0.26}^{+0.48}$	 & 	6.31$_{-0.66}^{+0.69}$	 & 	0.010	&	6.09$_{-0.25}^{+0.47}$	 & 	6.30$_{-0.62}^{+0.66}$	 & 	0.013	&	6.09$_{-0.27}^{+0.48}$	 & 	6.32$_{-0.67}^{+0.73}$	 & 	0.015	\\\hline
\rhalfstars\ [kpc]&	4.44$_{-0.78}^{+1.11}$	 & 	4.72$_{-1.40}^{+1.48}$	 & 	0.023	&	3.83$_{-0.51}^{+0.78}$	 & 	4.09$_{-1.07}^{+1.16}$	 & 	0.029	&	5.10$_{-0.80}^{+1.08}$	 & 	5.36$_{-1.40}^{+1.50}$	 & 	0.031	\\\hline
\roptc\ [kpc]&	13.90$_{-3.61}^{+6.45}$	 & 	16.32$_{-7.76}^{+8.15}$	 & 	0.119	&	11.60$_{-2.44}^{+4.88}$	 & 	13.83$_{-6.50}^{+7.10}$	 & 	0.183	&	16.91$_{-4.51}^{+7.01}$	 & 	18.81$_{-7.98}^{+8.50}$	 & 	0.157	\\\hline
\agest\ [Gyr]&	5.66$_{-1.42}^{+1.47}$	 & 	5.77$_{-2.03}^{+2.11}$	 & 	0.025	&	4.88$_{-1.34}^{+1.46}$	 & 	5.09$_{-1.93}^{+2.05}$	 & 	0.037	&	6.34$_{-1.30}^{+1.28}$	 & 	6.45$_{-1.87}^{+1.98}$	 & 	0.033	\\\hline
$n_{\text{S\'ersic}}$&	1.53$_{-0.26}^{+0.42}$	 & 	1.71$_{-0.68}^{+0.80}$	 & 	0.037	&	1.45$_{-0.21}^{+0.29}$	 & 	1.55$_{-0.44}^{+0.49}$	 & 	0.015	&	1.67$_{-0.35}^{+0.50}$	 & 	1.87$_{-0.81}^{+1.00}$	 & 	0.057	\\\hline
$\log_{10}(M_{\mathrm{H}_\mathrm{I}}$ [$M_{\odot}$])&	9.16$_{-0.37}^{+0.27}$	 & 	9.05$_{-0.56}^{+0.69}$	 & 	0.040	&	9.07$_{-0.39}^{+0.29}$	 & 	8.98$_{-0.51}^{+0.56}$	 & 	0.016	&	9.24$_{-0.34}^{+0.24}$	 & 	9.12$_{-0.58}^{+0.81}$	 & 	0.072	\\\hline
$\log_{10}(M_{\mathrm{H}_2}$ [$M_{\odot}$])&	8.59$_{-0.22}^{+0.29}$	 & 	8.60$_{-0.54}^{+0.81}$	 & 	0.083	&	8.62$_{-0.21}^{+0.30}$	 & 	8.65$_{-0.45}^{+0.69}$	 & 	0.073	&	8.57$_{-0.24}^{+0.28}$	 & 	8.55$_{-0.52}^{+0.97}$	 & 	0.136	\\\hline
(D/T)$_{\mathrm{gas}}$&	0.47$_{-0.09}^{+0.11}$	 & 	0.49$_{-0.16}^{+0.17}$	 & 	0.003	&	0.44$_{-0.09}^{+0.11}$	 & 	0.47$_{-0.16}^{+0.17}$	 & 	0.004	&	0.48$_{-0.08}^{+0.11}$	 & 	0.52$_{-0.15}^{+0.16}$	 & 	0.004	\\\hline
(D/T)$_{\mathrm{*}}$&	0.31$_{-0.09}^{+0.10}$	 & 	0.32$_{-0.13}^{+0.13}$	 & 	0.001	&	0.29$_{-0.08}^{+0.09}$	 & 	0.31$_{-0.12}^{+0.12}$	 & 	0.002	&	0.32$_{-0.10}^{+0.10}$	 & 	0.33$_{-0.13}^{+0.14}$	 & 	0.002	\\\hline
$\log_{10}(\text{SFR}$ [$M_{\odot}$ yr$^{-1}$])&	-0.40$_{-0.25}^{+0.30}$	 & 	-0.38$_{-0.40}^{+0.42}$	 & 	0.006	&	-0.35$_{-0.24}^{+0.31}$	 & 	-0.31$_{-0.39}^{+0.41}$	 & 	0.009	&	-0.47$_{-0.25}^{+0.31}$	 & 	-0.45$_{-0.39}^{+0.42}$	 & 	0.008	\\\hline
$\log_{10}(\text{sSFR}$ [yr$^{-1}$])&	-10.12$_{-0.18}^{+0.17}$	 & 	-10.18$_{-0.35}^{+0.39}$	 & 	0.011	&	-10.06$_{-0.19}^{+0.16}$	 & 	-10.11$_{-0.31}^{+0.36}$	 & 	0.015	&	-10.18$_{-0.16}^{+0.16}$	 & 	-10.25$_{-0.37}^{+0.42}$	 & 	0.016	\\\hline
$\log_{10}(j_{*}$ [kpc kms$^{-1}$])&	3.36$_{-0.17}^{+0.17}$	 & 	3.36$_{-0.24}^{+0.25}$	 & 	0.003	&	3.29$_{-0.16}^{+0.16}$	 & 	3.30$_{-0.23}^{+0.25}$	 & 	0.004	&	3.42$_{-0.16}^{+0.15}$	 & 	3.42$_{-0.23}^{+0.24}$	 & 	0.005	\\\hline
$\log_{10}(j_{\mathrm{gas}}$ [kpc kms$^{-1}$])&	-2.30$_{-0.16}^{+0.17}$	 & 	-2.28$_{-0.23}^{+0.25}$	 & 	0.004	&	-2.36$_{-0.15}^{+0.16}$	 & 	-2.34$_{-0.22}^{+0.25}$	 & 	0.007	&	-2.25$_{-0.15}^{+0.18}$	 & 	-2.23$_{-0.23}^{+0.24}$	 & 	0.004	\\\hline
$(v_{\mathrm{rot}}/\sigma)_{2\rhalfstars}$&	0.83$_{-0.31}^{+0.29}$	 & 	0.84$_{-0.39}^{+0.41}$	 & 	0.005	&	0.76$_{-0.26}^{+0.27}$	 & 	0.79$_{-0.35}^{+0.37}$	 & 	0.006	&	0.90$_{-0.35}^{+0.30}$	 & 	0.89$_{-0.42}^{+0.44}$	 & 	0.007	\\\hline
\ellTwost\ &	0.42$_{-0.13}^{+0.12}$	 & 	0.41$_{-0.16}^{+0.17}$	 & 	0.002	&	0.37$_{-0.12}^{+0.11}$	 & 	0.36$_{-0.15}^{+0.16}$	 & 	0.002	&	0.49$_{-0.14}^{+0.11}$	 & 	0.47$_{-0.16}^{+0.17}$	 & 	0.003	\\\hline
\lamTwost\ &	0.64$_{-0.19}^{+0.13}$	 & 	0.60$_{-0.20}^{+0.21}$	 & 	0.002	&	0.60$_{-0.17}^{+0.14}$	 & 	0.57$_{-0.19}^{+0.19}$	 & 	0.003	&	0.69$_{-0.21}^{+0.11}$	 & 	0.63$_{-0.21}^{+0.22}$	 & 	0.003	\\\hline
$t_{50\%,\mathrm{BH}}$ [Gyr]&	7.28$_{-1.90}^{+1.93}$	 & 	7.29$_{-2.48}^{+2.56}$	 & 	0.025	&	6.94$_{-1.87}^{+1.88}$	 & 	6.97$_{-2.43}^{+2.54}$	 & 	0.035	&	7.62$_{-1.74}^{+1.90}$	 & 	7.61$_{-2.45}^{+2.57}$	 & 	0.036	\\\hline
$t_{50\%,\mathrm{halo}}$ [Gyr]&	9.40$_{-1.17}^{+0.77}$	 & 	9.03$_{-1.61}^{+1.71}$	 & 	0.031	&	9.21$_{-1.25}^{+0.90}$	 & 	8.88$_{-1.61}^{+1.74}$	 & 	0.039	&	9.52$_{-1.01}^{+0.76}$	 & 	9.19$_{-1.56}^{+1.71}$	 & 	0.047	\\\hline
$t_{50\%,*}$ [Gyr]&	7.15$_{-1.33}^{+1.15}$	 & 	7.03$_{-1.78}^{+1.85}$	 & 	0.023	&	6.94$_{-1.44}^{+1.15}$	 & 	6.77$_{-1.80}^{+1.90}$	 & 	0.031	&	7.35$_{-1.14}^{+1.11}$	 & 	7.28$_{-1.68}^{+1.79}$	 & 	0.033	\\\hline
$\log_{10}(T_{\mathrm{SF}}$ [K])&	3.79$_{-0.05}^{+0.08}$	 & 	3.82$_{-0.11}^{+0.14}$	 & 	0.007	&	3.84$_{-0.06}^{+0.08}$	 & 	3.86$_{-0.11}^{+0.13}$	 & 	0.007	&	3.75$_{-0.03}^{+0.05}$	 & 	3.78$_{-0.10}^{+0.14}$	 & 	0.012	\\\hline
$\log_{10}(T_{\mathrm{NSF}}$ [K])&	5.21$_{-0.21}^{+0.29}$	 & 	5.27$_{-0.41}^{+0.43}$	 & 	0.006	&	5.24$_{-0.22}^{+0.28}$	 & 	5.29$_{-0.39}^{+0.41}$	 & 	0.008	&	5.18$_{-0.21}^{+0.30}$	 & 	5.25$_{-0.42}^{+0.45}$	 & 	0.009	\\
\hline
			 (i)		 & (ii)	 & (iii)	 & (iv)	  & (v) 	 & 	  (vi) 	  &  	 (vii) 	  & 	  (viii) 	& (ix) & (x) \\
		\end{tabular}
		\label{tab:dens6b_new_SB}
	\end{center}
\end{table*}

\end{document}